\documentclass[preprint]{aastex}

\usepackage{graphicx}        % For eps figures, newer & more powerfull
\usepackage{natbib}         % For citations: redefine \cite commands
\usepackage{amssymb}        % useful mathematical symbols
\usepackage{amsmath}
\usepackage{color}           % For color text: \color command
\usepackage{url}             % For breaking URLs easily trough lines
\usepackage{epstopdf}
\newcommand{\cz}{\rm cz}
\newcommand{\rhobar}{\bar{\rho}}
\newcommand{\tc}{\rm tc}
\newcommand{\ES}{\rm ES}
\newcommand{\sd}{\rm sd}
\newcommand{\qs}{\rm qs}

\begin{document}

\title{Spin-down dynamics of magnetized solar-type stars}
\author{\sc R. L. F. Oglethorpe$^1$ \& P. Garaud$^2$}
\date{}
\affil{$^1$Department of Applied Mathematics and Theoretical Physics, Centre for Mathematical Sciences, University of Cambridge, Wilberforce Road, Cambridge CB3 0WA, UK \\
$^2$Department of Applied Mathematics and Statistics, Baskin School of Engineering, University of California Santa Cruz, 1156 High Street, Santa Cruz, CA 95064, USA}

\begin{abstract}
It has long been known that solar-type stars undergo significant spin-down, via magnetic braking, during their Main-Sequence lifetimes.  However, magnetic braking only operates on the surface layers; it is not yet completely understood how angular momentum is transported within the star, and how rapidly the spin-down information is communicated to the deep interior.  In this work,  we use insight from recent progress in understanding internal solar dynamics to model the interior of other solar-type stars.  We assume, following Gough and McIntyre (1998), that the bulk of the radiation zone of these stars is held in uniform rotation by the presence of an embedded large-scale primordial field, confined below a stably-stratified, magnetic-free tachocline by large-scale meridional flows downwelling from the convection zone. We derive simple equations to describe the response of this model interior to spin-down of the surface layers, that are identical to the two-zone model of MacGregor and Brenner (1991), with a coupling timescale proportional to the local Eddington-Sweet timescale across the tachocline. This timescale depends both on the rotation rate of the star and on the thickness of the tachocline, and can vary from a few hundred thousand years to a few Gyr, depending on stellar properties. Qualitative predictions of the model appear to be consistent with observations, although depend sensitively on the assumed functional dependence of the tachocline thickness on the stellar rotation rate.  
 \end{abstract}

\keywords{ MHD --- Sun:~interior --- Sun:~magnetic fields --- Sun:~rotation }

\maketitle

\section{INTRODUCTION}
     \label{sec:introduction} 

It has long been known that there exists a remarkable dichotomy between stars of masses $M_\star > 1.5M_\odot$, which typically always remain rapid rotators, and stars of masses $M_\star < 1.5M_\odot$, which undergo significant spin-down during their Main-Sequence lifetimes. This dichotomy was first resolved by \citet{Schatzman62} (see also Schatzman 1959) who noted that the transition coincides with the disappearance of the outer convection zone with increasing stellar mass, and deduced that the latter must play an important role in the spin-down process. He argued, following \citet{Parker55}, that convection is necessary for dynamo action, and that magnetic activity drives the ejection of mass from the surface of the star into the interstellar medium. He then estimated the amount of angular momentum lost by the star assuming that the ejecta is forced to corotate with the surface magnetic field until a point where the field strength is no longer strong enough to act on the plasma. He found that for reasonable field strengths this point was sufficiently far out from the star that even a tiny amount of mass-loss could lead to very significant angular-momentum loss. The theory of stellar spin-down via magnetic braking was born.

Schatzman's original calculation has been refined further over the past five decades, following two parallel lines of investigation. The majority of the effort has been dedicated to improving stellar wind models, and the manner in which they act on the stellar surface \citep{WeberDavis67,Mestel68,Li99}. Questions of interest include how the spin-down process is affected by the strength and geometry of the stellar wind and how the latter depend on the star's mass and rotation rate \citep{Aibeoal07,Mattal12,ReinersMohanty12}.

However, magnetic braking as currently understood only operates on the surface layers of the star. A second line of investigation has therefore focused on modeling angular-momentum transport within the star, to see how the spin-down information is communicated to the deep interior. Unfortunately, and despite decades of research, only little is known about the subject. Part of the problem resides in the fact that the internal rotation profile of Main Sequence solar-type stars (the Sun excepted) remains extremely difficult to observe. 

Nevertheless, useful information can still be learned about the process by comparing simple idealized models with observations. Since turbulent convection can redistribute angular momentum across an entire convection zone within a matter of weeks to a few years (depending on the stellar type), the outer convective region is usually assumed to rotate more-or-less uniformly. Within the radiation zone, one can either assume that each layer individually conserves angular momentum \citep{Kawaler88}, or that the entire region is in a state of uniform rotation. In the latter scenario, an additional assumption must be made to determine how angular momentum is transported from the radiation zone to the convection zone. 

Following this idea,  \citet{MacGregorBrenner91} proposed a ``two-zone'' model, now commonly used in statistical comparisons of models with observations. The model assumes that both radiation and convection zones are rotating uniformly with rotation rates $\Omega_{\rm core}$ and $\Omega_{\rm cz}$ respectively, and that the angular-momentum redistribution between the two regions occurs on a coupling timescale $\tau_c$. The system then evolves according to the following coupled ordinary differential equations:
\begin{eqnarray}
\frac{{\rm d}J_{\rm core}}{{\rm d} t} = - \frac{\Delta J}{\tau_c}, \nonumber \\
\frac{{\rm d}J_{\rm cz}}{{\rm d} t} + \frac{{\rm d}J_{\rm core}}{{\rm d} t}  =  - \dot{J}_{\rm w}, \nonumber \\
\mbox{  where  } \Delta J = \frac{ I_{\rm cz} I_{\rm core} }{I_{\rm core} + I_{\rm cz}} (\Omega_{\rm core} - \Omega_{\rm cz}),
\label{eq:mb91}
\end{eqnarray}
where $J$ represents the angular momentum and $I$ the moment of inertia (such that $J = I \Omega$) of each region considered and $\dot{J}_{\rm w}$ is the rate of angular-momentum extraction from the entire star by the stellar wind. The model allows for analytical solutions in certain limits, and is very easy to integrate numerically in conjunction with stellar evolution. The real difficulty, if one wishes to use it for quantitative purposes, is to express $\tau_c$ and $\dot{J}_{\rm w}$ as functions of known stellar properties. 

Qualitatively speaking, however, the model provides a simple way of studying the difference in rotational evolution between stars in solid-body rotation and stars which can retain significant differential rotation between the core and the envelope, by suitably choosing $\tau_c$. With strong coupling (small $\tau_c$), the region effectively being spun down by the wind encompasses the entire star. By contrast, with weak coupling the region being spun down is at first limited to the convection zone, while the radiation zone only feels the effect of spin-down later, when $\Delta J$ has grown to be sufficiently large. Hence, for the same angular-momentum extraction rate, the apparent spin-down rate of the surface layers is at first much slower for solid-body rotators than for differential rotators. This model then predicts dramatically different rotational histories in the two cases, a prediction that can and has been tested against observations to estimate $\tau_c$ \citep{Allain98,Irwinal07,Denissenkoval10,Spadaal11,GalletBouvier13}.

Before we proceed to discuss these results, first note that stellar rotation rates in a given very young cluster and a given mass bin, usually exhibit a significant spread, with ``rapid rotators'' rotating up to a few tens of times the speed of the slower rotators \citep{Herbstal01,Lammal05,Irwinal08}. This spread is usually attributed to a spread in the initial pre-stellar core conditions, and the length of the initial pre-Main Sequence disk-locking phase \citep{Bouvieral97}. The spread of rotational velocities in these ``initial conditions'' then 
propagates to later ages, with rapid rotators and slow rotators in each mass bin having distinct evolutionary paths. 

By comparing observations to the predictions of a two-zone model, \citet{Irwinal07} found that the rotational periods of rapid rotators are well-explained by assuming solid-body rotation at all time, for any mass bin. By contrast, the solid-body rotation assumption is not consistent with observations for slow-rotators in the mass range $0.7M_\odot-1.1 M_\odot$. A coupling time $\tau_c$ as large as a Gyr appears to fit the data much better \citep[see also][]{Allain98}. This suggests that $\tau_c$ must depend on multiple factors, such as the mass and rotation rate of the star. Their conclusion was confirmed by subsequent work by \citet{Denissenkoval10}, \citet{Spadaal11} and \citet{GalletBouvier13}. Note that the rotation rates of low-mass stars ($M_\star < 0.7 M_\odot$), by contrast, are fairly well-explained by solid-body rotation. This is not surprising since the latter are fully convective for $M_\star<0.35M_\odot$ and nearly fully convective for $0.35M_\odot < M_\star < 0.7 M_\odot$ \citep[see also][for further work on the rotational histories of very low-mass stars in older clusters]{ReinersMohanty12}. 

Going beyond the two-zone model, however, and understanding from a more physical point of view what the source of the dynamical coupling between the radiative and convective regions is, and whether the radiation zone is indeed in uniform rotation, requires peering below the surface. Today, this can only be done for a single Main-Sequence star, the Sun\footnote{The recent asteroseismic detection of differential rotation in red-giants by \citet{Beckal12}, \citet{Mosseral12} and \citet{Deheuvelsal12} is an exciting development that may help test models of angular-momentum transport in later stages of stellar evolution.}. Our goal in this work is to use insight from recent progress in understanding internal solar dynamics to model the transport of angular momentum in the interior of other stars. This, of course, implicitly assumes that the Sun is representative of all solar-type stars, and that its internal workings are not fundamentally different from theirs. Comparing the predictions of any model based on this assumption with observations may, to some extent, help us establish whether it is valid or not. 
 
Thanks to helioseismology, we now have a good view of the large-scale internal dynamics of the Sun. Its outer convective region spans roughly a third of the solar radius and 2.5\% of the solar mass. 
It is rotating differentially, with an equatorial rotation rate $\Omega_{\rm eq} \simeq 2.9 \times 10^{-6}$s$^{-1}$, and a polar rotation rate about 25\% slower \citep{Thompson-etal96,Schou-etal98}. Meanwhile, the radiative interior is rotating uniformly, with an angular velocity similar to that of the surface at mid-latitudes. The shear layer between the two regions is called the tachocline. It is remarkably thin, with a thickness estimated at 2-4\% of the solar radius \citep{Kosovichev-etal97,Charbonneaual99,ElliottGough99}. Further to the question of how the spin-down of the surface layers is communicated to the interior, these observations raise new puzzles: why is the radiation zone rotating uniformly despite the latitudinal shear imposed by the overlying convection zone; and why is the tachocline so thin? Answering these questions in conjunction with the spin-down problem prompts a closer inspection of the various angular-momentum transport processes thought to take place in the Sun (and by proxy, in all solar-type stars).

Angular-momentum transporters in the solar radiation zone can be split into two categories: hydrodynamic processes and magnetohydrodynamic (MHD) processes. The former include stratified turbulence, large-scale meridional flows, and gravity waves. Purely hydrodynamic models of the solar interior based on all three types of processes, in isolation or in combination, have been studied at length \citep{SpiegelZahn92,Elliott97,Talonal02,CharbonnelTalon05,RogersGlatzmaier06,Brunal11,WoodBrummell12}. As discussed by \citet{McIntyre94}, \citet{GoughMcIntyre98}, \citet{Gough07} and \citet{Zahn07}, however, these models remain problematic. On the one hand, they tend to have difficulties explaining how to {\it maintain} uniform rotation, since they often rely on some level of differential rotation to be effective. This is particularly true of turbulent transport, and transport by large-scale flows. On the other hand, they also have difficulty explaining how the gradual extraction of angular momentum from the radiation zone during spin-down can proceed without concurrently inducing significant compositional mixing throughout the interior, which would be inconsistent with helioseismic inversions \citep{GoughKosovichev90}. 

MHD processes, which do not suffer from the same limitations, have recently gained popularity as a means to explain helioseismic and related observations of the dynamics of the solar interior \citep[see the review by][]{Garaud07}. \citet{MestelWeiss87} first discussed how the presence of a large-scale primordial field confined to the radiation zone would rapidly suppress any differential rotation within that region. Indeed, the magnetic diffusivity is low in conditions relevant for stellar interiors. If, in addition, meridional flows are weak (which is true in strongly stratified regions), then Ferraro's isorotation law \citep{Ferraro37} applies, which states that angular velocity must be constant on magnetic field lines. Although Ferraro's law does not preclude differential rotation entirely since different field lines could, in principle, rotate at different speeds, \citet{MestelWeiss87} argued that interactions between Alfv\'en waves along neighboring field lines, through a process called phase mixing, should rapidly suppress any remaining shear. 

\citet{CharbonneauMacGregor93} later used this idea to study the effect of spin-down in the solar interior. They considered a model similar to the one of  \citet{MestelWeiss87}, in which the Sun is threaded by a fixed poloidal field, confined beneath a fixed radius $r_f$. They assumed that the convection zone is gradually spun down by magnetic braking, while remaining at all time in a state of uniform rotation. They then studied the role of the magnetic field in promoting angular-momentum transport, both between the two zones and throughout the radiative interior, by solving simultaneously the axisymmetric azimuthal components of the momentum and magnetic induction equations. In their model, which ignores phase mixing and any meridional flows, (turbulent) viscosity had to be invoked to damp any remaining differential rotation across field lines, and to promote angular-momentum transport in regions where there is no field. 

They found that if the poloidal field has significant overlap with the convection zone (i.e. if $r_f > r_{\rm cz}$, where $r_{\rm cz}$ is the radius of the radiative--convective interface), then the spin-down is very rapidly communicated to the interior. For large enough viscosity, the entire star rotates more-or-less as a solid body at all times\footnote{By assuming that the convection zone is uniformly rotating, \citet{CharbonneauMacGregor93} avoided any complications related to the interaction between the primordial field and the convection zone shear, later studied by \citet{MacGregorCharbonneau99} for instance. In this sense, this model only treats the spin-down problem, but not that of the existence and maintenance of the thin solar tachocline.}. On the other hand, if the poloidal field is confined strictly below the convection zone (i.e. if $r_f < r_{\rm cz}$), then viscosity is also needed to communicate the spin-down from the convection zone to the magnetically-dominated part of the radiation zone. \citet{CharbonneauMacGregor93} showed that the system eventually evolves toward a quasi-steady state where both the convection zone and the deep interior are rotating uniformly, but at distinct angular velocities with $\Omega_{\rm core} > \Omega_{\rm cz}$. A shear layer separates the two zones, with a thickness that depends on $r_{\rm cz} - r_f$. In this quasi-steady state, the total viscous angular-momentum flux across this shear layer is equal to that extracted by the wind from the surface, and fixes the relative core-envelope lag $(\Omega_{\rm core}-\Omega_{\rm cz}) / \Omega_{\rm cz}$. 

These ideas have recently been revisited by \citet{Denissenkov10}, who, in addition to the question of spin-down, also attempted to address the issue of the solar tachocline and of the light-element abundances at the same time. Starting from the model proposed by \citet{CharbonneauMacGregor93} but ultimately allowing for a differentially rotating convection zone, he first argued that the high value of the viscosity 
needed to explain the uniform rotation of the radiation zone (to ensure that all field lines are rotating at the same rate) must be of turbulent origin. He then assumed that this turbulence would also transport chemical species at the same rate, and showed that this leads to inconsistent predictions for the surface Li and Be abundance in the Sun. To resolve the problem, he then invoked the work of \citet{SpiegelZahn92} and \citet{Zahn92} to argue that turbulence in the radiation zone must be highly anisotropic, redistributing angular momentum rapidly in the horizontal direction but very slowly in the radial direction -- and similarly for chemical species. Under this assumption, and using a related prescription for the turbulent transport coefficients, he was able to explain simultaneously both the uniform rotation of the solar interior, the thin solar tachocline and the surface light-element abundances.

However, while compelling in its ability to reproduce observations, this model suffers from a number of inconsistencies. First, note that as long as the field is confined beneath the convection zone, phase mixing is likely to drive the radiation zone towards uniform rotation without the need to invoke turbulence, and thus without causing significant concurrent compositional mixing. The added anisotropic turbulent transport central to the work of \citet{Denissenkov10} should thus be viewed a ``rapid fix'' for a problem that does not necessarily exist. Second, the fix is itself inconsistent, since it uses a prescription for turbulent transport that was created for a purely hydrodynamic system, in an environment dominated by magnetic fields. As shown by \citet{TobiasDiamondHughes07}, the turbulent viscosity prescription of \citet{SpiegelZahn92} is unlikely to apply in any circumstance. Indeed, in a purely hydrodynamic setting, strongly stratified turbulence drives a system away from, rather than towards uniform rotation. In the presence of a weak magnetic field, \citet{TobiasDiamondHughes07} showed that the fluid motions rapidly evolve into a state where Reynolds stresses and Maxwell stresses cancel out, and where the efficacy of turbulent angular-momentum transport is strongly quenched. Finally, note that all of these models \citep{MestelWeiss87,CharbonneauMacGregor93,Denissenkov10} assume the existence of a confined poloidal field, but do not explain how confinement is maintained. As it turns out, it is only by answering the fundamental question of magnetic confinement that new light can be shed on the problem. 

\citet{GoughMcIntyre98} (GM98 hereafter) were the first to put forward a global and self-consistent theory of the large-scale rotational dynamics of the solar interior, albeit in a steady state (that is, without spin-down). They first addressed the magnetic confinement question. They argued that large-scale meridional flows must be driven by gyroscopic pumping from the differentially-rotating convection zone down into the radiation zone, and that by pushing on the magnetic field lines they can confine the primordial field strictly below the base of the convection zone. The solar tachocline thus emerges as a ``magnetic-free" region (to be precise, a region where the Lorentz force is insignificant), lying between the base of the convection zone and the top of the magnetically-dominated, uniformly rotating radiative interior. GM98 then studied the tachocline dynamics more quantitatively, and estimated that these flows could transport angular momentum and chemical species across the tachocline on a local Eddington-Sweet timescale, which is a few Myr in the present-day Sun. The flows, however, do not penetrate below the base of the tachocline, hence satisfying observations of the light-element abundances. Angular-momentum extraction by the tachocline from the deeper regions is done by magnetic stresses, through a very thin magnetic boundary layer, now known as the tachopause (see Figure 1 for detail). 

Gough \& McIntyre's vision of the solar interior was recently confirmed by semi-analytical calculations \citep{WoodMcIntyre11,Woodal11} and by numerical simulations \citep{AAal13}. Given its success in explaining (at least qualitatively so far) existing observations for the Sun, we take the natural next step by assuming a similar structure for all solar-type stars, and studying its response to spin-down. 

Our complete model setup is presented in Section \ref{sec:model}. Our goal is to derive an analytical or semi-analytical description of the spin-down problem, and study how the latter varies with stellar parameters. To do so, we are forced to abandon the spherical geometry and model the star as a cylinder. This approximation is introduced and discussed in Section \ref{sec:model} and then tested on a simplified system for which analytical solutions exist for both cylindrical {\it and} spherical geometry, in Section \ref{sec:unstrat1}. The derivations presented in that Section also serve the pedagogical role of introducing our methodology. 

In Section \ref{sec:strat1}, we first consider the spin-down of a non-magnetic solar-type star. Our results recover many aspects of \citet{SpiegelZahn92}. Including the effects of a primordial magnetic field following the GM98 model requires modeling the transport of angular momentum out of the deep interior, across the tachopause and the tachocline. This is done in Section \ref{sec:blcase}. Finally,  in Section \ref{sec:discuss}, we re-interpret our findings in view of their application to stellar evolution, casting our mathematical results in a more astrophysical light, and discussing their limitations and possible implications in explaining observations. This final Section is written in such a way that it may be read independently of Sections \ref{sec:unstrat1}--\ref{sec:blcase}, for readers who, in a first pass, are primarily interested in the results rather than their derivation. 

\section{THE MODEL} \label{sec:model}

The principal difficulty involved in extending the GM98 model to study its behavior under spin-down is the nonlinear nature of the equations of motion and of the magnetic induction equation. In fact, following GM98, we do not even attempt to address it: the full nonlinear problem is so complex that it cannot be treated exactly analytically \citep[although see][for a first nonlinear solution of a simplified version of the GM98 model]{WoodMcIntyre11}. Instead, we shall make a number of assumptions and order-of-magnitude estimates to model the effects of the nonlinearites. These will be introduced and discussed in detail as they arise. 

A second difficulty lies in the geometry of the system. Since gravity (the vertical axis) is not parallel to the rotation axis in a star (except at the poles), even the restricted hydrodynamic linearized version of the governing equations does not usually have a simple analytical solution. Indeed, the misalignment between these two axes implies that eigen-solutions of the problem are not separable in the vertical and horizontal directions, which makes the problem much more complicated analytically.

\begin{figure}[t]
\includegraphics[width=\textwidth]{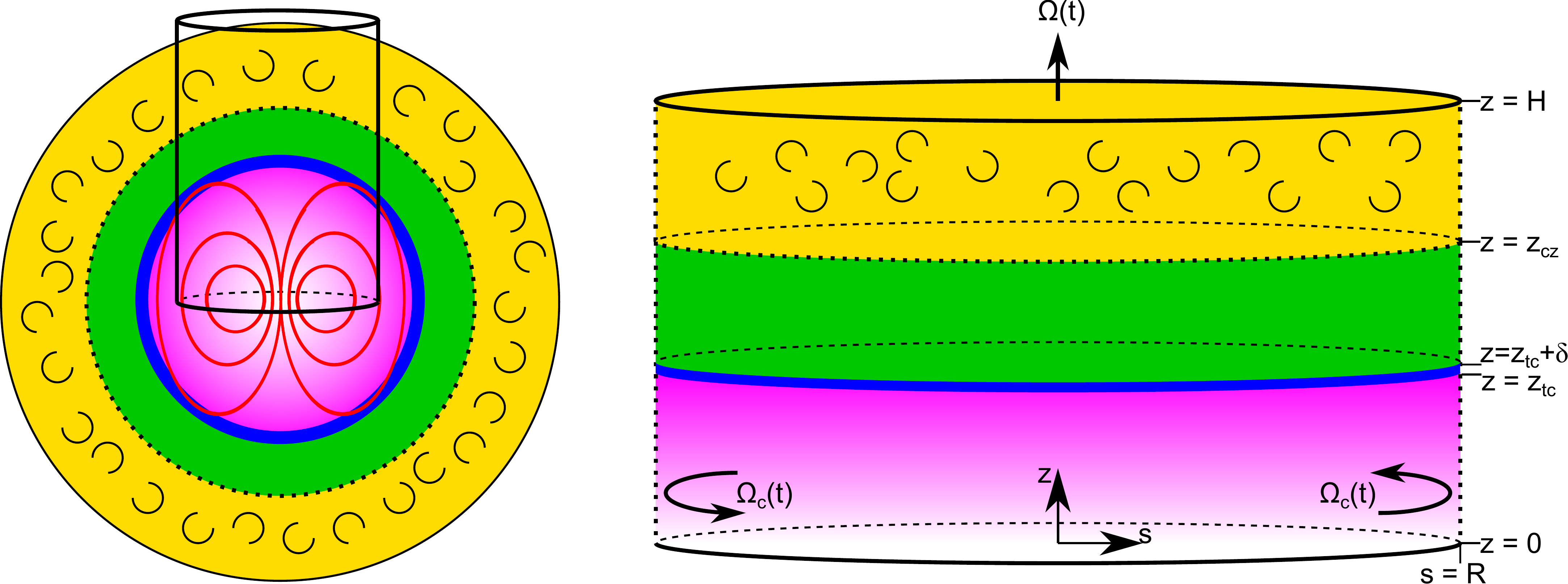}
 \caption{Figure contrasting two geometries: complete with convection zone and magnetic field.}
\label{fig:model}
\end{figure}

By contrast with the issue of nonlinearities, this second problem can be dealt with, at least approximately. To do so, we further simplify the model by considering a cylinder instead of a sphere, where gravity is {\it by construction} parallel to the rotation axis. This cylinder can, for instance, be viewed as the polar regions of the star (see Figure \ref{fig:model}). This simplification adds an order-unity geometrical error to all of our results, but on the other hand allows for fully analytical solutions of the linearized equations. We show in Section \ref{sec:unstrat1} that in some simple limit it is possible to find analytical solutions of the linearized system in both spherical and cylindrical geometries, and the solutions agree up to a geometrical factor, hence justifying our procedure.

\subsection{The ``cylindrical star"}

Our cylindrical model is presented in Figure \ref{fig:model}. As in GM98, we consider that the star is divided into four dynamically distinct regions, as shown in Figure \ref{fig:model}a: from the surface downward, the convection zone (yellow), the tachocline (green), the tachopause (blue), and the uniformly-rotating part of the radiation zone (purple), threaded by a primordial magnetic field (red). Figure \ref{fig:model}b shows the equivalent cylindrical model setup. 

The ``cylindrical star" has radius $R$ and total height $H$. The region with $z \in [z_{\cz},H]$ (yellow) represents the convection zone. The latter is assumed to rotate with a uniform angular velocity $\Omega_{\rm cz}(t)$ (as expressed in an inertial frame) which decreases over time via magnetic braking.  Henceforth, we work in a frame that is rotating with angular velocity $\Omega_{\rm cz}(t)$, so that by construction, the convection zone is at rest in the rotating frame. Note that we neglect here the possibility of any differential rotation in the convection zone. This is done in order to simplify our calculations, enabling us to neglect any meridional flows  driven by differential rotation (via gyroscopic pumping) in favor of those driven by the spin-down torque. We expect this assumption to hold for young stars, which undergo rapid spin-down, but not necessarily for older ones like the Sun (see Section \ref{sec:ccl3} for further discussion of this point). 

We use cylindrical coordinates $(s,\phi,z)$ aligned with the rotation axis (or equivalently, the vertical axis $\mathbf{e}_z$), and assume axial symmetry. In these coordinates, $\mathbf{u} = (u,v,w)$ is the velocity field relative to the rotating frame. The equations describing the dynamics of the system are the momentum equation, the mass conservation equation, the thermal energy equation, the induction equation and the solenoidal condition. Before studying them in more detail, we first proceed to describe each of the four regions listed above, and the assumptions made in each of them.

\subsection{The convection zone}

The dynamics of stellar convective zones are quite complex, and result from the nonlinear interplay between convection, rotation, large-scale flows and magnetic fields. We shall not attempt to model them in any detail here. Our main goal is merely to account for the rapid transport of angular momentum between the surface and the top of the radiation zone. To do this, we model the convection zone as in \citet{BrethertonSpiegel68}, who studied stellar spin-down by treating the effect of the convection on large-scale flows (and on momentum transport in general) through a Darcy friction term (i.e. a linear damping term). A similar method was used by \citet{GaraudAA09}, \citet{Woodal11} and \citet{AAal13} in their models of the solar interior. 
We therefore replace the Reynolds stress term $-\mathbf{u} \cdot \nabla  \mathbf{u}$ in the momentum equation (where ${\mathbf u}$ is the velocity field expressed in the rotating frame) with the term $ - \mathbf{u} / \tau$, where $\tau$ is a damping timescale. Dimensionally speaking, one can assume that $\tau$ is of the same order as the convective turnover timescale. For simplicity, we assume that $\tau$ is constant in the convective region, and that $1/\tau$ is zero in the radiative region.  
Finally, we also assume that the convection zone is adiabatically stratified, and transports heat very efficiently compared with all other timescales in the system. 

\subsection{The tachocline}

The region $z \in [z_{\rm tc}+\delta,z_{\cz}]$ (green) represents the tachocline. Its thickness, $\Delta = z_{\cz} - (z_{\rm tc}+\delta)$, depends on a number of factors including intrinsic properties of the star's structure, as well as its rotation rate and  the strength of the primordial field \citep[see GM98 or][for estimates of $\Delta$ in the steady-state solar case, and Appendix E for further discussion of $\Delta$ in the spin-down case]{Woodal11}. It is stably stratified, with a mean buoyancy frequency $\bar{N}_{\rm tc} \geq 0$ and mean background density $\bar{\rho}_{\rm tc}$. In this region we make several assumptions to simplify the equations of motion \citep[as in, for example,][]{SpiegelZahn92,GoughMcIntyre98,WoodMcIntyre11,Woodal11}. First, we assume that the thickness of the tachocline is small compared with a pressure and density scaleheight, and use the Boussinesq approximation \citep{SpiegelVeronis60} to study its dynamics. We also assume that, in the rotating frame, tachocline flows are slow enough to neglect the inertial term in the momentum equation, and also neglect the effect of viscous forces. These assumptions then imply that the tachocline is in hydrostatic and geostrophic equilibrium. We also assume that the flows are sufficiently steady and slow for the system to be in thermal equilibrium, with heat diffusion being balanced by the advection of the background entropy. Finally, as suggested by GM98, we assume that the Lorentz force is negligible in the tachocline.

\subsection{The tachopause and the deep interior}
 
The region $ z \in [z_{\rm tc},z_{\rm tc}+\delta]$ (blue) represents the tachopause. As discussed in Section \ref{sec:introduction}, the tachopause is a magnetized boundary layer through which the spin-down torque is ultimately transmitted to the interior. As in the case of the tachocline, its thickness $\delta$ depends on the local thermodynamical properties of the star, as well as its rotation rate and  the strength of the primordial field \citep[see GM98 or][]{Woodal11}. All of the assumptions concerning the dynamics of the tachocline made above apply here as well, with the exception of course of the one concerning the magnetic field -- the tachopause is by definition a layer in which the Lorentz force plays a fundamental role. In addition, the thickness of the tachopause is assumed to be much smaller than the thickness of the tachocline ($\delta \ll \Delta$) to allow for a boundary layer analysis of various balances within.

Finally, the region $0<z<z_{\rm tc}$ (pink) represents the magnetized radiative core, below the tachopause. We do not solve any equations to model that region. Instead, we merely assume that temperature perturbations and large-scale flows vanish for $z<z_{\rm tc}$, and that the entire region is held in uniform rotation with angular velocity $\Omega_{\rm c}(t)$ (expressed in the rotating frame, so $\Omega_{\rm c} = \Omega_{\rm core} - \Omega_{\rm cz}$). The latter is controlled by the spin-down rate of the regions above, through the torques acting within the tachopause. 

We now begin our theoretical investigation, which ultimately results in deriving an equation for the evolution of the total angular momentum of the core under this model that is identical to the two-zone model of \citet{MacGregorBrenner91}, but with a coupling timescale that is related to the local Eddington-Sweet timescale across the tachocline. The reader interested principally in a discussion of this result rather than its derivation may, as a first pass, skip directly to Section \ref{sec:discuss}. 

\section{CYLINDRICAL VS. SPHERICAL GEOMETRY}
\label{sec:unstrat1}

We begin our investigation by showing that results obtained using a cylindrical geometry are, within a geometrical factor of order unity, consistent with those obtained using a more realistic spherical geometry. 

To do so, we consider the simplest possible problem, of a non-magnetic star with an ``unstratified'' interior of constant density. While this case does not have any astrophysical relevance, it can easily be solved analytically in both cylindrical {\it and} spherical geometry \citep{BrethertonSpiegel68}. As such, it can serve as an illustration of the validity of the ``cylindrical star'' assumption, as well as a pedagogical tool to introduce the method of solution of the governing equations we shall use throughout. 

We also assume in this Section that there is no magnetic field. In this case, the tachopause does not exist, and the ``tachocline'' fills the entire radiation zone, extending all the way to the center of the star.  In this model, the only difference between the ``convective'' and ``radiative'' regions is the presence of the Darcy forcing term modeling angular-momentum transport by convective motions.  Under the assumptions discussed in Section \ref{sec:model}, regardless of geometry, the equations of motion are the momentum equation and incompressibility (the thermal energy equation is not needed):
\begin{equation}
 \frac{\partial\mathbf{u}}{\partial t} + 2\mathbf{\Omega_{\rm cz}}\times\mathbf{u} + \dot{\mathbf{\Omega}}_{\rm cz} \times\mathbf{r} + \frac{\mathbf{u}}{\tau} = -\frac{1}{\rhobar}\nabla p,  \quad \nabla . \mathbf{u} = 0,\label{eq:plugmom}
\end{equation}
where $\mathbf{r}$ is the position vector, $\bar \rho$ is the constant density of the fluid, 
and $p$ is the pressure perturbation away from hydrostatic equilibrium. 
 The second term in Equation \eqref{eq:plugmom} is the Coriolis force, the third is Euler's force (which is due to the deceleration of the frame) and the fourth term is the Darcy friction term (see Section \ref{sec:model}), which is assumed to be zero in the ``radiation zone". 

\subsection{Spherical geometry}

\citet{BrethertonSpiegel68} were the first to study the spin-down of such an unstratified star using the model described above. 
They found an analytical solution of the problem in a spherical geometry, assuming that the angular velocity of the convection zone $\Omega_{\rm cz}(t)$ (expressed in an inertial frame) decays exponentially over time. We now repeat their calculation and consider any functional form for $\Omega_{\rm cz}(t)$ for more generality.  

\citet{BrethertonSpiegel68} first assumed that the spin-down rate is low, so that $|\dot{\Omega}_{\rm cz}/\Omega_{\rm cz}|\ll \Omega_{\rm cz}$. 
They then looked for a quasi-steady solution in the rotating frame, requiring the term $\partial \mathbf{u}/\partial t$ in the momentum equation to be negligible. The resulting quasi-steady equations of motion in the ``radiative interior'' ($0<r<r_{\cz}$) are
\begin{equation}
 2\mathbf{\Omega_{\rm cz}}\times\mathbf{u} +\dot{\mathbf{\Omega}}_{\rm cz}\times\mathbf{r} = -\frac{1}{\rhobar}\nabla p,\quad \nabla . \mathbf{u} = 0. \label{eq:BSinterior}
\end{equation}
In the ``convection zone'' ($r_{\cz}<r<R$), they reduce to 
\begin{equation}
 \frac{1}{\tau}\mathbf{u} = -\frac{1}{\rhobar}\nabla p,\quad \nabla . \mathbf{u} = 0,
\label{eq:BSconv}
\end{equation}
assuming that the Darcy timescale $\tau$ is significantly smaller than the rotation timescale $\Omega_{\rm cz}^{-1}$.

Solving these equations, then matching $p$ and the normal velocity at the radiative--convective interface (at $r=r_{\cz}$), yields the shape of the streamlines (see Figure \ref{fig:streamlinesc}), as well the angular-velocity perturbation $\Omega(r,\theta,t)$ everywhere in the star (see their original work for the details of the calculation). The angular velocity in the radiative interior turns out to be uniform, with value $\Omega_{\rm c}$ such that
\begin{equation}
\frac{\Omega_{\rm c}}{\Omega_{\rm cz}} = - \frac{\dot{\Omega}_{\rm cz}}{4\Omega_{\rm cz}^{3}\tau}\frac{3r_{\cz}^{5} + 2R^{5}}{R^{5}-r_{\cz}^{5}}, \label{eq:BS}
\end{equation}
as expressed in the rotating frame, where $R$ is the radius of the star and $r_{\rm cz}$ that of the base of the convection zone.  Since $\dot{\Omega}_{\rm cz}<0$, we have $\Omega_{\rm c}>0$, implying that the interior is always rotating faster than the convection zone, or in other words, lagging behind in the context of the spin-down process. The relative lag is measured by $\Omega_{\rm c}/\Omega_{\rm cz}$; whether it increases or decreases with time depends on the behavior of $\dot{\Omega}_{\rm cz}/\Omega^3_{\rm cz}$. The implications of Equation (\ref{eq:BS}) are discussed in Section \ref{sec:discuss1}.

\subsection{Cylindrical geometry}

We now solve the same equations under the same assumptions, but this time in the cylindrical geometry presented in Section \ref{sec:model}. Inspection of the expression for the meridional flows given by \citet{BrethertonSpiegel68} (see their Equations 7 and 8) reveals that they have equatorial symmetry. To obtain solutions with the same symmetry in cylindrical geometry, we set the vertical velocity at $z=0$ to be zero. In addition, the vertical velocity at the surface of the star is also zero.  Following \cite{BrethertonSpiegel68}, we solve the equations for $z>z_{\cz}$ and $z<z_{\cz}$ separately, and match $p$ and the vertical velocity $w$ at the radiative--convective interface.  Our vertical boundary conditions are\footnote{Since we have assumed that viscosity is negligible, we do not match the radial ($u$) and azimuthal ($v$) velocities at the interface.} therefore: 
\begin{align}
 w = 0 \mbox{ at } z=0, z=H,  \quad  p(z=z_{\cz}^{-}) = p(z=z_{\cz}^{+}), &\quad w(z=z_{\cz}^{-}) = w(z=z_{\cz}^{+}). \label{eq:bcw}
\end{align}

Boundary conditions at the side wall ($s=R$) are more difficult to choose, as we want them to have as little influence as possible on the dynamics inside the cylinder.  For simplicity, we select
\begin{equation}
 p = 0 \mbox{ at } s=R. \label{eq:sidebc}
\end{equation}
This boundary condition allows a radial flow across the side wall, as required by analogy with the spherical solution (see Figure \ref{fig:streamlinesc}). By conservation of mass the fluid must somehow return to the convection zone, a process which necessarily occurs outside of the cylinder considered, and that we cannot model explicitly. In what follows, we assume that this return flow does not play any fundamental role in the spin-down process (in the sense that it does not change the solution beyond a factor of order unity). This assumption is verified a posteriori to be correct (see Section \ref{sec:discuss1}).

We first consider the radiative region ($z<z_{\cz}$). The various components of Equation (\ref{eq:plugmom}), when expressed in a cylindrical geometry, become
\begin{align}
\frac{\partial p}{\partial z} = 0, \quad \frac{\partial v}{\partial t} + 2\Omega_{\rm cz}u + \dot{\Omega}_{\rm cz} s = 0, \quad
 2\Omega_{\rm cz} v = \frac{1}{\rhobar}\frac{\partial p}{\partial s}. \label{eq:unstratmom}
\end{align}
where ${\mathbf u} = (u,v,w)$. We also have incompressibility:
\begin{equation}
 \frac{1}{s}\frac{\partial}{\partial s}(su) + \frac{\partial w}{\partial z} = 0. \label{eq:incompress}
\end{equation}
Following \cite{BrethertonSpiegel68}, we take $\partial v/\partial t = 0$, and combine \eqref{eq:unstratmom} and \eqref{eq:incompress} to find
\begin{equation}
 u = -\frac{\dot{\Omega}_{\rm cz}}{2\Omega_{\rm cz}}s,\quad w = \frac{\dot{\Omega}_{\rm cz}}{\Omega_{\rm cz}}z \mbox{  for } z < z_{\rm cz}  \mbox{  . } 
\label{eq:uw}
\end{equation}

In the convection zone, expanding (\ref{eq:BSconv}) we have
\begin{equation}
 \frac{1}{\rhobar}\frac{\partial p}{\partial s} = -\frac{u}{\tau}, \quad v = 0, \quad \frac{1}{\rhobar}\frac{\partial p}{\partial z} = -\frac{w}{\tau}. \label{eq:plugmom2}
\end{equation}
Combining these with incompressibility $\nabla . \mathbf{u} = 0$ gives
\begin{equation}
 \nabla^{2}p = 0,\quad \nabla^{2} w = 0.
\end{equation}
Using the boundary conditions \eqref{eq:bcw} and \eqref{eq:sidebc}, we can write $w$ and $p$ as
\begin{align}
 w &= \sum_{n}B_{n}\sinh\left(\lambda_n\frac{z-H}{R}\right) J_{0}\left(\lambda_{n}\frac{s}{R}\right), \label{eq:w1} \\
\frac{1}{\rhobar}p &= - \sum_{n} \frac{B_{n}}{\tau}\frac{R}{\lambda_{n}} J_{0}\left(\lambda_{n}\frac{s}{R}\right)\cosh\left(\lambda_{n} \frac{z-H}{R}\right), \mbox{  for } z > z_{\rm cz},  \label{eq:P1}
\end{align}
where $J_{0}$ is the zeroth-order regular Bessel function, the $\{\lambda_{n}\}$ are its zeros, and the coefficients
$\{B_{n}\}$ are integration constants. The latter are found by matching $w$ in Equations \eqref{eq:uw} and \eqref{eq:w1}
at $z=z_{\cz}$, so that 
\begin{equation}
 B_{n} = \frac{\dot{\Omega}_{\rm cz}}{\Omega_{\rm cz}} \frac{2z_{\cz}}{\lambda_{n}J_{1}(\lambda_{n})\sinh\left(\lambda_{n}\frac{z_{\cz}-H}{R}\right)}. 
 \label{eq:Bn1}
\end{equation}
where $J_1$ is the first order Bessel function. 
Equation \eqref{eq:unstratmom} shows that, for $z<z_{\cz}$, $p$ is constant with height, so that matching $p$ at $z=z_{\cz}$ and using (\ref{eq:Bn1}) gives
\begin{equation}
\frac{1}{\rhobar}p = \frac{\dot{\Omega}_{\rm cz}}{\Omega_{\rm cz} \tau}\sum_{n}J_{0}\left(\lambda_{n}\frac{s}{R}\right)\left[\frac{2z_{\cz}R}{\lambda_{n}^{2}J_{1}(\lambda_{n})\tanh\left( \lambda_{n}\frac{H-z_{\cz}}{R}\right)}\right], \mbox{  for } z < z_{\rm cz},
\end{equation}
and, using (\ref{eq:unstratmom}), we find that the angular velocity of the radiative region, expressed in the rotating frame, becomes
\begin{equation}
\frac{\Omega(s,z)}{\Omega_{\rm cz}} = \frac{v(s)}{s\Omega_{\cz}} = - \frac{\dot{\Omega}_{\rm cz}}{\tau\Omega_{\rm cz}^{3}}\sum_{n}\frac{J_{1}(\lambda_{n}\frac{s}{R})}{s\lambda_{n}J_{1}(\lambda_{n})}\frac{z_{\cz}}{\tanh\left( \lambda_{n} \frac{H-z_{\cz}}{R}\right)}. \label{eq:angvelunstrat}
\end{equation}
The latter is always positive, and is a function of $s$ only, as expected from the Taylor-Proudman constraint.

% 4(1+beta - alpha^-1) -2 = 2 + 4beta - 4alpha^-1

\subsection{Discussion}
\label{sec:discuss1}

\begin{figure}[t]
\begin{center} 
\includegraphics[width=0.6\textwidth]{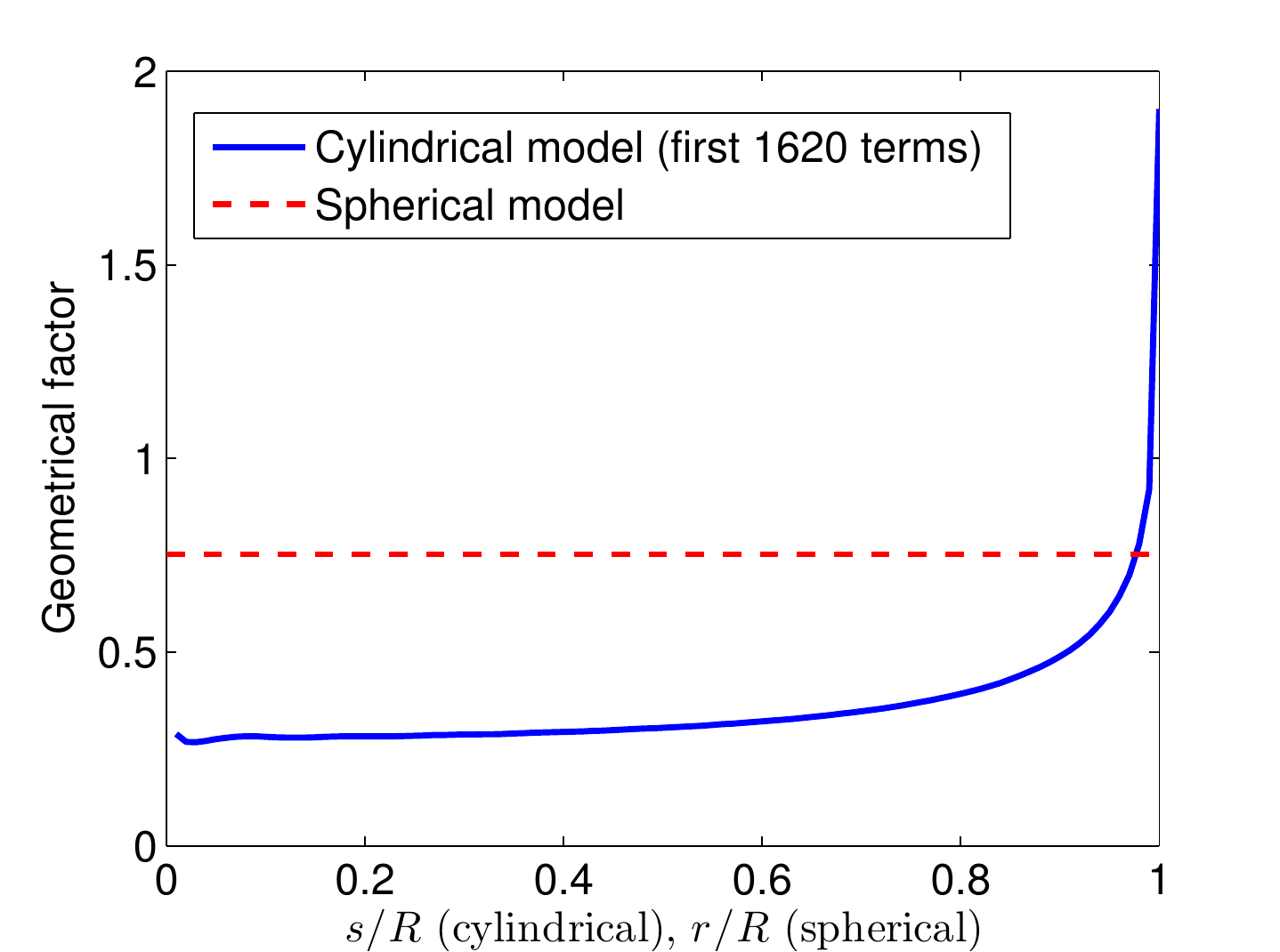}
\end{center} 
\caption{Comparison of the two geometrical factors multiplying $-\dot{\Omega}_{\rm cz}/\tau\Omega_{\rm cz}^{3}$ in Equations \eqref{eq:BS} and \eqref{eq:angvelunstrat}.  The lengths are scaled such that $z_{\cz}/H = z_{\cz}/R = 0.7$ for the cylindrical geometry, and $r_{\cz}/R = 0.7$ for the spherical geometry. }
\label{fig:geometricalfactor}
\end{figure}

The solutions in spherical and cylindrical geometries bear strong similarities. In both cases, we find that the azimuthal velocity is constant along the rotation axis, as expected from the Taylor-Proudman constraint since the fluid is unstratified. Furthermore, the radiation zone is always rotating more rapidly than the convection zone. The relative lag between the two regions, in both cases, is equal to the prefactor $-\dot{\Omega}_{\rm cz}/{\tau\Omega_{\rm cz}^{3}}$, times a non-dimensional term that depends only on the geometry of the system. This term is shown as a function of $s$ in Figure \ref{fig:geometricalfactor} for both cylindrical and spherical geometries. It is constant in the spherical case, and increases with $s$ in the cylindrical case. However, the two are consistent (within a factor of order unity). This shows that the difference in the results obtained in the two geometries is not dramatic.

\begin{figure}[t]
\includegraphics[width=0.5\textwidth]{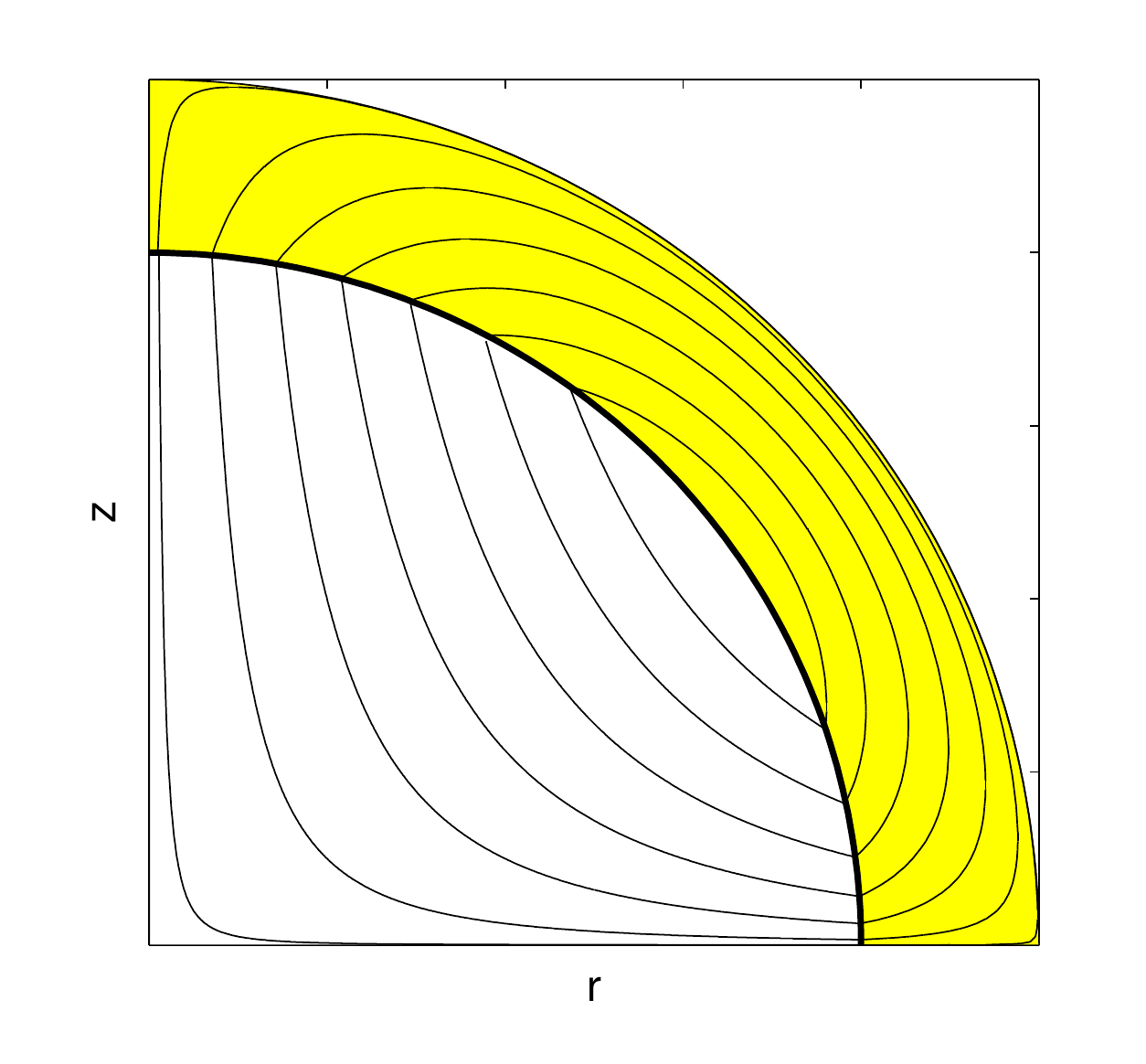} \quad \includegraphics[width=0.5\textwidth]{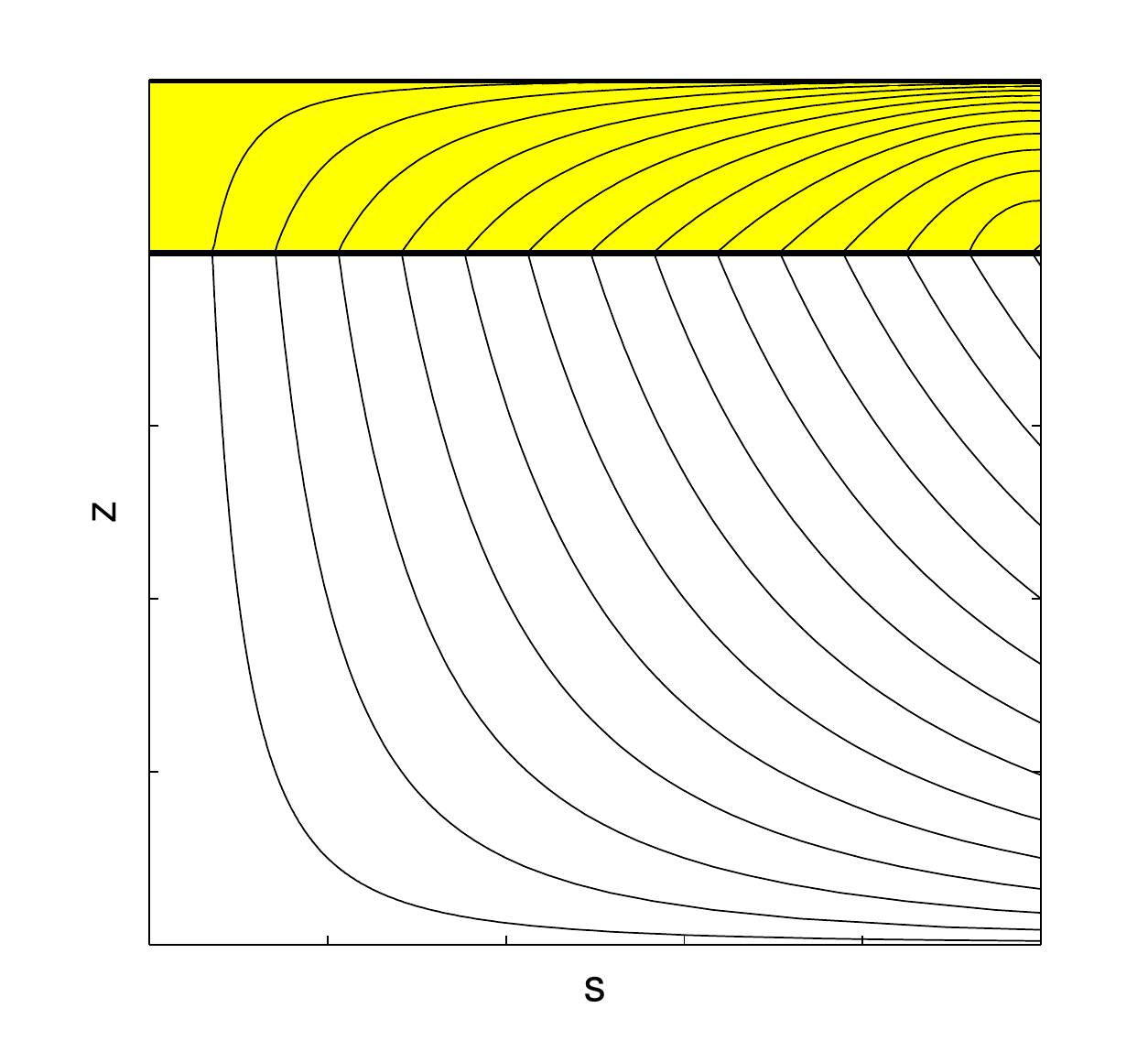} 
 \caption{Comparison of the streamlines in the spherical and cylindrical solutions. The cylindrical case (right) can be viewed as a distorted version of the spherical case (left), as long as the solution of the latter is truncated at a cylindrical radius roughly equal to $s=0.6 R_\star$. }
\label{fig:streamlinesc}
\end{figure}

The structure of the meridional circulation is also very similar in both cases, as shown in Figure \ref{fig:streamlinesc}. In the cylindrical case, one could imagine the streamlines closing back on themselves outside of the domain, as they do in the spherical case. This changes the global angular momentum balance somewhat, but as discussed above, does not affect the outcome by more than an order unity factor.

While $\Omega_{\rm cz}(t)$ should in principle be calculated self-consistently from a stellar wind model, it is informative to look at specific ``plausible" spin-down laws. 
For the purpose of the following discussion, we either assume that $\Omega_{\rm cz}(t)$ decays exponentially, with 
\begin{equation}
\Omega_{\rm cz}(t) = \Omega_{0}\exp(-k(t-t_{0})),
\label{eq:omegaexp}
\end{equation}
or as a power law, with 
\begin{equation}
\Omega_{\rm cz}(t) = \Omega_{0}(t/t_{0})^{-\alpha},
\label{eq:omegapow}
\end{equation}
for some $\alpha>0$. 
In these laws, $\Omega_0 = \Omega_{\rm cz}(t_0)$, where $t_0$ is the initial timescale considered (e.g. the end of the disk-locking phase, for instance, or the Zero Age Main Sequence). 
The parameters $\alpha$ and $k$ are unspecified here, but can be fitted to any desired spin-down model. 

The $-\dot{\Omega}_{\rm cz}/\Omega_{\rm cz}^{3}$ prefactor in Equations (\ref{eq:BS}) and (\ref{eq:angvelunstrat}) shows that the relative lag between the radiative and convective regions always increases exponentially with time for an exponential spin-down law. This suggests a break-down of the quasi-steady approximation for this case\footnote{This was, oddly enough, not discussed by \citet{BrethertonSpiegel68}.}. Using a power-law to model spin-down reveals that $\Omega(s)/\Omega_{\rm cz}$ increases with time if $\alpha > 1/2$, is constant if $\alpha = 1/2$, and decreases with time if $\alpha < 1/2$. This, again, suggests a break-down of the quasi-steady approximation if spin-down occurs faster than the Skumanich power-law, which has $\alpha = 1/2$ \citep{Skumanich72}. Although interesting, we believe that the correspondence between the critical power-index $\alpha = 1/2$ and the Skumanich law is a coincidence, given the simplistic nature of this particular unstratified model.

To conclude this Section, since the solutions in \eqref{eq:BS} and \eqref{eq:angvelunstrat} are the same up to a purely geometrical factor, and since that geometrical factor is typically of order unity, we now make the assumption that results from our cylindrical model may carry across to the spherical geometry, up to a geometrical error, for more complex systems such as ones including stratification or a magnetic field. Hence from now on, all our analysis will be restricted to cylindrical geometry.

\section{SPIN-DOWN IN A NON-MAGNETIC STAR} \label{sec:strat1}

We now consider the spin-down of a non-magnetic but more realistically stratified star. As in Section \ref{sec:unstrat1}, we study only solar-type stars, with an outer convection zone ($z_{\rm cz}<z<H$) and an inner radiation zone ($0<z<z_{\rm cz}$). In preparation for the implementation of the GM98 model, which is our ultimate goal, we assume that the radiation zone is sub-divided into two regions, a uniformly rotating core (for $z < z_{\rm tc}$) and a tachocline (for $z_{\rm tc} < z < z_{\rm cz}$). Here, we do not specify the mechanism by which the core is held in solid-body rotation. Furthermore, we assume that this core is not dynamically connected to the tachocline, but instead, merely acts as a passive boundary whose only role is to be impenetrable to the fluid. Within that approximation, note the core cannot be spun-down, as no angular momentum can be extracted from it. Finally, since the tachopause does not exist, we take $\delta = 0$. 

While this setup may seem odd at first, note that one could use it to represent a normal non-magnetic star (that is, with no rigidly rotating core) simply by taking the regular limit $z_{\rm tc} \rightarrow 0$; the impenetrability of the boundary at $z = z_{\rm tc}$ simply becomes an equatorial symmetry condition (see Section \ref{sec:unstrat1}). However, it is important to remember that this limit is generally inconsistent with the Boussinesq approximation, which requires $\Delta = z_{\rm cz} - z_{\rm tc}$ to be smaller than a pressure scaleheight. We therefore advise the reader against indiscriminately using our results in this fashion. Instead, we note that the solutions derived in this Section with $z_{\rm tc} \ne 0$ will be applicable (with a few modifications) to model the complete stellar spin-down problem in Section \ref{sec:blcase}. In this sense, our work in this section should be viewed once again as a pedagogical step toward understanding the final result.

\subsection{Model equations}

The convection zone is assumed to be very nearly adiabatic with a buoyancy frequency $\bar{N}=0$, and entropy perturbations are assumed to be negligible. For more realism, and in this Section only, we allow the background density $\bar{\rho}$ to vary with height in the convection zone, and use the anelastic approximation. We shall show that the results are identical to those derived in the Boussinesq case. The equations governing the dynamics within the convection zone thus become, under similar assumptions to the ones discussed in Section \ref{sec:model}: 
\begin{equation}
 \frac{1}{\tau}\mathbf{u} = -\frac{1}{\bar{\rho}}\nabla p,\quad \nabla \cdot (\bar{\rho} \mathbf{u} ) = 0, \quad  T = 0 \mbox{ ,}
\label{eq:BSconv2}
\end{equation}
where $T$ is the temperature perturbation.

Since the tachocline is thought to be thin, we use the Boussinesq approximation \citep{SpiegelVeronis60} to model it, and assume that the buoyancy frequency, gravity, density and temperature do not depart significantly from their mean tachocline values  $\bar N_{\rm tc}$, $\bar{g}_{\rm tc}$, $\bar{\rho}_{\rm tc}$ and $\bar{T}_{\rm tc}$. In this approximation, density and temperature perturbations are formally related through a linearized equation of state, in which pressure perturbations are negligible. Hence
\begin{equation}
\frac{\rho}{\bar \rho_{\rm tc}} = - \frac{T}{\bar T_{\rm tc}}\mbox{   . } 
\end{equation}
The momentum equation in the tachocline becomes
\begin{equation}
 \frac{\partial\mathbf{u}}{\partial t} + 2\mathbf{\Omega_{\rm cz}}\times\mathbf{u} + \dot{\mathbf{\Omega}}_{\rm cz}\times\mathbf{r} = -\frac{1}{\bar{\rho}_{\rm tc}}\nabla p + \frac{\bar{g}_{\rm tc}}{\bar{T}_{\rm tc}} T \hat{\mathbf{e}}_{z}. \label{eq:bulkmom}
\end{equation}
Combining the radial and vertical components of \eqref{eq:bulkmom}, with the assumptions discussed in Section \ref{sec:model} and above, yields the well-known thermal-wind equation
\begin{equation}
 2\Omega_{\rm cz} \frac{\partial v}{\partial z} = \frac{\bar{g}_{\rm tc}}{\bar{T}_{\rm tc}}\frac{\partial T}{\partial s}. \label{eq:thermalwind}
\end{equation}
The tachocline is also assumed to be in thermal equilibrium, so
\begin{equation}
 \frac{\bar{N}_{\rm tc}^2\bar{T}_{\rm tc}}{\bar{g}_{\rm tc}}w = \kappa_{\rm tc} \nabla^2 T, \label{eq:thermaleq}
\end{equation}
where $\kappa_{\rm tc}$ is its (constant) thermal diffusivity. 

The boundary conditions for velocity and pressure are similar to those of the previous Section, but with the lower boundary raised to $z = z_{\rm tc}$. We require impermeability ($w = 0$) at $z = z_{\rm tc}$ and $z = H$, $p = 0$ at $s=R$, and that $p$ and $w$ must be continuous at $z = z_{\rm cz}$. The temperature perturbations are assumed to vanish at $z=z_{\rm tc}$, and at $z=z_{\rm cz}$. 

We now first study this system of equations in the same ``quasi-steady'' state discussed in the previous Section, and then solve for the complete time-dependence of the system to determine under which conditions this quasi-steady state is valid. 

\subsection{Quasi-steady solution} \label{sec:steadystate}

As in Section \ref{sec:unstrat1}, we define the ``quasi-steady'' state as the solution of the governing equations in which $\partial\mathbf{u}/\partial t$ is neglected. We find solutions separately for $z>z_{\rm cz}$ and $z<z_{\rm cz}$, and match $w$ and $p$ at the radiative--convective interface. The full calculation is given in Appendix A.  We find that the azimuthal velocity in the tachocline is 
\begin{equation}
v(s,z,t) = - \frac{\dot{\Omega}_{\rm cz}}{\Omega_{\rm cz}^2 }\sum_{n} \frac{J_{1}(\lambda_{n}\frac{s}{R})}{\lambda_{n}J_{1}(\lambda_{n})}  \left[  \frac{\Delta}{\tau\tanh \left( \lambda_{n}\frac{H-z_{\cz}}{R}\right) } + \frac{\bar{N}_{\rm tc}^{2}}{\kappa_{\rm tc}} G_n(z) \right] ,
\label{eq:v1}
\end{equation}
where $\Delta = z_{\rm cz} - z_{\rm tc}$ is the thickness of the tachocline (since $\delta = 0$), and where $G_n$ is the geometrical factor 
\begin{align}
G_n(z) &= \frac{R}{\lambda_n} \left\{  \frac{ \Delta R }{\lambda_{n}\sinh\left(\lambda_{n}\frac{\Delta}{R}\right)}    \left[ \cosh \left(\lambda_{n}\frac{z-z_{\rm tc}}{R} \right) - \cosh\left(\lambda_{n}\frac{\Delta}{R}\right) \right] - \frac{(z-z_{\rm tc})^{2}-\Delta^{2}}{2} \right\} .
\end{align}
Equation \eqref{eq:v1} reduces to \eqref{eq:angvelunstrat} when the tachocline is unstratified ($\bar{N}_{\rm tc}=0$), regardless of the position of the lower boundary $z_{\rm tc}$. This is not surprising: the Taylor-Proudman constraint requires $v$ to be independent of $z$ in that limit. It is also shown in Appendix A that the density variation in the convection zone has no effect on the angular velocity within the tachocline, as long as the density is continuous across $z = z_{\rm cz}$. In what follows, we can therefore equivalently use $\nabla \cdot {\bf u} = 0$ in the convection zone for mathematical simplicity even though the latter does not actually satisfy the Boussinesq approximation.

\begin{figure}[h]
\centerline{\includegraphics[width=0.5\textwidth]{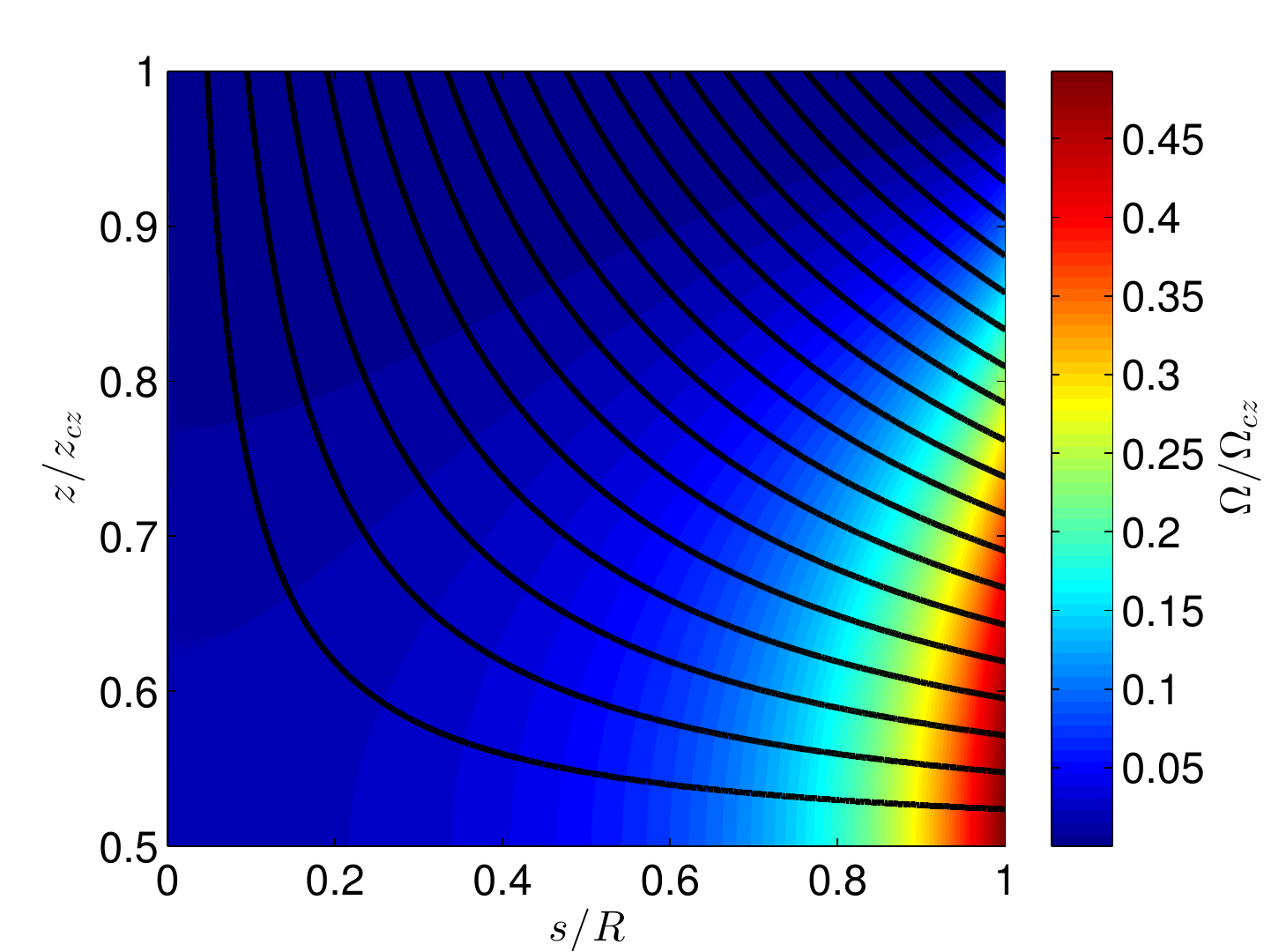}}
\caption{Contour plot of the quasi-steady solution for $\Omega(s,z) = v(s,z)/s$ given in \eqref{eq:v1}, with $z_{\rm tc} = 0.5 z_{\rm cz}$, where $\Omega_{\cz} = \Omega_0(t/t_0)^{-1/2}$, and where $R$, $H$, $\tau$, $\bar N_{\rm tc}$, $\kappa_{\rm tc}$, $\bar \rho_{\rm tc}$, $t_0$ and $\Omega_0$ are otherwise given in Table \ref{tab1}. Note that $\Omega$ is strictly positive everywhere, and increases with depth and cylindrical radius. The solid black lines show flow streamlines, with the flow direction being downward and outward.}
\label{fig:Omegastrat}
\end{figure}

The angular velocity and meridional circulation profile for $\bar N_{\rm tc} \ne 0$ is shown in Figure~\ref{fig:Omegastrat}. To understand its properties, first note that since $G(z_{\rm cz}) = 0$, the angular velocity just below the base of the convection zone is the same in the stratified and unstratified cases. However, $v$ is no longer independent of $z$, but instead increases with depth. This effect is due to the added buoyancy force in the momentum equation, which relaxes the Taylor-Proudman constraint. The latter is replaced by the thermal-wind constraint given in Equation (\ref{eq:thermalwind}), which relates any variation of angular velocity along the rotation axis to gradients of temperature perpendicular to it. The relative lag between the base of the tachocline and the convection zone is thus controlled simultaneously by thermal-wind balance and by thermal equilibrium within the tachocline. 

The combination of these two constraints yields a simple estimate for the relative angular-velocity shear across the tachocline as a function of input stellar parameters. Indeed, first note that the downwelling flow velocity across the tachocline is primarily controlled by the tachocline thickness, and by the spin-down rate through: 
\begin{equation}
w = \frac{\dot{\Omega}_{\rm cz}}{\Omega_{\rm cz}}(z-z_{\rm tc}), 
\end{equation}
(see Appendix A for detail). An order-of-magnitude approximation of this equation yields
\begin{equation}
|w| \sim - \Delta  \frac{\dot{\Omega}_{\rm cz}}{\Omega_{\rm cz}}.
\end{equation}
In thermal equilibrium (see Equation \ref{eq:thermaleq}), however, the advection of the background entropy stratification by these flows must balance the diffusion of the induced temperature perturbations $T$. This sets the typical amplitude of $T$ in the tachocline to be 
\begin{equation}
T \sim \frac{\Delta^2}{\kappa_{\rm tc}}  \frac{\bar N_{\rm tc}^2 \bar T_{\rm tc}}{\bar g_{\rm tc}} |w| .
\end{equation}
Finally, by thermal-wind balance, latitudinal variations in $T$ control the allowable shear across the tachocline and therefore the total angular-velocity difference $({\Omega}_{\rm b}-0)$ between the base of the tachocline (where $\Omega \sim \Omega_{\rm b}$) and the radiative-convective interface (where $\Omega = 0$): 
\begin{equation}
2 \Omega_{\rm cz} R \frac{\Omega_{\rm b}}{\Delta} \sim \frac{\bar g_{\rm tc}}{\bar T_{\rm tc}} \frac{T}{R} .
\end{equation}
Combining all these estimates yields
\begin{equation}
t_{\rm ES}(\Omega_{\rm b}) \equiv  \frac{ \bar N_{\rm tc}^2  }{2 \Omega_{\rm cz} \Omega_{\rm b}}  \left( \frac{\Delta}{R} \right)^2  \frac{\Delta^2 }{\kappa_{\rm tc} } \sim - \frac{\Omega_{\rm cz}}{\dot{\Omega}_{\rm cz}}  \equiv t_{\rm sd}(\Omega_{\rm cz}),
\label{eq:tesdef}
\end{equation}
which states that the system adjusts itself (by selecting $\Omega_{\rm b}$) such that the Eddington-Sweet timescale\footnote{An Eddington-Sweet timescale is, by definition, the timescale for mixing (of chemical species and angular momentum) by meridional flows that are constrained by thermal equilibrium and thermal-wind balance. It is commonly used to describe mixing across the entire radiative region of a star, in which case it is typically much longer than the star's age unless the star is very rapidly rotating. However, when used to described mixing across a thin region such as the tachocline, the {\it local} Eddington-Sweet timescale can be much shorter. } based on that angular-velocity lag and on the thickness of the tachocline (expressed in the left-hand-side of this equation) is equal to the spin-down timescale of the star (expressed in the right-hand-side). We then have
\begin{equation}
\frac{\Omega_{\rm b}}{\Omega_{\rm cz}} \sim - \frac{\dot{\Omega}_{\rm cz}}{\Omega^3_{\rm cz}}  \left( \frac{\Delta}{R} \right)^2 \frac{ \bar N_{\rm tc}^2 \Delta^2 }{2 \kappa_{\rm tc} }  = \frac{t_{\rm ES}(\Omega_{\rm cz})}{t_{\rm sd}(\Omega_{\rm cz})},\; \mbox{ where } t_{\rm ES}(\Omega_{\rm cz}) = \frac{ \bar N_{\rm tc}^2  }{2 \Omega^2_{\rm cz}}  \left( \frac{\Delta}{R} \right)^2  \frac{\Delta^2 }{\kappa_{\rm tc} }.
\label{eq:oom}
\end{equation}
This shows that the relative angular-velocity shear between the top and the bottom of the tachocline is equal to the ratio of the local Eddington-Sweet timescale (based, this time, on the rotation rate of the convection zone) to the spin-down timescale. 
We then expect the shear across the tachocline to be larger (i) if the background stratification is larger (ii) if the local thermal diffusivity is smaller, (iii) if the spin-down rate is larger or (iv) if the tachocline is thicker. This is indeed what the exact Equation (\ref{eq:v1}) and its order of magnitude approximation (\ref{eq:oom}) both show. 

Finally, note that the relative lag is also proportional to $- \dot{\Omega}_{\rm cz}/\Omega^3_{\rm cz}$,  as in the unstratified case. This is not entirely surprising, since the unstratified case is a regular limit of this stratified problem as $\bar N_{\rm tc} \rightarrow 0$. As such, as long as $\Delta$ is constant, $\Omega_{\rm b}/\Omega_{\rm cz}$ diverges with time for an exponential spin-down law, or for any power-law with $\alpha > 1/2$, as discussed in Section \ref{sec:discuss1}. However, since $\Delta$ likely depends on time as well through its dependence on $\Omega_{\rm cz}$ (see Appendix E), other criteria apply (see Sections \ref{sec:blcase} and \ref{sec:discuss} for detail). 

\subsection{Transient solution}
\label{sec:transient}

Having found a quasi-steady solution to the spin-down problem, we now revisit the original time-dependent equations to determine when that solution is valid, and how rapidly the system relaxes to it.  Guided by the steady-state solution, we expand the azimuthal velocity $v$ on the same basis of Bessel functions, namely
\begin{equation}
 v(s,z,t) = \sum_{n}\frac{dJ_{0}(\lambda_{n}\frac{s}{R})}{ds}v_{n}(z,t).
\end{equation}
Combining \eqref{eq:bulkmom}, \eqref{eq:thermalwind} and incompressibility with this ansatz, and retaining the time-derivative in the azimuthal component of the momentum equation, gives
\begin{equation}
 \frac{\partial v_{n}}{\partial t} + \left(\frac{2\Omega_{\rm cz}(t)R}{\bar{N}_{\rm tc} \lambda_{n}}\right)^{2}\kappa_{\rm tc}  \frac{\partial^{4}v_{n}}{\partial z^{4}}- \left(\frac{2\Omega_{\rm cz}(t)}{\bar{N}_{\rm tc} }\right)^{2}\kappa_{\rm tc} \frac{\partial^{2}v_{n}}{\partial z^{2}}  =\dot{\Omega}_{\rm cz}(t) \frac{4R^2}{\lambda^3_{n}J_{1}(\lambda_{n})}, \label{eq:v3}
\end{equation}
where we have explicitly written in the time-dependence of $\Omega_{\rm cz}$ and $\dot{\Omega}_{\rm cz}$ to remember that it must be taken into account. This equation is quite similar to the one derived by~\cite{SpiegelZahn92} in the context of the evolution of the differential rotation profile within the solar tachocline. This is not surprising, as our underlying assumptions (thermal-wind balance, thermal equilibrium) are essentially the same. 
The first and second terms on the left-hand-side of \eqref{eq:v3} are the same as theirs (see their Equation 4.10). The hyperdiffusion term arises from the advection of angular momentum by  local Eddington--Sweet flows, a transport process that ~\cite{SpiegelZahn92} called ``thermal spreading''. The third term on the left-hand side is also part of the thermal spreading process, but 
was neglected by ~\cite{SpiegelZahn92} on the grounds that it is quite small when $\Delta \ll R$. For consistency with the Boussinesq approximation, which requires $\Delta$ to be smaller than a pressure scaleheight, we also neglect it from here on. Finally, the right-hand side contains the global forcing term arising from Euler's force. Since viscous and turbulent transport are neglected here, the evolution of the angular momentum in the tachocline has two contributions only: transport by meridional flows, and global extraction by Euler's force. As such, we expect the system to behave in rather different ways if the spin-down timescale is much larger or much smaller than local Eddington--Sweet mixing timescale. 

In what follows, we introduce the new variable $x = (z-z_{\rm tc})/\Delta$. Together with the simplification discussed above, Equation \eqref{eq:v3} becomes 
\begin{equation}
 \frac{\partial v_{n}}{\partial t} +  \frac{4\Omega^2_{\rm cz}(t)R^2}{\bar{N}^2_{\rm tc} \lambda^2_{n}} \frac{\kappa_{\rm tc}}{\Delta^4(t)}  \frac{\partial^{4}v_{n}}{\partial x^{4}} =\dot{\Omega}_{\rm cz}(t) \frac{4R^2}{\lambda^3_{n}J_{1}(\lambda_{n})}, \label{eq:v4}
\end{equation}
where we have also explicitly written in the time-dependence of $\Delta$. 
To solve Equation \eqref{eq:v4}, we note that it is further separable in $x$ and $t$, and write 
\begin{equation}
v_{n}(x,t) = \sum_{m}V_{nm}(t)Z_{nm}(x),
\label{eq:vnxt}
\end{equation}
where the vertical eigenmodes $Z_{nm}(x)$ satisfy
\begin{equation}
{\cal{L}}(Z_{nm}) \equiv \frac{d^{4} Z_{nm}}{d x^{4}} = \mu_{nm}^{4} Z_{nm}, \label{eq:Znm}
\end{equation}
for some constants $\mu_{nm}$. It can be shown that the $Z_{nm}$ functions form an orthogonal set, so that projecting \eqref{eq:v4} onto each of them individually gives, for each $n$ and $m$,
\begin{equation}
\frac{dV_{nm}}{dt} + \frac{V_{nm}}{ \tau^{\rm ES}_{nm}(t) } = \dot{\Omega}_{\rm cz}(t)\frac{4 R^{2}}{\lambda^3_{n}J_1(\lambda_{n})}\frac{\int_{0}^{1}Z_{nm}(x)dx}{\int_{0}^{1}Z_{nm}^{2}(x)dx} \equiv F_{nm}(t) \mbox{  ,} \label{eq:Tnm1}
\end{equation}
where additional information on $Z_{nm}$ and $\mu_{nm}$, are given in Appendix B. The quantity  
\begin{equation}
 \tau^{\rm ES}_{nm}(t) = \frac{\bar N^2_{\rm tc} \Delta^4(t) }{4\Omega^2_{\rm cz}(t) R^2 \kappa_{\rm tc} }\frac{\lambda_{n}^{2}}{\mu_{nm}^{4}} = \frac{\lambda_{n}^{2}}{2\mu_{nm}^{4}}  t_{\rm ES}\left(\Omega_{\rm cz}(t)\right) \mbox{  ,}  
 \label{eq:tESnm}
\end{equation}
where $t_{\rm ES}(\Omega_{\rm cz})$ was defined in Equation \eqref{eq:oom}, naturally emerges from this calculation, and can be interpreted as a local Eddington-Sweet timescale based on the typical geometry of the spatial eigenmode considered. 

Equation~\eqref{eq:Tnm1} can easily be solved using an integrating factor, 
\begin{equation}
\mu(t) = \exp\left( \int_{t_0}^t  \frac{1}{\tau^{\rm ES}_{nm}(t')} dt' \right) \mbox{ ,}
\end{equation}
which yields
\begin{eqnarray}
V_{nm}(t) &=& \frac{1}{\mu(t) } \left[ V_{nm}(t_0)  + \int_{t_0}^t  \mu(t')F_{nm}(t') dt' \right] \nonumber \\
&=&  \exp\left( - \int_{t_0}^t  \frac{1}{\tau^{\rm ES}_{nm}(t')} dt' \right) V_{nm}(t_0)  + \int_{t_0}^t    \exp\left( \int_{t}^{t'}  \frac{1}{\tau^{\rm ES}_{nm}(t'')} dt'' \right)      F_{nm}(t') dt' \mbox{  .}
\label{eq:Vnmsol}
\end{eqnarray}

As expected, we see that $V_{nm}(t)$ contains two terms, one that depends on the initial conditions (the first term on the right-hand side of Equation \ref{eq:Vnmsol}) and one that depends on the forcing applied to the system (the second term on the right-hand side of Equation \ref{eq:Vnmsol}). 
For $V_{nm}(t)$ to tend to the quasi-steady solution discussed in Section \ref{sec:steadystate} as $t\rightarrow +\infty$ (see Equation \ref{eq:v1}), 
the effect of the initial conditions must decay, since \eqref{eq:v1} is independent of the initial differential rotation profile of the star. Furthermore, the terms containing the forcing in Equation \eqref{eq:Vnmsol}, when recombined as in Equation \eqref{eq:vnxt} must eventually recover \eqref{eq:v1}. Whether this occurs or not clearly depends on the behavior of the integrating factor $\mu(t)$. We now study the latter in more detail. 

We first re-write the integral in $\mu(t)$ as
\begin{equation}
 \int_{t_0}^t  \frac{1}{\tau^{\rm ES}_{nm}(t')} dt' =  - \frac{2\mu_{nm}^4}{\lambda_{n}^2}\int_{t_0}^t  \frac{t_{\rm sd}(t')}{t_{\rm ES}(t')} \frac{\dot \Omega_{\rm cz}(t')}{\Omega_{\rm cz}(t')} dt'  =  - \frac{2\mu_{nm}^4}{\lambda_{n}^2} \int_{\Omega_0}^{\Omega_{\rm cz}(t)} \frac{t_{\rm sd}(\Omega_{\rm cz})}{t_{\rm ES}(\Omega_{\rm cz})} \frac{d \Omega_{\rm cz}}{\Omega_{\rm cz}} \mbox{  ,}
 \label{eq:integral}
\end{equation}
where the spin-down timescale $t_{\rm sd}$ was defined in Equation \eqref{eq:tesdef}.
Writing it in this form enables us to study the long-term behavior of this integral for a fairly broad class of problems. 

First, note that for an exponential spin-down law (see Equation \ref{eq:omegaexp}), $t_{\rm sd}$ is constant (and equal to $1/k$) while for a power-law spin-down rate (see Equation \ref{eq:omegapow}), then
\begin{equation}
t_{\rm sd}(\Omega_{\rm cz})= \frac{t_0}{\alpha} \left(\frac{\Omega_{\rm cz}}{\Omega_0}  \right)^{-\frac{1}{\alpha}}  = t_{\rm sd}(\Omega_0) \left(\frac{\Omega_{\rm cz}}{\Omega_0}  \right)^{-\alpha^{-1}}   \mbox{  .}
\label{eq:tsdocz}
\end{equation}
The second form of $t_{\rm sd}$ written above can actually be used to describe both exponential and power-law spin-down models if one views the exponential case as having $\alpha^{-1} = 0$. 

Next, note that both GM98 and \citet{Woodal11} found $\Delta$ to be a power-law function of the mean stellar rotation rate. While their findings do not directly apply here, since they were derived assuming that the tachocline circulation is driven by the latitudinal shear in the convection zone rather than by spin-down, we may nevertheless safely assume a similar functional dependence\footnote{While $\Delta$ must be an input of the problem in the non-magnetic case, it is intrinsically related to stellar properties, to the spin-down rate and to the assumed internal magnetic field strength in the magnetic case, as shown in Appendix E. Here, we take the simplest possible form for $\Delta$, and assume that any time-dependence not included in $\Omega_{\rm cz}$ is much slower and can thus be neglected.}, taking 
\begin{equation}
\Delta = \Delta_0 \left( \frac{\Omega_{\rm cz}}{\Omega_0}\right)^{\beta} ,
\label{eq:tachoindex}
\end{equation}
(where we anticipate that $\beta \ge 0$, and $\Delta_0$ is by construction $\Delta$ at $t=t_0$), so that 
\begin{equation}
t_{\rm ES}(\Omega_{\rm cz}) = t_{\rm ES}(\Omega_0) \left( \frac{\Omega_{\rm cz}}{\Omega_0}\right)^{4\beta - 2} \mbox{  .}
\label{eq:tesocz}
\end{equation}
Combining Equations \eqref{eq:tesocz} and \eqref{eq:tsdocz} with \eqref{eq:integral}, we find that 
\begin{eqnarray}
 \int_{t_0}^t  \frac{1}{\tau^{\rm ES}_{nm}(t')} dt' &=&   -  \frac{t_{\rm sd}(\Omega_0)  }{ \tau_{nm}^{\rm ES}(\Omega_0)  }  \int_{\Omega_0}^{\Omega_{\rm cz}(t)} \left(\frac{\Omega_{\rm cz}}{\Omega_0}  \right)^{1 - \alpha^{-1} - 4\beta   }  \frac{d \Omega_{\rm cz} }{\Omega_0} \nonumber \\ 
 &=& \begin{cases} \displaystyle{-  \frac{1}{ q }  \frac{t_{\rm sd}(\Omega_0)}{\tau^{\rm ES}_{nm}(\Omega_0)}   \left[ \left(\frac{\Omega_{\rm cz}}{\Omega_0}  \right)^{q}  - 1 \right]}  & \mbox{ if } q\ne 0, \\
 \displaystyle{ - \frac{t_{\rm sd}(\Omega_0)}{\tau^{\rm ES}_{nm}(\Omega_0)}  \ln \left(\frac{\Omega_{\rm cz}}{\Omega_0}  \right)} & \mbox{ if } q = 0 , \end{cases}
\label{eq:theintegral}
 \end{eqnarray}
where $q  = 2 -  \alpha^{-1} - 4\beta$. We therefore see that the behavior of the integrating factor $\mu(t)$ depends {\it only} on the sign of $q$. 

If $q>0$, then the right-hand-side of Equation \eqref{eq:theintegral} tends to a constant as the star spins down. In that case, $\mu(t)$ also tends to a constant as $t \rightarrow +\infty$, which implies that the contribution of the initial conditions to $V_{nm}(t)$ does not disappear in Equation \eqref{eq:Vnmsol}. In other words, the star cannot relax to the state described by the quasi-steady solution discussed in Section \ref{sec:steadystate}, and the latter becomes irrelevant to the spin-down problem.

On the other hand, if $q\le0$ then the right-hand-side of Equation \eqref{eq:theintegral} is positive, and increases towards $+\infty$ as the star spins down. In that case, the integrating factor $\mu(t)$ also increases with time (super-exponentially when $q<0$, and as $\Omega_{\rm cz}^{-1}(t)$ when $q=0$), which implies that the contribution of the initial conditions to $V_{nm}(t)$ rapidly disappears. For $q<0$, the latter decays exponentially roughly on the local Eddington-Sweet timescale across the tachocline. This timescale decreases with time and rapidly becomes much smaller than the age of the star except if the tachocline is very thick (which we explicitly assumed was not the case). 

Furthermore, it can be shown with additional algebra that the complete transient solution given by Equation \eqref{eq:vnxt} actually tends to the quasi-steady solution \eqref{eq:v1} when $t \rightarrow +\infty$, when $q \le 0$ (see Appendix B2).  In other words, the quasi-steady solution derived and discussed in Section \ref{sec:steadystate} is a meaningful description of stellar spin-down, after a transient phase which is short compared with the age of the star, during which all knowledge of the initial rotation profile disappears. 

The physical interpretation of $q$ as a critical value of this problem is quite straightforward given that $q$ is effectively defined so that
\begin{equation} 
\frac{t_{\rm sd}(\Omega_{\rm cz}) }{t_{\rm ES}(\Omega_{\rm cz}) } \propto \left(\frac{\Omega_{\rm cz}}{\Omega_0} \right) ^q \mbox{  .}
\label{eq:ratio}
\end{equation}
 If $q >  0$, then the ratio of the spin-down timescale to the local angular-momentum transport timescale across the tachocline decreases as the star spins down. This implies that the meridional flows are less and less efficient at extracting angular momentum from the tachocline relative to the rate at which it is removed from the envelope. The lag between the envelope and the bottom of the tachocline then increases with time, until such a point where new dynamics not taken into account here, such as turbulent transport, must come into play. In other words, the quasi-steady solution derived in the previous Section is only of limited validity. When $q \le 0$ on the other hand, the converse is true: the system rapidly tends to the quasi-steady solution {\it regardless} of the initial conditions. %In this quasi-steady state, the tachocline shear adjusts itself in such a way as to drive meridional flows that extract exactly the same amount of angular momentum from the core per unit time that is removed from the system overall by the wind. 
 
\subsection{Discussion}
\label{sec:discuss2}

Our findings regarding the dynamics of both transient and quasi-steady spin-down solutions can easily be summarized as follows. If, for a given spin-down law $\Omega_{\rm cz}(t)$ and a given tachocline structure (characterized by its local thermodynamic properties and its thickness), the system is such that the ratio of the spin-down timescale to the local Eddington--Sweet timescale $ t_{\rm sd}(\Omega_{\rm cz})/t_{\rm ES}(\Omega_{\rm cz})$ (where $t_{\rm sd}$ is given in Equation \ref{eq:tsdocz} and $t_{\rm ES}$ is given in Equation \ref{eq:tesocz}) monotonically decreases as the star spins down, then angular-momentum transport by large-scale meridional flows across the tachocline is not sufficient to maintain dynamical coupling with the envelope. After some time, the shear across the tachocline is likely to become large enough to be unstable to shearing instabilities. Angular-momentum transport will then be dominated by turbulent motions, and must be described using an entirely different formalism (not discussed here).   

If, on the other hand, the system is such that $ t_{\rm sd}(\Omega_{\rm cz})/t_{\rm ES}(\Omega_{\rm cz})$ increases or remains constant as the star spins down, then the tachocline remains coupled to the envelope, and the shear simply adjusts itself geostrophically so that the angular momentum flux transported by the meridional flows out of the tachocline is, at all times, equal to angular-momentum flux removed from the star by the wind. The relative lag $\Omega_{\rm b}/\Omega_{\rm cz}$ between the base and the top of the tachocline, in this case, is correctly given by the quasi-steady solution, and is roughly equal to $t_{\rm ES}(\Omega_{\rm cz}) / t_{\rm sd}(\Omega_{\rm cz}) $ (which is either constant, or decreases with time), see Equation \eqref{eq:oom}.

We therefore find that the quasi-steady solution is conveniently valid whenever it makes sense, that is, whenever $\Omega_{\rm b}/\Omega_{\rm cz} \propto t_{\rm ES}(\Omega_{\rm cz}) / t_{\rm sd}(\Omega_{\rm cz})$ decreases (or at least remains constant and much smaller than one) as the star spins down. In other words, we can actually avoid the calculation of the transient solution entirely, since the quasi-steady solution itself provides all the information needed as to the limits of its own validity.

\section{SPIN-DOWN OF A MAGNETIZED STAR}
\label{sec:blcase}

We now finally return to the originally-posed problem and investigate the manner in which the tachocline spin-down  is finally communicated to the deep radiative interior. In the GM98 model, this process is mediated by magnetic torques within the tachopause, a thin boundary layer that separates the tachocline from the magnetically-dominated, uniformly-rotating region below (see Figure \ref{fig:model}a). These torques are generated as the primordial magnetic field, confined below the tachocline by downwelling meridional flows, is wound up into a significant toroidal field by the rotational shear. 
In order to model angular-momentum transport across the tachopause exactly, one should therefore solve the magnetic induction equation in addition to the previously discussed equations describing the tachocline dynamics (i.e. Equations \ref{eq:bulkmom} and \ref{eq:thermaleq}), and include the Lorentz force in the momentum balance. The nonlinear nature of the added terms, unfortunately, makes it impossible to derive exact analytical solutions of the problem without further assumptions. 

\citet{Woodal11}, however, were able to derive analytical solutions for a simplified version of the GM98 model. They found that the mathematical equations describing tachopause are, in many ways, analogous to those describing a viscous Ekman layer, with the viscous drag force replaced by a magnetic one. In other words, they showed that one can develop physical insight into the problem {\it and} obtain quantitatively meaningful results by considering a thought experiment in which the uniformly rotating part of the radiative interior is simply a massive, impenetrable solid sphere (or, in our case, a cylinder), whose rotation rate is gradually spun-down by the fluid lying above through friction. 

Our final model is thus constructed as follows. We consider the same setup as the one described and studied in Section \ref{sec:strat1}, but the base of the tachocline is now no longer passive. Instead, it hosts a thin tachopause of thickness $\delta$, which communicates the tachocline spin-down to the rigidly-rotating, impermeable interior via magnetic torques. The latter will be modeled using a boundary layer jump condition. We thus recover the picture first presented in Figure \ref{fig:model}b.

\subsection{Global angular-momentum balance}
\label{sec:globalbalance}

Let $\Omega_{\rm c}$ be the angular velocity of the rigidly rotating ``core" region of the radiation zone, expressed in the rotating frame. Rigid-body rotation throughout the entire star implies that $\Omega_{\rm c} \simeq 0$, while $\Omega_{\rm c} > 0$ expresses a lag between the core and the convection zone. In our cylindrical model, the core spans the region $z<z_{\rm tc} $, $s<R$ (see Figure \ref{fig:model}). Recall that the tachopause spans the interval $[z_{\rm tc} ,z_{\rm tc} + \delta]$, and is assumed to be thinner than the tachocline, which lies above (with $ z \in [z_{\rm tc} +\delta, z_{\rm cz}])$. 

Although negligible in the tachocline, magnetic stresses are significant within the tachopause, and must be included when studying the global angular-momentum balance. To find an evolution equation for $\Omega_{\rm c}(t)$, we thus begin by writing the complete angular-momentum conservation equation as
\begin{equation}
 \frac{\partial}{\partial t}(\bar \rho sv + \bar \rho s^{2}\Omega_{\rm cz}) + \nabla \cdot \left(\bar \rho \mathbf{u} s^{2}  \Omega_{\rm cz} - \frac{sB_\phi {\bf B} }{4\pi} \right) = 0, \label{eq:angmomcons}
%Checked
\end{equation}
where ${\bf B} = (B_s,B_\phi,B_z)$ is the magnetic field, and where we have ignored viscous stresses on the grounds that they are most likely negligible. Note that we assume that the system is laminar, and ignore the contribution of turbulent transport  -- this assumption is discussed in Sections \ref{sec:EkmanMagneticbl} and \ref{sec:discuss}.

Integrating Equation \eqref{eq:angmomcons} over the volume $V$ of the core up to the top of the tachopause, and using the divergence theorem then yields 
\begin{align}
&  2\pi \int_{0}^{z_{\rm tc}+\delta}\int_{0}^{R}\frac{\partial}{\partial t}(\bar \rho s v + \bar \rho s^{2}\Omega_{\rm cz}) s ds  dz + 2\pi R \int_{z_{\rm tc}}^{z_{\rm tc}+\delta} \left(\bar \rho u s^{2}  \Omega_{\rm cz} - \frac{sB_\phi B_s }{4\pi} \right)_{s=R} dz  \nonumber \\
& \quad \quad \quad + 2\pi \int_{0}^{R} \left( \bar \rho w s^{2}  \Omega_{\rm cz} - \frac{sB_\phi B_z }{4\pi} \right)_{z=z_{\rm tc}+\delta}  s ds 
  = 0 , \label{eq:appl0} \end{align}
%Checked
where the second integral is a surface integral through the side of the tachopause, and the third integral is a surface integral through the top of the tachopause. To derive this equation, we have used the fact that the angular-momentum flux through the bottom and side boundaries of the core is zero. Indeed, $\bf{u}$ and $B_\phi$  disappear since the core is assumed to be rigidly rotating and impermeable.   

By definition of the tachocline, the magnetic  torque becomes negligible just above the tachopause, and thus disappears from the surface integral at $z = z_{\rm tc} + \delta$. We assume that it also disappears from the integral on the side-boundary. This assumption is somewhat difficult to justify a priori, and will require verification when full numerical solutions of the problem, in a spherical geometry, are available. However, it is consistent with the assumption that the dynamics occurring beyond the sides of the cylinders do not directly affect spin-down. We argue that it is at least plausible as long as the magnitude of the toroidal field $B_\phi$ on the side-wall of the tachopause remains small, which requires in turn that the radial angular-velocity gradient $\partial \Omega / \partial s$ at the same location be small. 

The remaining terms in Equation (\ref{eq:appl0}) can be expressed as  
\begin{equation}
\frac{d}{dt} \left[ I_{\rm core} (\Omega_{\rm c}+\Omega_{\rm cz})\right] + 2\pi \bar{\rho}_{\rm tc}\Omega_{\rm cz} R^{3}\int_{z_{\rm tc}}^{z_{\rm tc}+\delta}u(R,z,t)dz + 2 \pi \bar{\rho}_{\rm tc}\Omega_{\rm cz} \int_{0}^{R}w(s,z_{\rm tc}+\delta,t)s^{3}ds  \approx 0, \label{eq:appI1}
%Checked
\end{equation}
where, as in the previous Section, $\bar \rho_{\rm tc}$ is defined as the mean density of the tachocline and tachopause region, and where $I_{\rm core} $ is the moment of inertia of the core and tachopause combined, defined as 
\begin{equation}
 I_{\rm core} = \int_{V}\bar \rho(z)  s^{2}dV = \frac{\pi}{2}R^{4}\int_{0}^{z_{\rm tc}+\delta}\bar \rho(z)dz = \frac{M_{\rm core}}{2}R^{2}, \label{eq:Ib}
%Checked
\end{equation}
where $M_{\rm core}$ is the mass of the cylinder included in the volume $V$. Note that in order to derive \eqref{eq:appI1}, we have used the fact that $v = s\Omega_{\rm c}$ within the core, and assumed that $v \simeq s \Omega_{\rm c}$ in the tachopause as well. Since the tachopause is very thin, the error made  has negligible impact on the result.

Within the scope of these assumptions, angular momentum is extracted from the core through a series of channels, which unfold as follows. The surface layers are spun-down by the magnetized wind torque, and then communicate the spin-down information to the rest of the convection zone by turbulent stresses. The spin-down torque also drives large-scale meridional flows in the convection zone, which extract angular momentum from the tachocline and the top of the tachopause (by flowing downward from the convection zone and then outward in the tachopause), and at the same time confine the internal magnetic field. The tachopause then finally spins the core down via magnetic stresses, generated as the local core-tachopause shear winds the internal poloidal field into a toroidal one. Equation (\ref{eq:appI1}) describes only the hydrodynamic processes in the tachocline, but the other channels are implied in the assumptions that (1)  the convection zone rotates at the velocity $\Omega_{\rm cz}$, (2) the tachopause rotates at a velocity close to $\Omega_{\rm c}$, and (3) the vertical flow at the base of the tachocline is given by the magnetic jump condition that describes the tachopause dynamics. We study the latter in Section \ref{sec:EkmanMagneticbl}.

To estimate the second term in Equation (\ref{eq:appI1}), note that  the mass flux entering the tachopause through the surface $z=z_{\rm tc}+\delta$ must be the same as that leaving through the side wall, since the core is impermeable. Hence: 
\begin{equation}
 2\pi\int_{0}^{R}w(s,z_{\rm tc}+\delta,t) s ds = -2\pi R\int_{z_{\rm tc}}^{z_{\rm tc}+\delta}u(R,z,t) dz.
%Checked
\end{equation}
Combining this with \eqref{eq:appI1} gives
\begin{align}
\frac{d J_{\rm core}}{dt} +  2 \pi \bar{\rho}_{\rm tc} \Omega_{\rm cz} \int_{0}^{R} s w_{0}(s,t) (R^2 - s^2) ds \simeq 0 ,
 \label{eq:amw0}
%Checked
\end{align}
where 
\begin{equation}
J_{\rm core} = I_{\rm core}(\Omega_{\cz} + \Omega_{\rm c})
\end{equation}
is the total angular momentum of the core\footnote{Technically, $J_{\rm core}$ includes the angular momentum of the tachopause, but the latter is negligible in comparison.} in the inertial frame, and where we have introduced $w_0(s,t) = w(s,z_{\rm tc}+\delta,t)$ for simplicity. Equation \eqref{eq:amw0} thus shows that the rate of angular-momentum transport between the surface layers and the core is fully determined once $\bar{\rho}_{\rm tc} w_0(s,t)$, the vertical mass flux downwelling from the tachocline into tachopause, is known. 

To find $w_0(s,t)$, we must once more solve the equations describing the dynamics of the tachocline and of the convection zone, and match them to one another at the radiative--convective interface. The main difference with the work presented in Section \ref{sec:steadystate} lies in the treatment of the lower boundary of the tachocline, which is no longer passive nor strictly impermeable, but must instead be modified to take the presence of the tachopause into account. This is done by replacing the impermeability condition used in Section \ref{sec:steadystate} by a ``jump condition", that relates $w_0$ to $v_0(s,t) = v(s,z_{\tc} + \delta,t)$ at the bottom of the tachocline, and depends on the magnetohydrodynamics of the tachopause. This jump condition is now derived. 

\subsection{Tachopause jump condition} 
\label{sec:EkmanMagneticbl}

Guided by the analogy between the tachopause and an Ekman layer suggested by the work of \citet{Woodal11}, we begin by deriving a tachopause jump condition assuming that its dynamics are dominated by viscous torques only. This assumption greatly facilitates our derivation, and the result can then be used without any further algebra to deduce the equivalent jump condition for a magnetized tachopause. 

While well-known in the context of a steadily-rotating frame, the derivation of the Ekman jump condition has not yet, to our knowledge, been done in a frame that is spinning down. The steps of the calculation are essentially identical, however, and are presented in Appendix C. We find that the quasi-steady vertical and azimuthal velocity profiles at the base of the tachocline, $w_0(s)$ and $v_0(s)$, are related via: 
\begin{equation}
 w_{0}= \frac{\dot{\Omega}_{\rm cz}}{\Omega_{\rm cz}}\delta_{\rm E} \left( 1 - \frac{1}{\sqrt{2}} \right)  - \frac{\delta_{\rm E}}{\sqrt{2}s}\frac{\partial}{\partial s}(s^{2} \Omega_{\rm c} - sv_{0}), \label{eq:jump}
\end{equation}
where $\delta_{\rm E} = \sqrt{\nu_{\rm tc}/2\Omega_{\rm cz}}$ is the thickness of the Ekman layer that mimics the tachopause, and is based on the local viscosity $\nu_{\rm tc}$ of the star at $z = z_{\rm tc}$. 
The second term on the right-hand-side is the standard jump condition (expressed in a cylindrical coordinate system), while the first term is the correction arising from Euler's force. 

In stars, however, the tachopause transmits the spin-down torque via magnetic rather than viscous friction. \citet{Woodal11} found that its thickness\footnote{The tachopause studied by \citet{Woodal11} differs from that of GM98. The \citet{Woodal11} tachopause is assumed to be isothermal, while the GM98 tachopause is in thermal equilibrium. This results in rather different structures and thicknesses. We note that the GM98 tachopause does not lead to a jump condition that is strictly analogous to an Ekman jump condition, but instead, contains an additional term that depends on the temperature perturbation, see \citet{Woodal11}. For simplicity, we restrict our analysis to the case of an isothermal tachopause.} is given by 
\begin{equation}
 \delta = \sqrt{\frac{2\pi\bar{\rho}_{\rm tc} \eta_{\rm tc} \Omega_{\rm cz} R^{2}}{B_{0}^{2}}},  \label{eq:delta}
\end{equation}
where $B_0$ is the strength of the confined magnetic field (just below $z_{\rm tc}$), and $\eta_{\rm tc}$ is the local magnetic diffusivity of the star near $z = z_{\rm tc}$. They also found that the jump condition relating the vertical and azimuthal velocities in the tachocline is exactly of the same mathematical form as that of an Ekman layer, albeit with the numerical constant $1/\sqrt{2}$ replaced by $\pi/4$, and $\delta_{\rm E}$ replaced by $\delta$. The magnetic jump condition thus becomes 
\begin{equation}
 w_{0} = \frac{\dot{\Omega}_{\rm cz} }{\Omega_{\rm cz} }\delta \left( 1 -  \frac{\pi}{4} \right) - \frac{\pi}{4}\frac{\delta}{s}\frac{\partial}{\partial s}\left(s^{2} \Omega_{\rm c} - sv_{0}\right). \label{eq:Bw}
\end{equation}

In both cases discussed above, the tachopause is assumed to be laminar. The angular-momentum transport budget within this region involves only large-scale flows, and either magnetic or viscous stresses. The reason behind the similarity of the two jump conditions comes from the fact that viscous and magnetic stresses are directly proportional to the local angular-velocity shear\footnote{This is by definition in the case of viscous stresses, and indirectly through the linearized Lorentz force and induction equation in the magnetic case. }. As a result, one may conjecture that any boundary layer in which the angular-momentum balance relies on large-scale meridional flows and some form of stress that is proportional to the angular-velocity shear will also result in the same type of jump condition, {\it even if that boundary layer is not laminar.} From this argument, we propose that the general form of the jump condition should be
\begin{equation}
 w_{0} = \frac{\dot{\Omega}_{\rm cz} }{\Omega_{\rm cz} }\delta (1-C) - C\frac{\delta}{s}\frac{\partial}{\partial s}\left(s^{2} \Omega_{\rm c}  -s v_{0}\right), \label{eq:jumpgen}
\end{equation}
where $C$ is a constant of order unity which depends on the type of stresses acting in the tachopause, and $\delta$ is its thickness, which is no longer necessarily related to $B_0$ via \eqref{eq:delta} specifically, but likely depends on $B_0$ in some form or another. 

In any case, as we demonstrate below, the long-term behavior of the spin-down problem is ultimately controlled only by the {\it slowest} timescale in the sequence of processes responsible for angular-momentum transport from the core to the surface. In most cases, this turns out to be the Eddington--Sweet timescale across the tachocline, rather than any timescale intrinsic to the tachopause. In this sense, the global spin-down timescale is usually independent of the exact nature and structure of the tachopause (at least, in an explicit sense, see Section \ref{sec:ccl3} and Appendix E), unless the latter is very thick -- which we have previously assumed is not the case.

\subsection{Evolution of the core angular momentum}
\label{sec:globalsol}

We can now solve the system of equations governing the convection zone and the tachocline as in Section \ref{sec:strat1}, using this time the jump condition \eqref{eq:jumpgen} as a lower boundary condition for the tachocline flows. At this junction we have two possibilities: to solve the full time-dependent tachocline dynamics as in Section \ref{sec:transient}, or to study them using a quasi-steady approximation as in Section \ref{sec:steadystate}. Having proved in Section \ref{sec:transient} that the time-dependent solution very rapidly tends to the quasi-steady solution (when it is well-behaved), and given that the latter is much more easily derived, we assume here that the tachocline dynamics are in a quasi-steady state\footnote{This assumption is equivalent to requiring that the tachocline be in thermal equilibrium and in force balance at all times during spin-down, which is likely to be true as long as the tachocline is thin, and its local Eddington-Sweet timescale is always short compared with the spin-down timescale.}. The derivation of this solution is presented in Appendix D, and eventually yields the mass flux into the tachopause, $\bar{\rho}_{\rm tc} w_0(s)$. Using the expression obtained into Equation \eqref{eq:amw0}, we then find that for $\delta \ll \Delta$ (i.e. when the tachopause is much thinner than the tachocline), then 
\begin{equation}
 \frac{dJ_{\rm core}}{dt} =  -\dot{\Omega}_{\cz}I_{\rm tc} + \frac{\Omega_{\rm cz} }{\Delta} I_{\rm tc}\sum_{n}\frac{32}{\lambda_{n}^{4}}\frac{a_{n}}{b_{n}}, \label{eq:Omegabstrat}
\end{equation}
where 
\begin{equation}
I_{\rm tc} = \frac{\bar \rho_{\rm tc} \pi \Delta R^4}{2}
\label{eq:tachoinertia}
\end{equation}
is the moment of inertia of the tachocline, and with
\begin{align}
 a_{n} &= \frac{\dot{\Omega}_{\rm cz}}{\Omega_{\rm cz}} \Delta  - 2C\delta \Omega_{\rm c} - \frac{C\delta \bar{N}_{\rm tc}^{2}\Delta}{2\Omega_{\rm cz} \kappa_{\rm tc}}\frac{\dot{\Omega}_{\rm cz}}{\Omega_{\rm cz}}\left(\frac{R}{\lambda_{n}}\frac{\cosh\left(\lambda_{n}\frac{\Delta}{R}\right) - 1}{\sinh\left(\lambda_{n}\frac{\Delta}{R}\right)} - \frac{\Delta}{2}\right), \label{eq:an} \\
 b_{n} &= 1+\frac{C\delta\lambda_{n}}{2\Omega_{\rm cz} R}\left[\frac{2 \bar{N}_{\rm tc}^{2}R^{2}}{\kappa_{\rm tc} \lambda_{n}^{2}}\frac{1-\cosh\left( \lambda_{n}\frac{\Delta}{R}\right) }{\sinh\left( \lambda_{n}\frac{\Delta}{R}\right)} + \frac{\bar{N}_{\rm tc}^{2}\Delta R}{\kappa_{\rm tc} \lambda_{n}} + \frac{1}{\tau \tanh\left(  \lambda_{n}\frac{H-z_{\cz}}{R}\right)}\right], \label{eq:bn1}
\end{align}
where $\Delta \simeq z_{\cz}-z_{\rm tc}$ is the thickness of the tachocline. Note that for most stars whose outer convection zone is not too thin, and where $\tau$ is not too small, the last term in the square brackets of Equation \eqref{eq:bn1} is negligible compared with the first two. In what follows, we neglect it, but bear in mind that it may be important for solar-type stars whose mass approach the critical mass above which the outer convection zone disappears. 

While somewhat obscure at first, Equation \eqref{eq:Omegabstrat} has a simple limit. Indeed, when $\delta = 0$, 
\begin{equation}
a_{n} = \frac{\dot{\Omega}_{\rm cz}}{\Omega_{\rm cz}}\Delta  \mbox{   and   }    b_{n} = 1 .
\end{equation}
Using properties of Bessel Functions, it can be shown that $\sum_n 32/ \lambda_n^{4} = 1$. As a result,
\begin{equation}
\frac{dJ_{\rm core}}{dt} = 0,  
\end{equation}
which implies that, when viewed in an inertial frame, the core retains its initial angular momentum. This result is as expected, since $\delta = 0$ means that the tachopause is absent, and without it the tachocline cannot exert any torque on the core. In other words, the convective envelope and the tachocline both spin down exactly as described in Section \ref{sec:steadystate}, but the core does not.  

\begin{figure}[h]
\begin{center}
(a) \includegraphics[width=0.45\textwidth]{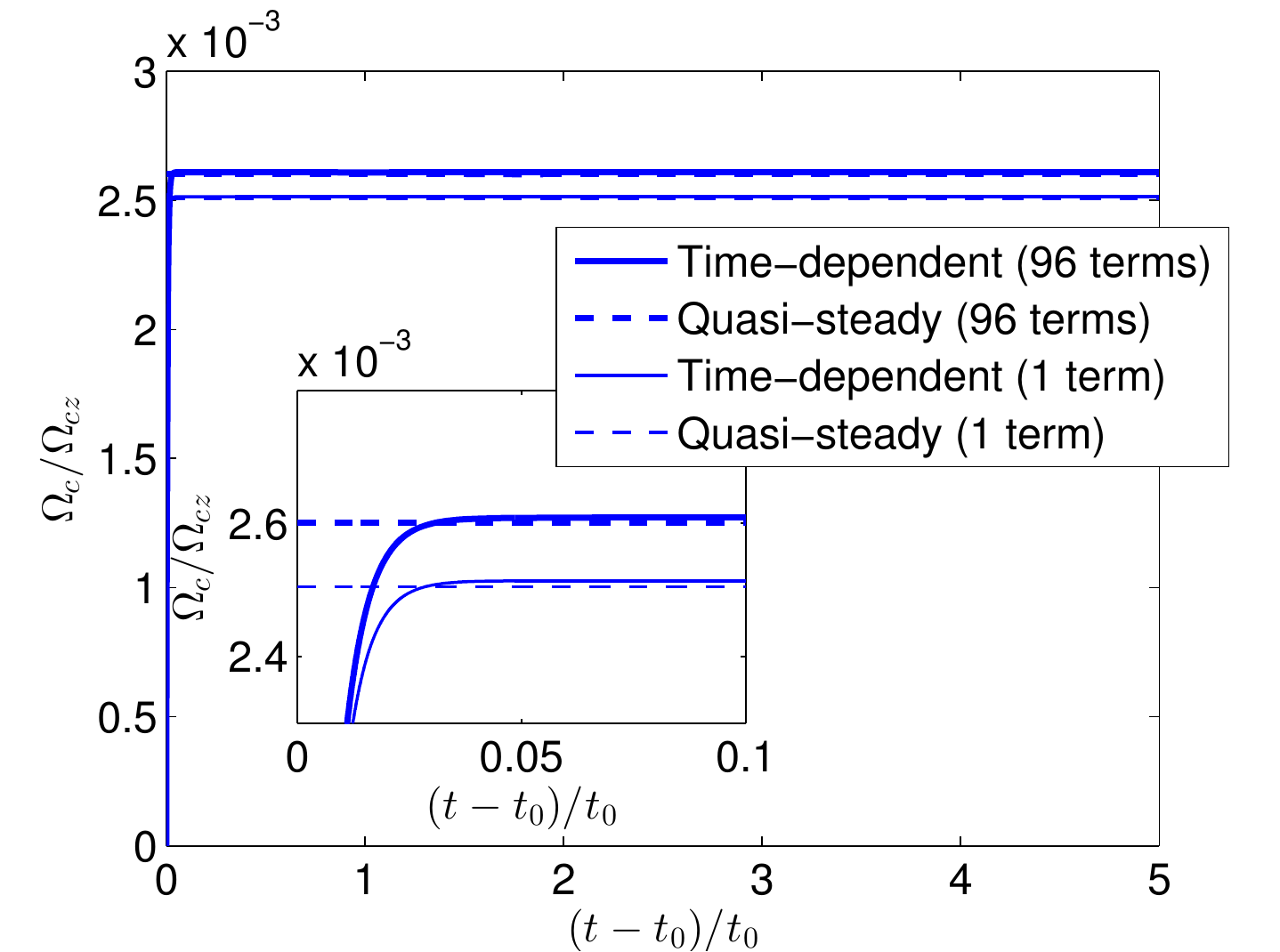}
(b) \includegraphics[width=0.45\textwidth]{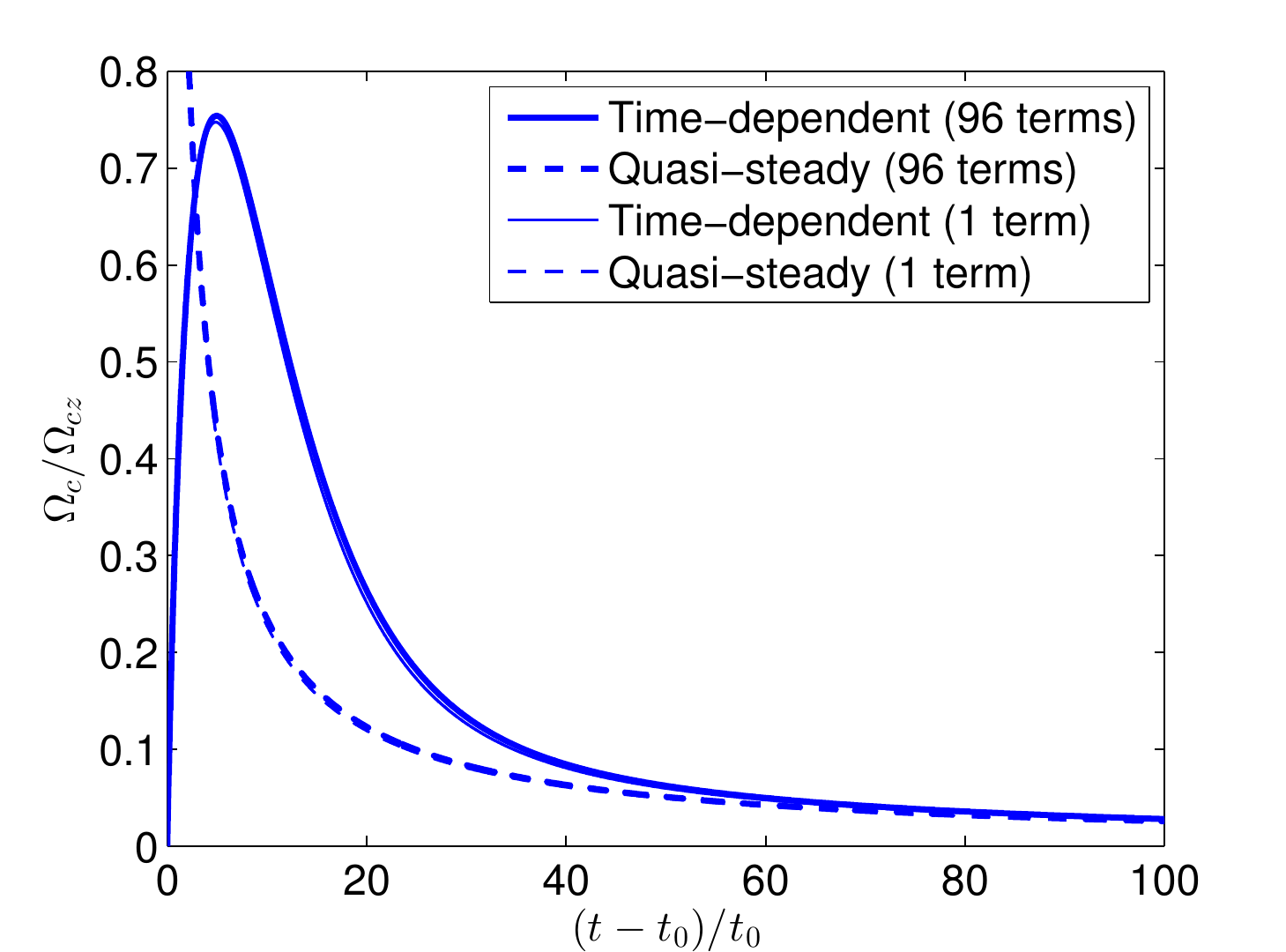}
\end{center}
\caption{ (a) Evolution of $\Omega_{\rm c}/\Omega_{\rm cz}$ for a star with parameters shown in Table \ref{tab1}, undergoing spin-down following Skumanich's law. The solid (red) line is the exact solution of Equation \eqref{eq:Omegabstrat} using 96 terms in the sum. The dot-dashed (blue) line is the exact solution of Equation \eqref{eq:Omegac_simple}, which keeps only one term in the sum. The latter is a fairly good approximation to the ``true" solution. The dashed and dotted lines are the quasi-steady approximations to the full solutions, obtained by dropping the time-derivative of $\Omega_{\rm c}$ in Equation \eqref{eq:Omegabstrat} and solving for $\Omega_{\rm c}$ algebraically. % (see Section XXX for detail). 
The 1-term quasi-steady solution is given by Equation \eqref{eq:finalcorescaling}. They are a good approximation to the full solution after a transient period whose duration depends on the thickness of the tachocline. (b) Same figure, but for a tachocline whose thickness varies as $(\Omega_{\rm cz}/\Omega_0)^{2/3}$ (see main text for detail). Because the initial tachocline thickness in this model is much thicker, the true solution takes much longer to approach the quasi-steady solution.}
\label{fig:Omegab6}
\end{figure}

When $\delta>0$, by contrast, the angular velocity of the core evolves with time in response to the spin-down torque communicated by the tachopause. This is illustrated in Figure~\ref{fig:Omegab6}, which shows a plot of the relative core-envelope lag $\Omega_{\rm c}/\Omega_{\rm cz}$, as a function of time, in two idealized test cases. Note that the moment of inertia of the core is assumed to be constant to simplify the interpretation of the results. At time $t = t_0$ we also assume for simplicity that $\Omega_{\rm c}(t_{0}) = 0$ (or in other words, that the star is uniformly rotating). In Figure~\ref{fig:Omegab6}a, we show $\Omega_{\rm c}/\Omega_{\rm cz}$ for a ``reference" star whose parameters are summarized in Table \ref{tab1}, using a Skumanich spin-down law with $\Omega_{\rm cz}(t) = \Omega_{0}(t/t_0)^{-1/2}$ (i.e. $\alpha = 1/2$), and with a tachocline of constant thickness (setting $\beta = 0$ in Equation \ref{eq:tachoindex}).  In Figure~\ref{fig:Omegab6}b, we evolve the same star but allow for a tachocline whose thickness varies with $\Omega_{\rm cz}$, taking $\beta = 2/3$ in Equation \eqref{eq:tachoindex} and choosing $\Delta_0$ such that the values of $\Delta(t)$ in both Figure \ref{fig:Omegab6}a and Figure \ref{fig:Omegab6}b agree at $t = 10^3 t_0$ (which more-or-less represents the ``present day" time for the Sun). In both figures the relative core-envelope lag first increases rapidly, then eventually converges to a {\it global} quasi-steady state\footnote{We differentiate here between a {\it global} quasi-steady state where {\it all} layers of the star are evolving concurrently, and the {\it local} quasi-steady state discussed in Section \ref{sec:steadystate} where only the tachocline dynamics are quasi-steady.} whose time-dependence can be predicted analytically (see Section \ref{sec:quasisteady2}). The time taken to reach this global quasi-steady state, however, depends sensitively on the thickness of the tachocline (see Section \ref{sec:transient2}), and is much larger in Figure \ref{fig:Omegab6}b, which has a much larger initial tachocline thickness, than in Figure \ref{fig:Omegab6}a.

\begin{table}
\begin{center}
\caption{Parameters for reference model. }
\label{tab1}
\begin{tabular}{cccc}
\\
\tableline\tableline
Global parameters & Value & Tachocline parameters\tablenotemark{a} & Value \\
\tableline
$M_{\rm core}$ (g) &  $10^{33,}\tablenotemark{a} $ & $\Delta$ (cm) & $1.5\times10^{9,}\tablenotemark{g}$  \\  
$H$ (cm)  & $7\times10^{10,}\tablenotemark{b}$ &  $C \delta$ (cm) & $1.05 \times10^{7,}\tablenotemark{h}$  \\
$R$ (cm)  & $5\times10^{10,}\tablenotemark{c}$ &  $\bar N_{\rm tc}$ (s$^{-1}$) &$8\times10^{-4,}\tablenotemark{i}$ \\
$z_{\rm cz}$ (cm)  & $5\times10^{10,}\tablenotemark{c}$ & $\kappa_{\rm tc}$ (cm$^{2}$/s) & $1.4\times10^{7,}\tablenotemark{i}$  \\ 
$t_{0}$ (s) &  $10^{14,}\tablenotemark{d}$ & $\bar{\rho}_{\rm tc}$ (g/cm$^{3}$) & $0.21\tablenotemark{i}$  \\
$\Omega_{0}$ (s$^{-1}$)  & $9.5\times10^{-5,}\tablenotemark{d}$ & $z_{\rm tc}$ &  $z_{\rm cz} - \Delta - \delta$  \\
$\tau$ (s) & $3\times10^{4,}\tablenotemark{e}$ & & \\ 
         \hline
\tableline
\end{tabular}
\tablenotetext{a}{Half the mass of the solar radiation zone (we consider one hemisphere only to comply with the cylindrical geometry).}
\tablenotetext{b}{The solar radius $r_\odot$.}
\tablenotetext{c}{The radius of the solar radiation zone $r_{\rm cz}$.}
\tablenotetext{d}{These values are selected somewhat arbitrarily, but are consistent with typical ZAMS properties. }
\tablenotetext{e}{$\tau$ is selected so that $\tau \Omega_\odot \simeq 0.1$, where $\Omega_\odot \simeq 3 \times 10^{-6}$s$^{-1}$ is the mean rotation rate of the Sun today. }
\tablenotetext{f}{$\Delta$ is selected to be 0.03$r_{\rm cz}$, a value somewhat in between upper limits from direct helioseismic determinations \citep{Charbonneaual99} and estimates from chemical mixing \citep{ElliottGough99}. }
\tablenotetext{h}{$\delta$ is selected to be 0.01$\Delta$, a value somewhat smaller than the one recommended by GM98 (who argue that $\delta \simeq 0.04\Delta$), but nevertheless consistent with it within an order unity, and $C \simeq 0.7$.}
\tablenotetext{i}{Typical values in the solar tachocline, from \citet{Gough07}.}
\end{center}
\end{table}

To understand the results, first note that a good approximation to Equation \eqref{eq:Omegabstrat} can be obtained by keeping {\it a single term} in the sum over all spatial eigenmodes, that is, by using
\begin{equation}
 \frac{dJ_{\rm core}}{dt} = -\dot{\Omega}_{\rm cz}I_{\rm tc} + \frac{\Omega_{\rm cz}}{\Delta} I_{\rm tc}  \frac{a_1}{b_{1}},
\label{eq:Omegac_simple}
\end{equation}
instead, where we have replaced $32/\lambda_1^4$ in the second term on the right-hand-side by 1, to ensure that the core is not spun-down by the tachocline when $\delta = 0$ (see discussion above). Since $\lambda_1 \simeq 2.4$, this is a fairly good approximation anyway. In addition, as long as the tachocline and tachopause are thin, $a_1$ and $b_1$ further simplify by Taylor expansion to 
\begin{align}
 a_{1} &= \frac{\dot{\Omega}_{\rm cz}}{\Omega_{\rm cz}} \Delta  - 2C\delta \Omega_{\rm c} + \frac{C\lambda^2_{1} }{24} \dot{\Omega}_{\cz}t_{\rm ES}(\Omega_{\cz})  \delta,  \nonumber \\
 b_{1} &= 1+ \frac{C\lambda_{1}^{2}}{12}\Omega_{\cz}t_{\rm ES}(\Omega_{\cz})\frac{\delta}{\Delta}.  \label{eq:an_simple}
\end{align}
Substituting these terms into Equation \eqref{eq:Omegac_simple}, reducing the right-hand side to the same denominator and simplifying, yields 
\begin{equation}
 \frac{dJ_{\rm core}}{dt} = - \frac{C\delta}{\Delta} \Omega_{\rm cz} I_{\rm tc} \left(   \frac{  2 \Omega_{\rm c} + \frac{\lambda^2_{1} }{24} \dot{\Omega}_{\cz}t_{\rm ES}(\Omega_{\cz}) }{1+ \frac{C\lambda_{1}^{2}}{12}\Omega_{\cz}t_{\rm ES}(\Omega_{\cz})\frac{\delta}{\Delta}} \right) \simeq -   I_{\rm tc} \left(   \frac{  24 \Omega_{\rm c} }{\lambda_{1}^{2}t_{\rm ES}(\Omega_{\cz})}  + \frac{\dot{\Omega}_{\cz}}{2} \right)   . 
\label{eq:Omegac_simple2}
\end{equation}
In writing the second expression, we have simplified the denominator further by noting that the second term in $b_1$ is usually larger than 1 by many orders of magnitude for any physically meaningful values of $\delta/\Delta$ (except of course in the strict limit $\delta \rightarrow 0$ discussed earlier, which we do not consider here). The resulting expression for the rate of change of $J_{\rm core}$ is now completely independent of $C\delta$, or in other words, independent of the detailed nature and structure of the tachopause. This property of the solution is discussed in more detail in Section \ref{sec:discuss}, but essentially stems from the fact that the tachopause can propagate the spin-down torque near-instantaneously to the core when it is thin, and thus does not introduce any new timescale in the problem.

Finally, note that since $dJ_{\rm core}/dt = d( I_{\rm core} \Omega_{\rm c} )/dt + d( I_{\rm core} \Omega_{\rm cz})/dt$, and since the assumption of a thin tachocline fundamental to this work implies that $I_{\rm tc} \ll I_{\rm core}$, we can neglect $I_{\rm tc} \dot{\Omega}_{\rm cz}$ in the right-hand side of Equation \eqref{eq:Omegac_simple2} in comparison with the $I_{\rm core} \dot{\Omega}_{\rm cz}$ term on its left-hand side, so that:
\begin{equation}
 \frac{dJ_{\rm core}}{dt} \simeq  -     \frac{  24}{\lambda_{1}^{2}   } \frac{ \Omega_{\rm c} I_{\rm tc} }{t_{\rm ES}(\Omega_{\cz})}  =  -   K \frac{( \Omega_{\rm core} - \Omega_{\rm cz}  ) I_{\rm tc} }{t_{\rm ES}(\Omega_{\cz})} , 
\label{eq:Omegac_simple3}
\end{equation}
where $K$ is a constant of order unity, and $\Omega_{\rm core}$ is the angular velocity of the core in an inertial frame. Written in this final form, our model bears some obvious similarities with the two-zone model of \citet{MacGregorBrenner91}. In fact, it can be cast exactly as in Equation \eqref{eq:mb91} provided we define the coupling timescale between the core and the envelope to be 
\begin{equation}
\tau_c =  \frac{t_{\rm ES}(\Omega_{\rm cz}) }{K} \frac{I_{\rm core} I_{\rm cz}}{I_{\rm tc} (I_{\rm core} + I_{\rm cz})} \mbox{   ,}
\end{equation}
where $I_{\rm cz}$ is the moment of inertia of the convection zone. This result is discussed in more detail in Section \ref{sec:ccl1}.
 
The comparison between the solution of the exact Equation \eqref{eq:Omegabstrat}, and that the much simpler Equation \eqref{eq:Omegac_simple3} is shown in Figure~\ref{fig:Omegab6}. The two are within ten percent of one another at all times. Given that our cylindrical geometry solutions approximate a real star to within a geometrical factor of order unity at best anyway, the error made in using \eqref{eq:Omegac_simple3} instead of  \eqref{eq:Omegabstrat} is of a similar nature, and can be incorporated in the former. In what follows, we therefore advocate the use of \eqref{eq:Omegac_simple3} as a much simpler and more physically meaningful, and yet equivalent description of the spin-down problem. 

\subsection{Properties of the quasi-steady solution}
\label{sec:quasisteady2}

We now use this simpler expression to derive a global quasi-steady approximation to the solution, which gives insight into the long-term behavior of the system. By analogy with Section \ref{sec:steadystate}, we derive it by neglecting the acceleration, but keeping Euler's force in the momentum equation. This yields the following algebraic equation instead,
\begin{equation}
I_{\rm core} \dot\Omega_{\rm cz} = -   \frac{  24}{\lambda_{1}^{2}   }  \frac{ \Omega_{\rm c}   I_{\rm tc} }{t_{\rm ES}(\Omega_{\cz})} , 
\label{eq:Omegac_simpleqs}
\end{equation}
which can then be solved for $\Omega_{\rm c}$, and thus yields its quasi-steady approximation $\Omega^{\rm qs}_{\rm c}$. Note that we have again assumed here that $I_{\rm core}$ varies sufficiently slowly with time that its derivative can be neglected. This is done for simplicity of interpretation of the results but is not necessarily valid during all evolutionary stages of the star. We then find that:
\begin{equation}
 \frac{\Omega_{\rm c}^{\rm qs}}{\Omega_{\cz}} =  \frac{\lambda_{1}^{2}}{24} \frac{ t_{\rm ES}(\Omega_{\cz}) }{ t_{\rm sd} (\Omega_{\rm cz})} \frac{I_{\rm core}}{I_{\rm tc} }   = \frac{\lambda_{1}^{2}}{24} \frac{ t_{\rm ES}(\Omega_0) }{ t_{\rm sd} (\Omega_0)} \frac{\Delta I_{\rm core}}{\Delta_0 I_{\rm tc} } \left( \frac{\Omega_{\rm cz}}{\Omega_0} \right)^{-(q+\beta)} , \label{eq:finalcorescaling}
\end{equation}
where $q + \beta  = 2 -  \alpha^{-1} - 3\beta $.

The scaling shown in Equation (\ref{eq:finalcorescaling}) is similar to that discussed in Section \ref{sec:steadystate} (see Equation~\ref{eq:oom}), and can therefore also be understood using  order-of-magnitude arguments based on thermal-wind balance, thermal equilibrium and mass conservation. However, it contains new factor which is the ratio $I_{\rm core}/I_{\tc}$. The presence of this factor can be understood physically by noting that this time, the core is spun-down as well as the tachocline. A much larger torque is needed to spin down a more massive core, and as a consequence, if the same spin-down torque is applied, then the core-envelope lag is proportionally larger for a more massive core. Furthermore, since $I_{\rm core}/I_{\tc}$ is inversely proportional to the thickness of the tachocline, $\Omega^{\rm qs}_{\rm c}/\Omega_{\rm cz}$ now scales with the third instead of the fourth power of $\Delta$, and varies as $(\Omega_{\rm cz}/\Omega_0)^{-(q + \beta
)}$ instead of $(\Omega_{\rm cz}/\Omega_0)^{-q}$.  

We can now apply the same reasoning as in Section \ref{sec:transient} as to the limits of validity of our laminar solution. If $q + \beta > 0$ then $\Omega^{\rm qs}_{\rm c}/\Omega_{\rm cz}$ increases as the star spins down. In this case, the quasi-steady solution is not a good approximation to the actual time-dependent problem, and the laminar solution probably eventually breaks down to a turbulent one with different scalings instead. If $q + \beta  \le 0$ on the other hand  then $\Omega^{\rm qs}_{\rm c}/\Omega_{\rm cz}$ remains constant or decreases, and the laminar solution is likely always valid. When $q + \beta < 0$, the system always tends to solid-body rotation as the star spins down. By contrast, if $q + \beta = 0$ then $\Omega^{\rm qs}_{\rm c} /\Omega_{\rm cz}$ tends to a constant, which implies that the system eventually maintains a non-zero core-envelope lag as $t \rightarrow \infty$. 

\begin{figure}
\begin{center}
\includegraphics[width=0.6\textwidth]{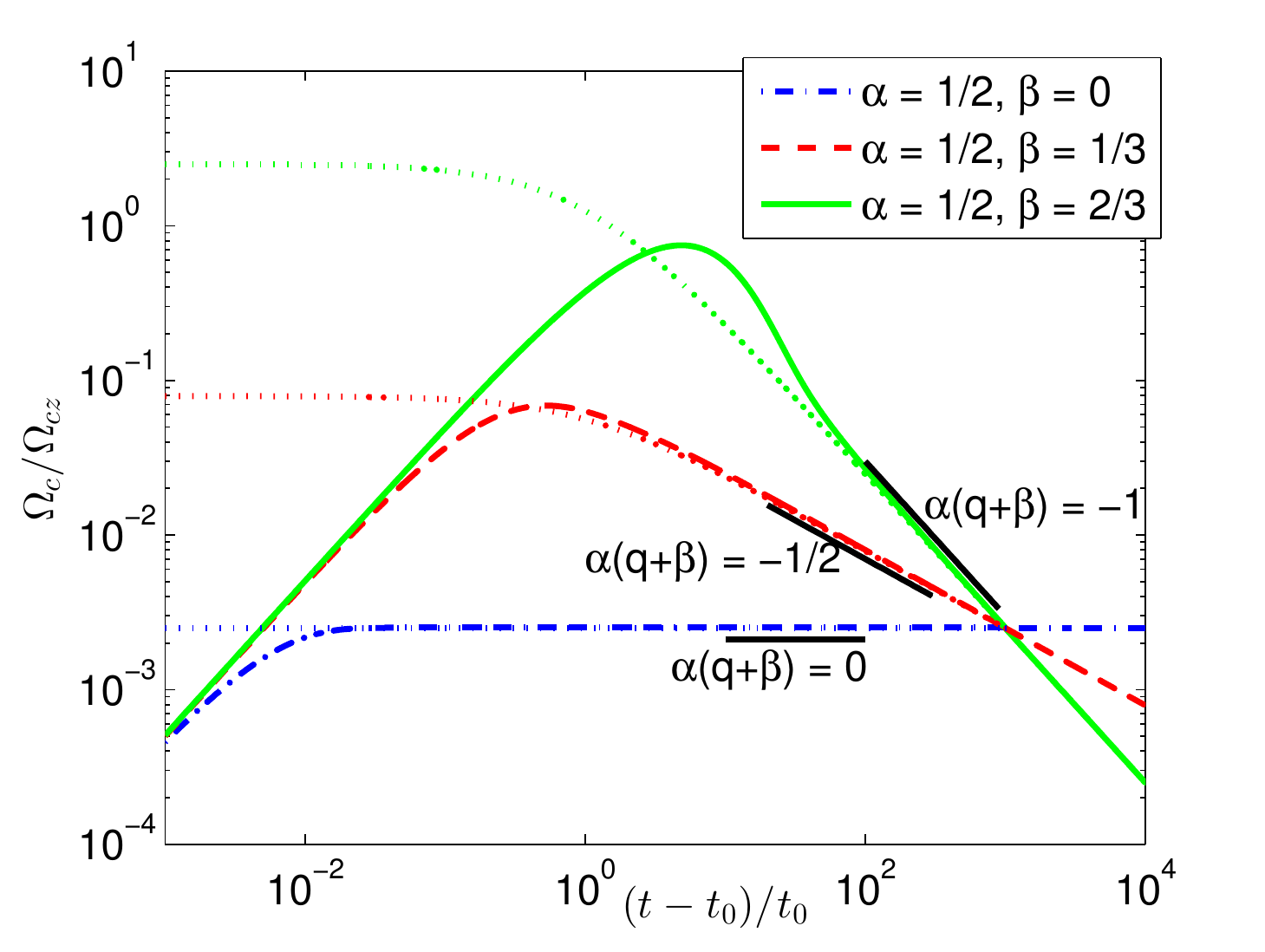}
\end{center}
\caption{Evolution of $\Omega_{\rm c}/\Omega_{\rm cz}$ for the reference star described in Table 1, undergoing a Skumanich law spin-down. The tachocline thickness varies with $\Omega_{\rm cz}$ as in Equation \eqref{eq:tachoindex}, with $\beta=0$, $1/3$ and $2/3$ respectively. In all cases $\Delta_0$ is chosen such that $\Delta = 1.5\times10^9$ cm at $t = 10^3 t_0$ (see Table \ref{tab1}), so $\Delta_0 = 4.7\times10^9$ cm for $\beta = 1/3$, and $\Delta_0 = 1.5\times10^{10}$ cm for $\beta  =2/3$. The exact solution to Equation (\ref{eq:Omegac_simple}) is shown by the solid, dashed and dot-dashed lines, and the quasi-steady state approximation to that solution (given by Equation (\ref{eq:finalcorescaling})) is shown by the dotted lines. For large $t$, $\Omega_{\rm c}/\Omega_{\rm cz} \propto t^{\alpha(q+\beta)}$ (see Section \ref{sec:quasisteady2}).}
\label{fig:longtime}
\end{figure}

This behavior is illustrated in Figure~\ref{fig:longtime}, which shows the long-term evolution of the ``reference" star, for $\alpha = 1/2$ and $\beta = 0$, $\beta = 1/3$ and $\beta = 2/3$ respectively.  We see that, for large times, the exact solution for $\Omega_{\rm c}/\Omega_{\rm cz}$ does indeed tend to the quasi-steady solution as expected. The latter is either constant or decreases with time, depending on the value of $q + \beta $. 

We now briefly examine under which conditions $q + \beta  = 2 -  \alpha^{-1} - 3\beta \le 0 $. Recall that $\alpha$ is the spin-down law index (with the convention that $\alpha^{-1} = 0$ for an exponential spin-down law, see Equation \ref{eq:tsdocz}) and $\beta$ is the index of the power-law describing the variation of the tachocline thickness with $\Omega_{\rm cz}$ (see Equation \ref{eq:tachoindex}). The condition $2 - \alpha^{-1} - 3\beta \le 0$ is then automatically satisfied whenever $\alpha \le 1/2$, as long as $\beta > 0$, or whenever $\beta \ge 2/3$, regardless of $\alpha$. 

The fact that the observationally-favored Skumanich law \citep{Skumanich72}, which has $\alpha = 1/2$, lies in the region of parameter space for which our quasi-steady laminar solutions are valid {\it regardless} of $\beta$ is a very nice -- if somewhat unexpected -- feature of our model, and speaks to its relevance for observations.  The fact that it is also a {\it critical} parameter value, on the other hand, is probably just a coincidence. 

In general, however, $\alpha$ is not known a priori -- it is an outcome of the complete spin-down problem \citep[see][for instance, who found both exponential and power-law spin-down solutions depending on their assumed wind model]{ReinersMohanty12}. Our theoretical results suggest that if $\beta \ge 2/3$, then again a laminar quasi-steady solution always exists regardless of the spin-down law. If $\beta < 2/3$, on the other hand, whether the laminar solution holds or not also depends on the actual spin-down rate of the convection zone, and must therefore be determined ``on the fly" while the solution of \eqref{eq:Omegac_simple3} is being computed.

Unfortunately, it is difficult to constrain $\beta$ from theory alone (see Appendix E). Indeed, while GM98 and \citet{Woodal11} both found that $\beta > 2/3$ in a related but distinctly different model system, their results cannot be used here. In this particular instance, help in constraining $\beta$ comes from observations instead. The requirement that fast rotators should be in solid-body rotation at all times immediately constrains $\beta$ to be strictly {\it smaller} than $2/3$ (see Section \ref{sec:discuss} for detail), which does imply that our model could break down if the spin-down rate is too rapid. For further constraints, we note that surface abundances of light elements and other tracers can potentially be used to estimate the depth of the chemically mixed tachocline, and study its variation with stellar rotation rate (see Section \ref{sec:ccl3}). 

\subsection{Properties of the initial transient solution}
\label{sec:transient2}

The initial, nearly linear increase in the core-envelope lag seen in Figures~\ref{fig:Omegab6} and \ref{fig:longtime} can be understood by noting that the angular-momentum transport rate across the tachopause depends on the local torques, which in turn depend on the local angular-velocity shear. At first, the latter is small, so the torques are not strong enough to spin the core down. When viewed in an inertial frame, the latter continues to spin at its original rate, so that
\begin{equation}
\Omega_{\rm c}(t) = - \Omega_{\rm cz}(t) \simeq - \dot \Omega_{\rm cz}(t_0) (t-t_0) ,
\label{eq:transientsol} 
\end{equation}
for $t - t_0$ smaller than the spin-down timescale.  

As the core-envelope lag increases, so does the shear, until a point where the torque exerted is just strong enough to communicate the surface spin-down to the interior. When this happens, the core locks on to the tachocline, and the angular-momentum flux becomes independent of radius. The system reaches a quasi-steady state in the spinning-down frame, in which both convective zone, tachocline and core concurrently spin-down at more-or-less the same rate. Equating the early-time solution \eqref{eq:transientsol} with the quasi-steady solution \eqref{eq:finalcorescaling}, we find that this happens (very roughly) when 
\begin{equation}
 t-t_0  = \frac{\lambda_{1}^{2}}{24}t_{\rm ES}(\Omega_{\cz}(t)) \frac{I_{\rm core}}{I_{\rm tc}(t) } \frac{ \dot\Omega_{\rm cz}(t) }{\dot \Omega_{\rm cz}(t_0)} . 
\label{eq:waittime}
\end{equation}
Although implicit for $t$, and therefore difficult to solve algebraically, this expression readily shows that the duration of the initial transient phase is proportional to the local Eddington--Sweet timescale across the tachocline, times $I_{\rm core}/I_{\rm tc}$, a quantity that is overall proportional to $\Delta^3/\Omega_{\rm cz}^2$. In other words, it is much longer if the initial tachocline thickness is larger, which explains the results of Figure \ref{fig:longtime}.

In what follows, we now summarize our results, importing them into a spherical geometry and casting them in a more astrophysically relevant terminology, and discuss their implications for stellar spin-down and related observations.

\section{SUMMARY AND DISCUSSION} 
\label{sec:discuss}

\subsection{Summary of the results}
\label{sec:ccl1}

In this work, we have studied the impact of spin-down on solar-type stars whose internal dynamics are assumed to be analogous to that of the Sun as first introduced by GM98. More specifically, we considered stars with a radiation zone held in uniform rotation by the presence of a large-scale primordial magnetic field, that is confined strictly below the base of the convection zone by large-scale meridional flows. The geometry of such stars was shown in Figure \ref{fig:model}. 

Separating the convection zone and the bulk of the radiation zone (the ``core", hereafter), which are rotating with angular velocities  $\Omega_{\rm cz}$ and $\Omega_{\rm core}$ respectively (as expressed in an inertial frame), lie two thin nested shear layers \citep[GM98;][]{WoodMcIntyre11,Woodal11,AAal13}: the tachocline, which resides just beneath the radiative--convective interface, and is -- dynamically speaking at least -- magnetic free, and the tachopause, which lies below the tachocline, and connects the latter magnetically to the core. 

The extraction of angular momentum from the star by the stellar wind in our model takes a rather different form across each of these regions. Magnetic braking exerts a torque on the surface layers, which is nearly instantaneously communicated down to the base of the convection zone by the turbulence. Angular-momentum transport (and chemical mixing) across the underlying tachocline, on the other hand, is mediated principally by large-scale meridional flows and roughly takes place on a local Eddington-Sweet timescale:
\begin{equation}
t_{\rm ES}(\Omega_{\rm cz}) = \frac{ \bar{N}_{\rm tc}^2  }{2 \Omega^2_{\rm cz}  }  \left( \frac{\Delta}{r_{\rm cz}} \right)^2  \frac{\Delta^2 }{\kappa_{\rm tc} }  \mbox{  ,}
\label{eq:couplingtime}
\end{equation}
where $\bar N_{\rm tc}$, $\kappa_{\rm tc}$ are the buoyancy frequency and thermal diffusivity of the fluid within the tachocline, $\Delta$ is its thickness, and $r_{\rm cz}$ is the radius of the base of the convection zone.
The spin-down torque is then finally communicated through the tachopause down to the deep interior primarily by magnetic torques. The mostly-dipolar primordial field is wound-up by the radial shear, which generates a significant toroidal field. The resulting Lorentz force reacts against the shear, and thus extracts angular momentum from the core. This happens  on an Alfv\'enic timescale which is more or less instantaneous compared with the spin-down timescale or the tachocline mixing timescale. 

The tachocline is clearly the ``bottleneck" of this angular-momentum extraction sequence, and therefore controls the overall rotational evolution of the star. As a result, the timescale $t_{\rm ES}(\Omega_{\rm cz})$ introduced above plays a role that is similar (but not identical, see below) to the coupling timescale between the core and the envelope in the two-zone model of \citet{MacGregorBrenner91} (see $\tau_c$ in Equation \ref{eq:mb91}). 

Thanks to the help of an idealized model for which exact solutions exist, we have formally shown that the concurrent evolution of the rigidly-rotating core and the convective envelope can typically (i.e. under reasonable assumptions usually valid in most solar-type stars) be modeled as: 
\begin{align}
 \frac{dJ_{\rm core}}{dt} = & - \frac{\Delta J}{\tau_c}   \mbox{  ,}  \label{eq:jcore} \\ 
 \frac{dJ_{\rm cz}}{dt} +  \frac{dJ_{\rm core}}{dt}  =&  - \dot J_{\rm w}  \mbox{  ,} \label{eq:jcz} \mbox{ where  } \\
\Delta J  =& (\Omega_{\rm core} - \Omega_{\rm cz}) \frac{I_{\rm core} I_{\rm cz}}{I_{\rm core} + I_{\rm cz}}  \mbox{  ,}
\end{align}
that is, exactly as in the two-zone model of \citet{MacGregorBrenner91} (see Equation \ref{eq:mb91}), with a coupling timescale $\tau_c$ given by
\begin{equation}
\tau_c = \frac{t_{\rm ES}(\Omega_{\rm cz}) }{K} \frac{I_{\rm core} I_{\rm cz}}{I_{\rm tc} (I_{\rm core} + I_{\rm cz})} \mbox{  ,}
\label{eq:ourtauc}
\end{equation}
where $K$ is a positive geometrical constant of order unity, and where $I_{\rm tc}$ is the moment of inertia of the tachocline (which, to a good approximation, is  $I_{\rm tc} = 4\pi \bar \rho_{\rm tc} r_{\rm cz}^4 \Delta$, where $\bar \rho_{\rm tc}$ is the local density within the tachocline). Equation \eqref{eq:jcore} was derived in Section \ref{sec:globalsol}. 

Inspection of Equations \eqref{eq:jcore} -- \eqref{eq:ourtauc} shows that they do not explicitly depend on the magnetohydrodynamics of the tachopause\footnote{Note that this does not imply that magnetic stresses are not important in our model, on the contrary -- they are crucial to the tachopause dynamics. However, by virtue of adjusting immediately to any perturbation, they do not appear in the evolution equation directly. On the other hand, if the tachopause is not much thinner than the tachocline, then Equation (\ref{eq:Omegac_simple}) must be used instead of Equation \eqref{eq:jcore}, and now explicitly depends on the tachopause dynamics.}, a result inherently tied to the assumption that the tachopause is much thinner than the tachocline, and that it is at all time in complete dynamical and thermal equilibrium. When this is the case, the tachopause responds near-instantaneously to any perturbation and thus cannot introduce any additional timescale in the system (see Section \ref{sec:globalsol} for detail). This property of our model turns out to be quite convenient: as discussed by \citet{AAal13}, the specific nature and dynamical properties of the tachopause are arguably the ``weakest link" of the GM98 model, being the most sensitive to any dynamics that were purposefully neglected (turbulence, gravity waves, etc.). But as Equations \eqref{eq:jcore} -- \eqref{eq:ourtauc} show, this model-dependence does not have any {\it direct} impact on the long-term evolution of the angular velocity of the core. 

On the other hand, Equation \eqref{eq:jcore} is very sensitive to the properties of the tachocline, and in particular to its thickness $\Delta$. The latter presumably depends on the star's mean rotation rate, on its spin-down rate, on the strength of the internal primordial field, and on the position and local thermodynamical properties of the base of the convection zone. Both GM98 and \citet{Woodal11} propose scalings for $\Delta$ as a function of these quantities in the solar case, where the large-scale meridional flows are driven by the {\it latitudinal} shear within the convection zone rather than by spin-down. Unfortunately, as discussed in Appendix E, these scalings do not directly apply here. Furthermore, any attempt to estimate $\Delta$ from first principles necessarily yields a result that depends sensitively on the structure of the tachopause, which we have just argued is both poorly constrained and strongly dependent on the model considered. In this sense, while the dynamics of the tachopause do not explicitly participate in Equations \eqref{eq:jcore} -- \eqref{eq:ourtauc}, they nevertheless indirectly influence the rotational evolution of the star by controlling $\Delta$ (which appears in $t_{\rm ES}(\Omega_{\rm cz})$ and therefore in $\tau_c$). 

For these reasons, instead of proposing a carefully derived, mathematically correct but highly model-dependent formula for the tachocline thickness $\Delta$, we suggest the following simple parametric prescription: 
\begin{equation}
\Delta = \Delta_0(B_0) \left( \frac{\Omega_{\rm cz}}{\Omega_0}\right)^{\beta} \left( \frac{r_{\rm cz}}{r_{\rm cz}(t_0)}\right)^{\gamma}  \mbox{  .} 
\label{eq:deltaparam}
\end{equation}
In this model, we have hidden all information about the unknown (and non-observable) internal field strength $B_0$ into $\Delta_0(B_0)$. Any information about the time-dependence induced by spin-down is contained in the second term, and any information about the local properties of the tachocline is contained in the third term\footnote{Indeed, as the radiative interior can be grossly modeled as a polytrope, all thermodynamic quantities at the base of the convection zone can presumably be modeled as powers of $r_{\rm cz}$, and by proxy, so can the local buoyancy frequency $\bar N_{\rm tc}$, and the local magnetic and thermal diffusivities $\eta_{\rm tc}$ and $\kappa_{\rm tc}$.}. While $\beta$ and $\gamma$ are difficult to estimate from theory alone (see Appendix E for detail), we hope that they can, in the future, be constrained observationally by studying simultaneously the rotational histories of solar-type stars in young clusters and their light-element surface abundances (see Section \ref{sec:ccl3}). 

In general, Equations \eqref{eq:jcore}--\eqref{eq:deltaparam} have to be solved -- and should be solved -- numerically, in conjunction with the equations for stellar evolution which yield $I_{\rm core}$, $I_{\rm cz}$ and $\tau_{\rm c}$ at each point in time. However, good insight into the long-term behavior of the solutions can be obtained by considering a ``quasi-steady" approximation, in which (1) we assume that the spin-down rate $\dot{\Omega}_{\rm cz}$ is known, (2) $I_{\rm core}$ does not vary too rapidly with time and (3) $\Omega_{\rm core} - \Omega_{\rm cz}$ is not too large. Using all three approximations implies that  $d J_{\rm core} / dt \simeq I_{\rm core} \dot{\Omega}_{\rm cz}$ is known, and one can then simply solve Equation \eqref{eq:jcore} analytically for $\Omega_{\rm core}$. The relative core-envelope lag, in this quasi-steady approximation, is given by
\begin{equation} 
\frac{\Omega_{\rm core}-\Omega_{\rm cz}}{ \Omega_{\rm cz}} =  K \frac{ t_{\rm ES}(\Omega_{\rm cz})  }{t_{\rm sd} (\Omega_{\rm cz}) } \frac{ I_{\rm core}}{ I_{\rm tc} }  =  - K \frac{\dot{\Omega}_{\rm cz}}{\Omega_{\rm cz}^3}  \frac{\bar N_{\rm tc}^2 \Delta^4}{r_{\rm cz}^2 \kappa_{\rm tc}}  \frac{I_{\rm core}}{I_{\rm tc}} 
\label{eq:celag1}
\end{equation}
where $t_{\rm sd} = | {\Omega}_{\rm cz}/ \dot{\Omega}_{\rm cz}| $ is the spin-down timescale of the convective envelope. Physically speaking, this formula is equivalent to stating that the star adjusts itself in such a way that the integrated angular-momentum flux out of each spherical shell is constant with radius, and equal to that extracted from the star by the stellar wind.

We have shown in Sections \ref{sec:transient} and \ref{sec:quasisteady2} that our model is only technically valid when this solution is bounded (in the sense that the quasi-steady relative core-envelope lag remains constant or decreases with time). This happens when $2 - \alpha^{-1} - 3\beta \le 0$, where $\beta$ is defined in \eqref{eq:deltaparam}, and where $\alpha$ is defined such that $t_{\rm sd} \propto \Omega_{\rm cz}^{-\alpha^{-1}}$. When the model applies, then the quasi-steady solution is {\it also} an attracting solution of the governing equations (which means that the system relaxes to this state regardless of its initial conditions). All stars satisfying this condition are therefore expected to have a core-envelope lag given by \eqref{eq:celag1}, after a transient period whose duration is of the order of $t_{\rm ES}(\Omega_{\rm cz})$ evaluated at $t = t_0$ (see Section \ref{sec:transient2} for detail).

\subsection{Caveats of the model}
\label{sec:ccl2}

Before we proceed to discuss the observational implications of our model, let us briefly address  its caveats and limitations. In many ways, they are the same as those of the GM98 model, listed and discussed at length by GM98 and by \citet{AAal13}. 

Central to our calculation is the assumption that the star has a dynamical structure similar to the Sun, with an outer convection zone and a uniformly rotating magnetized core both in solid-body rotation, separated by a thin magnetic-free tachocline, and an even thinner tachopause which, by contrast, is essentially magnetic in nature. As discussed by \citet{AAal13}, a necessary condition for such a layered model to exist is 
\begin{equation}
\frac{\bar N_{\rm tc}}{\Omega_{\rm cz}} \sqrt{\frac{\nu_{\rm tc}}{\kappa_{\rm tc}}} \ll \frac{r_{\rm cz}}{\Delta} \mbox{  ,} 
\label{eq:sigma}
\end{equation}
where $\nu_{\rm tc}$ is the viscosity in the tachocline region. If this condition is not satisfied, then the meridional flows downwelling from the convection zone are unable to confine the magnetic field, and a different model must be used \citep[see][for details]{AAal13}. However, since the Sun satisfies this property, we expect that most young and thus more rapidly rotating stars are likely to satisfy it as well.

Even if \eqref{eq:sigma} is satisfied, the existence of such a layered structure is not yet guaranteed. While it has now been revealed in fully nonlinear, full-sphere, steady-state simulations of the solar interior for the first time \citep{AAal13}, one should still verify that it can also be achieved in a spin-down problem. Indeed, as mentioned in Section \ref{sec:ccl1}, the mechanisms driving the large-scale tachocline flows, which are responsible for confining the internal magnetic field within and below the tachopause, are subtly different in the solar steady-state case and in the spin-down case. %It is the same issue that prevents us from using the scalings derived by GM98 and \citet{Woodal11} relating the thickness of the tachocline to other properties of the star in this work. 
We defer the task of running full-sphere numerical simulations of the spin-down problem to a future publication. Beyond the question of existence of a tachocline and tachopause, such a calculation could furthermore yield a first estimate of the possible relationships between their thicknesses and other stellar parameters, a result that cannot be robustly obtained from linear theory alone here (see Appendix E). 

The next major assumption we need to verify is whether the effects of turbulence can indeed be neglected while modeling the tachocline. As discussed by \citet{AAal13}, thermal-wind balance and thermal equilibrium -- two key balances in the system -- are both still likely to hold even in the presence of turbulence (given reasonable assumptions as to its source). On the other hand, angular-momentum balance is much more sensitive to any added effects, and must be studied carefully. For our model to hold, radial angular-momentum transport across the tachocline must be dominated by advection by large-scale flows, rather than by turbulence.  Naturally, and as discussed throughout this work, this assumption can only hold as long as the shear across the tachocline is ``weak enough" not to cause significant turbulent transport -- the question being what ``weak enough" means in this context. 

One may first ask under which conditions the tachocline is linearly unstable to shear instabilities. As studied by \citet{Lignieresal99}, this could depend both on the Richardson number $Ri \simeq \bar N_{\rm tc}^2 \Delta^2  / r_{\rm cz}^2 (\Omega_{\rm cz} - \Omega_{\rm core})^2$, and on the P\'eclet number $Pe = r_{\rm cz} (\Omega_{\rm cz} - \Omega_{\rm core})\Delta/\kappa_{\rm tc}$. In this particular problem, however, the P\'eclet number is typically so large that the relevant criterion for global, tachocline-scale shear instabilities is the standard $Ri < O(1)$ rather than $Ri Pe< O(1)$ advocated by \citet{Zahn74} (which is only applicable to the small $Pe$ limit). Using solar values as guidance (see Table 1), we find that our model is expected to break down completely only when 
\begin{equation}
2,300 \left( \frac{\bar N_{\rm tc}}{8 \times 10^{-4}}\right)^2 \left( \frac{\Delta/r_{\rm cz}}{0.03} \right)^2 \left( \frac{ 5 \times 10^{-7}}{\Omega_{\rm core} - \Omega_{\rm cz}} \right)^2  < O(1)  \mbox{  .}
\end{equation} 
For this inequality to hold, we therefore see that a substantial shear is required. Although unlikely in older stars, this could happen during the early stages of the spin-down process where $\Omega_{\rm cz}$ is much larger, and where rapid core-contraction can result in significant core-envelope shear. One should therefore monitor $Ri$ carefully in the process of time-stepping Equations \eqref{eq:jcore}-\eqref{eq:deltaparam}, and use a turbulent coupling timescale instead should $Ri$ drop below 1.

Alternatively, ignoring linear stability considerations, one could simply assume that the system becomes unstable to finite amplitude perturbations for much weaker shearing rates. In that state, \citet{PratLignieres13} \citep[see also][]{Zahn74,Zahn92} suggest that turbulent transport can be described using the vertical turbulent diffusion coefficient\footnote{Although this expression is technically only valid in the limit of low P\'eclet number \citep{PratLignieres13}, one may argue that the strong stratification only permits motion with a very short vertical scale, thus ensuring that the turbulent P\'eclet number based on that scale is small. }
\begin{equation}
D_t \simeq 0.05 \kappa_{\rm tc} Ri^{-1}  \mbox{  .}
\end{equation}
In the presence of this kind of stratified shear turbulence, our model holds provided the timescale for advection of angular momentum across the tachocline by large-scale flows (given more-or-less by $t_{\rm ES}(\Omega_{\rm cz}$)) is shorter than the timescale for the turbulent diffusion of angular momentum (given more-or-less by $\Delta^2 / D_t$). This implies that our model is expected to apply whenever  
\begin{equation}
\frac{\bar N_{\rm tc}^2 \Delta^4 }{\Omega_{\rm cz}^2 r_{\rm cz}^2 \kappa_{\rm tc}} \frac{ D_t }{ \Delta^2}  < O(1)  \Leftrightarrow  0.05 \frac{ (\Omega_{\rm core} - \Omega_{\rm cz})^2  }{\Omega_{\rm cz}^2  } <  O(1) \mbox{   .}
\end{equation}
This criterion, rather interestingly and perhaps surprisingly, depends only on the relative core-envelope lag iself. We then find that the use of our model is also always justified unless $(\Omega_{\rm core} - \Omega_{\rm cz})$ is unrealistically large (that is, much larger than $\Omega_{\rm cz}$ itself). It thus appears that neglecting turbulent angular-momentum transport in the tachocline could well be justified, except perhaps very early on in the spin-down process if the Richardson number $Ri$ ever drops below 1. 

\subsection{Observational implications}
\label{sec:ccl3}

While a complete discussion of the observational implications of our model will have to be done by applying it in conjunction with stellar evolution, and statistically comparing its predictions against observations \citep[as in][for instance]{Allain98,Irwinal07,Denissenkoval10,Spadaal11,ReinersMohanty12,GalletBouvier13}, we can nevertheless already discuss its prospects in the light of previous work. 

As reported in Section \ref{sec:introduction}, these previous studies found that stars that begin their lives as rapid rotators can, at all times, be modeled assuming solid-body rotation, while the rotation rates of stars in the mass-range $0.7M_\odot - 1.1M_\odot$ that are initially slow rotators are best modeled with the two-zone model of \citet{MacGregorBrenner91} assuming a rather long core-envelope coupling timescale (of the order of hundreds of Myr up to a Gyr). It is very difficult to explain such long timescales using a magnetic model, unless rather dramatic assumptions are made concerning the degree of confinement of the field (which must then also be explained). It is also difficult to explain observations with a purely turbulent model, since the latter does not easily explain why the core should be mostly in solid-body rotation. By contrast, our model naturally results in a system that behaves like the two-zone model, with a coupling timescale that depends both on the stellar structure and on the rotation rate of the star (see Equation \ref{eq:ourtauc}), and that can be very substantial for slower rotators with fairly thick tachoclines.

Specifically, we find that the coupling timescale is proportional to $\Delta^3 / r_{\rm cz}^2 \Omega_{\rm cz}^2$. As such, for similar tachocline thicknesses, it is naturally much shorter for fast rotators than for slow rotators, and for higher-mass stars (which have a larger radiation zone) than for lower-mass stars. Both results are qualitatively consistent with the aforementioned observations. Of course, the unknown dependence of $\Delta$ on $\Omega_{\rm cz}$ and $r_{\rm cz}$ makes it difficult at this point to give strict estimates of how strong this effect may be. Nevertheless, we can already infer from the data (whereby fast rotators should also be solid-body rotators) that $\Delta^3 / r_{\rm cz}^2 \Omega_{\rm cz}^2$ must be a decreasing function of $\Omega_{\rm cz}$. This constrains $\beta$ (see Equation \ref{eq:deltaparam}) to be strictly smaller than $2/3$, and quite possibly substantially smaller than that.
 
Additional information on the tachocline thickness $\Delta$ may be obtained by studying the relative differences in surface chemical abundances of stars within the same cluster. Light-element such as lithium and beryllium  undergo significant Main-Sequence depletion, that can only be explained by extra mixing below the base of the convection zone \citep[see for instance the review by][]{Pinsonneault97}. Within the scope of our model, we expect their respective depletion rates to depend sensitively on $\Delta(t)$, so that present-day surface abundances of a given star provide an integrated view of the variation of the tachocline depth with time (and therefore with rotation rate). Concurrently fitting rotational histories with Li and Be abundances may thus help constrain both $\beta$ and $\gamma$. 

For stars in older clusters ($>$ few hundred Myr), we generally expect Equation \eqref{eq:celag1} to hold. Indeed, as long as the condition $2 - \alpha^{-1} - 3\beta \le 0$ is satisfied (see \ref{sec:quasisteady2} and \ref{sec:ccl1} for detail), stars should have relaxed to their quasi-steady state by that age. In that state, aside from the surface abundances which depend on the rotational history of the star (as discussed above), all dynamical information about the star's initial conditions is lost, and the core-envelope lag only depends on present-day parameters. We then see from \eqref{eq:celag1} that {\it everything else being equal},  their core-envelope lag should be much larger (1) if the spin-down rate is larger or if the star is rotating more slowly; (2) for lower-mass solar-type stars; (3) for stars with thicker tachoclines. While asteroseismology has not yet been able to detect any core-envelope lag in solar-type stars other than the Sun, one can only hope that such detection may be possible at some point in the future, and will independently help constrain our model. 
 
Finally, note that our model predicts that stars in this quasi-steady state, with $2 - \alpha^{-1} - 3\beta < 0$, always eventually reach solid-body rotation (at least in a radial sense) and that $\Delta \rightarrow 0$ as $\Omega_{\rm cz} \rightarrow 0$ and $\dot{\Omega}_{\rm cz} \rightarrow 0$. However, substantial latitudinal differential rotation is likely to persist in their convection zones, as it does in the Sun. This latitudinal shear by itself also drives large-scale meridional flows by gyroscopic pumping. As studied by GM98, these flows transport angular momentum and interact with the embedded primordial field, leading to a finite tachocline thickness (see Appendix E) even when $\dot{\Omega}_{\rm cz} = 0$. Although we have ignored this effect here for simplicity, and on the grounds that these shear-induced flows are presumably weaker than those driven by the spin-down torque for young stars, it can no longer be ignored for much older stars. In future work, we shall attempt to model simultaneously the effects of spin-down and of latitudinal shear in the convection zone in driving the tachocline flows, so as to present an integrated model of the rotational evolution of solar-type stars that can be used all the way from the Zero-Age Main Sequence to the present-day Sun.

%More generally, the strong dependence of our model on the unknown tachocline thickness is both bad news and good news. From a theoretical point of view, it makes any prediction of the actual core-envelope coupling timescale very sensitive to actual value of $\beta$ used in Equation \eqref{eq:deltaparam}. One must therefore be extremely careful in using it {\it on its own}. However, if the value of $\beta$ can somehow be otherwise constrained -- by concurently studying the effect of tachocline mixing on light-element or He abundances for instance -- then it becomes much more powerful. 

\section*{Acknowledgements}

This work originated from R. L. F. Oglethorpe's summer project at the Woods Hole GFD Summer Program in 2012. We thank the NSF and the ONR for supporting this excellent program. R. L. F. Oglethorpe acknowledges funding from an EPSRC studentship. 
P. Garaud acknowledges funding from the NSF (CAREER-0847477). We thank Nic Brummell, Douglas Gough, Subhanjoy Mohanty, Nigel Weiss and Toby Wood for fruitful discussions. 
\appendix

\section{QUASI-STEADY SOLUTION FOR THE SPIN-DOWN OF A NON-MAGNETIC STAR}

In this Appendix, we derive the result presented in Equation (\ref{eq:v1}). Assuming a quasi-steady state, we drop the time-derivative in the momentum equation (\ref{eq:bulkmom}). 
Its azimuthal component then reduces to
\begin{equation}
 u(s) = - \frac{\dot{\Omega}_{\rm cz}}{2\Omega_{\rm cz}}s, 
 \label{eq:u1app}
\end{equation}
implying (using mass conservation) that $w$ must be a linear function of $z$, exactly as in Section \ref{sec:unstrat1}. To satisfy impermeability at $z=z_{\rm tc}$, we must therefore have
\begin{equation}
w(z) = \frac{\dot{\Omega}_{\rm cz}}{\Omega_{\rm cz}}(z-z_{\rm tc}).
\end{equation}
The boundary condition \eqref{eq:sidebc}, combined with the vertical component of the momentum equation,
\begin{equation}
 \frac{1}{\bar{\rho}_{\rm tc}}\frac{\partial p}{\partial z} = \frac{\bar{g}_{\rm tc}}{\bar{T}_{\rm tc}}T, \label{eq:appP1}
\end{equation}
implies that $T=0$ on $s=R$.  Solving the thermal energy equation \eqref{eq:thermaleq} with this boundary condition, along with
$T = 0$ at $z=z_{\rm tc}$ 
then gives
\begin{equation}
T(s,z) = \sum_{n}J_{0}\left(\lambda_{n}\frac{s}{R}\right)\left[\alpha_{n}\sinh\left(\lambda_{n}\frac{z-z_{\rm tc}}{R} \right)  - \frac{C_{n}(z-z_{\rm tc}) R^{2}}{\lambda_{n}^{2}}\right],
\end{equation}
where the constants $\{ \lambda_{n} \}$ are the zeros of the Bessel function $J_{0}$, and 
\begin{equation}
 C_{n} = \frac{\bar{N}_{\rm tc}^{2}\bar{T}_{\rm tc}}{\bar{g}_{\rm tc}\kappa_{\rm tc}}\frac{\dot{\Omega}_{\cz}}{\Omega_{\cz}} \frac{2}{\lambda_{n}J_{1}(\lambda_{n})}.
\end{equation}
Using the fact that $T =0$ at $z=z_{\cz}$ determines the $\{\alpha_n\}$ coefficients to be 
\begin{equation}
\alpha_{n} = \frac{R^{2}}{\lambda_{n}^{2}}\frac{C_{n}\Delta}{\sinh\left( \lambda_{n}\frac{\Delta}{R}\right) }  ,
\end{equation}
where $\Delta = z_{\rm cz} - z_{\rm tc}$ is the thickness of the tachocline. 

Equations \eqref{eq:appP1} and  \eqref{eq:thermalwind} can then be used to derive $p$ and $v$, up to the unknown set of integration constants $\{p_{n}\}$ (which are the same in both equations):
\begin{align}
p(s,z) &= \frac{\bar{\rho}_{\rm tc} \bar{g}_{\rm tc}}{\bar{T}_{\rm tc}} \sum_n J_0\left(\lambda_n \frac{s}{R}\right) \left[ \frac{\alpha_nR}{\lambda_n} \cosh \left(\lambda_n\frac{z-z_{\rm tc}}{R} \right) - \frac{C_n(z-z_{\rm tc})^2R^2}{2\lambda_n^2} + p_n \right], \\
 v(s,z) &= \frac{\bar{g}_{\rm tc} }{2\Omega_{\rm cz} \bar{T}_{\rm tc}} \sum_{n}\frac{d}{ds} J_{0}\left(\lambda_{n}\frac{s}{R}\right) \left[\frac{\alpha_nR}{\lambda_n}\cosh \left(\lambda_n\frac{z-z_{\rm tc}}{R} \right) - \frac{C_n(z-z_{\rm tc})^2 R^2}{2\lambda_n^2} + p_{n}\right]. 
\label{eq:vbulk1}
\end{align}
To find $\{p_{n}\}$, we need to solve for the dynamics of the convection zone, and match them onto the tachocline solution. Using the same method as the one outlined in Section \ref{sec:unstrat1}, but this time with the anelastic mass conservation equation (see Equation \ref{eq:BSconv2}), we find that 
\begin{align}
 \bar{\rho}(z) w(s,z) &= \sum_{n}\tilde{B}_{n}\sinh\left(\lambda_{n} \frac{z-H}{R}\right) J_{0}\left(\lambda_{n}\frac{s}{R}\right), \label{eq:w1an} \\
p(s,z)  &= - \sum_{n}J_{0}\left(\lambda_{n}\frac{s}{R}\right) \frac{\tilde{B}_{n}}{\tau}\frac{R}{\lambda_{n}}\cosh\left (\lambda_{n} \frac{z-H}{R}\right) , \mbox{  for } z > z_{\rm cz},  \label{eq:P1an}
\end{align}
with $\{\tilde{B}_{n}\}$ given again by matching $w$ at $z_{\rm cz}$ so that 
\begin{equation}
 \tilde{B}_{n} = \bar{\rho}(z_{\rm cz}) \frac{\dot{\Omega}_{\rm cz}}{\Omega_{\rm cz}} \frac{2\Delta}{\lambda_{n}J_{1}(\lambda_{n})\sinh \left( \lambda_{n}\frac{z_{\cz}-H}{R}\right)}. \label{eq:Bn}
\end{equation}
Matching $p$ at $z=z_{\rm cz}$, and using the fact that $\bar{\rho}_{\rm tc}$ in the tachocline is, to a first approximation, equal to $\bar{\rho}(z_{\rm cz})$ gives
\begin{equation}
 p_{n} = - \frac{\bar{T}_{\rm tc}}{\bar{\rho}_{\rm tc}\bar{g}_{\rm tc}}\frac{\tilde{B}_{n}R}{\tau\lambda_{n}}\cosh\left( \lambda_{n} \frac{H-z_{\cz}}{R}\right)  - \frac{\alpha_{n}R}{\lambda_{n}}\cosh\left(\lambda_{n}\frac{\Delta}{R}\right)+ \frac{C_n  \Delta^{2}R^{2}}{2\lambda_{n}^{2}},
\end{equation}
so that the azimuthal velocity \eqref{eq:vbulk1} leads to Equation \eqref{eq:v1}.

\section{TRANSIENT SOLUTION FOR THE SPIN-DOWN OF A NON-MAGNETIC STAR}

\subsection{Vertical eigenmodes}

In this Section we derive solutions of Equation \eqref{eq:Znm} in the form of the eigenfunctions $Z_{nm}(x)$ and their associated eigenvalues $\mu_{nm}$, where $x = (z-z_{\tc})/\Delta$. To do this, we first need to specify the various boundary conditions on $Z_{nm}$.  The boundary conditions at $z=z_{\tc}$ ($x = 0$) are
\begin{align}
 T = 0 & \Rightarrow \frac{\partial v_{n}}{\partial z} = 0 \Rightarrow  \frac{d Z_{nm}}{d x} = 0, \label{eq:bcT2}\\
 w = 0 & \Rightarrow \frac{\partial^{3} v_{n}}{\partial z^{3}} = 0 \Rightarrow  \frac{d^3 Z_{nm}}{d x^3} = 0. \label{eq:bcw2}
\end{align}
In the convection zone, by contrast, we assume that the dynamics always relax to the steady state on a very rapid timescale. Furthermore, we have shown in Section \ref{sec:steadystate} that, within the context of a Darcy friction model, one can interchangeably use the Boussinesq approximation or the more realistic anelastic approximation. Here we adopt the former for the convection zone, hence \eqref{eq:w1} and \eqref{eq:P1} hold. The boundary conditions at $z=z_{\cz}$ ($x = 1$) are then
\begin{align}
 T \mbox{ continuous} &\Rightarrow T = 0 \Rightarrow \frac{\partial v_{n}}{\partial z} = 0 \Rightarrow  \frac{d Z_{nm}}{d x} = 0,  \label{eq:T2} \\
 p \mbox{ continuous} &\Rightarrow \frac{\partial p}{\partial s} \mbox{ continuous} \Rightarrow 2\Omega_{\rm cz} v_{n} = -\frac{B_{n}R}{\tau\lambda_{n}}\cosh\left(\lambda_{n} \frac{z_{\cz}-H}{R}\right), \label{eq:P2} \\
 w \mbox{ continuous} &\Rightarrow \frac{2\Omega_{\rm cz} \kappa_{\tc}}{\bar{N}_{\tc}^{2}}\frac{\partial^{3} v_{n}}{\partial z^{3}} = B_{n}\sinh\left(\lambda_{n} \frac{z_{\cz}-H}{R}\right),  \label{eq:w2}
\end{align}
using \eqref{eq:T2}, where $\{B_n\}$ remain to be determined. Equations \eqref{eq:P2} and \eqref{eq:w2} can finally be combined to give
\begin{equation}
 Z_{nm} = \frac{R}{\tau\lambda_{n}}\frac{1}{\tanh(\lambda_{n}\frac{H-z_{\cz}}{R})} \frac{\kappa_{\tc}}{\bar{N}_{\tc}^{2}\Delta^{3}} \frac{d^{3}Z_{nm}}{dx^{3}} \equiv K_{n}\frac{d^{3}Z_{nm}}{dx^{3}} 
 \quad \mbox{at}\quad x = 1, \label{eq:bctop}
\end{equation}
which defines the constants $\{K_n\}$, and shows them to be positive. We then see that the eigenvalue problem defined by Equation \eqref{eq:Znm} and associated boundary conditions listed above is homogeneous. It can easily be shown that the operator ${\cal L}$ on the left-hand-side of Equation \eqref{eq:Znm} is self-adjoint with these boundary conditions, which implies that the vertical eigenmodes are orthogonal, with 
\begin{equation}
\int_{0}^{1}  Z_{nm}(x)  Z_{nm'}(x) dx = \delta_{mm'} \int_{0}^{1}  Z^2_{nm}(x)  dx .
\end{equation}
It can also be shown by considering the integral $\int_{0}^{1} Z_{nm}{\cal L}(Z_{nm}) dx$, suitably integrating it by parts, and applying the boundary conditions, that the eigenvalues associated with the operator ${\cal L}$ and our boundary conditions must be strictly positive, hence our choice of writing them as $\mu_{nm}^4$ in Equation \eqref{eq:Znm}. 

Since \eqref{eq:Znm} is an equation with constant coefficients, we seek solutions of the form $\text{e}^{\sigma_{nm} z}$
and find four solutions for $\sigma_{nm}$: $\pm \mu_{nm}$ and $\pm i \mu_{nm}$. Using this information, solutions of \eqref{eq:Znm} that satisfy $Z'_{nm}(0) = 0$ and $Z'''_{nm}(0) = 0$ can be written as a linear combination of $\cosh(\mu_{nm}x)$ and $\cos(\mu_{nm}x)$
Applying the boundary condition \eqref{eq:T2} we then have
\begin{equation}
 Z_{nm}(x) = \frac{\sin(\mu_{nm})}{\sinh(\mu_{nm})}\cosh(\mu_{nm}x) + \cos(\mu_{nm}x),
\end{equation}
while the $\mu_{nm}$ coefficients can be found by applying \eqref{eq:bctop}. They are the solution of 
\begin{equation}
\frac{1}{\tanh(\mu_{nm})} + \frac{1}{\tan(\mu_{nm})} = 2K_{n}\mu_{nm}^{3} .
\end{equation}

\subsection{Limit of the time-dependent solution}

We now seek to show that the solution to the time-dependent problem given in Equation (\ref{eq:Vnmsol}) tends to the quasi-steady solution derived in Section \ref{sec:steadystate} in the limit of large time for $q\leq 0$.  

Since the contribution of the initial conditions disappear as $t\rightarrow +\infty$ for $q\leq 0$, we find that
\begin{equation}
 V_{nm}(t)\rightarrow \frac{1}{\mu(t)}\int_{t_{0}}^{t}\mu(t')F_{nm}(t')dt'.
\end{equation}
For ease of notation, we write
\begin{equation}
 F_{nm}(t) = f_{n}\dot{\Omega}_{\cz}(t) = -f_{n}\frac{\Omega_{\cz}}{t_{\sd}(\Omega_{\cz})}, \; \mbox{ where } f_{n} = \frac{4R^{2}}{\lambda_n^3 J_{1}(\lambda_{n})}\frac{\int_0^1 Z_{nm}(x)dx}{\int_0^1 Z_{nm}^{2}(x)dx}, \label{eq:Fsimple}
\end{equation}
so that
\begin{equation}
 V_{nm}(t) \rightarrow \frac{f_n}{\mu(\Omega_{\cz}(t))}\int_{\Omega_{0}}^{\Omega_{\cz}(t)}  \mu(\Omega_{\cz})d\Omega_{\cz}. \label{eq:Vlim1}
\end{equation}

In the case that $q<0$, Equation (\ref{eq:theintegral}) gives
\begin{equation}
 V_{nm}(t) \rightarrow \frac{f_n}{\mu(\Omega_{\cz}(t))}\int_{\Omega_{0}}^{\Omega_{\cz}(t)}\exp\left[-\frac{1}{q}\frac{t_{\sd}(\Omega_{0})}{\tau_{nm}^{\ES}(\Omega_{0})}\left[\left(\frac{\Omega_{\cz}}{\Omega_{0}}\right)^{q}-1\right]\right] d\Omega_{\cz}.
\end{equation}
Using the method of steepest descent \citep[as in][for instance]{RileyHobsonBence}, noting that $\Omega_{\cz}<\Omega_{0}$ and that $q<0$, we find that
\begin{equation}
 V_{nm}(t) \rightarrow -\frac{f_n}{\mu(\Omega_{\cz}(t))}\exp\left[-\frac{1}{q}\frac{t_{\sd}(\Omega_{0})}{\tau_{nm}^{\ES}(\Omega_{0})}\left[\left(\frac{\Omega_{\cz}(t)}{\Omega_{0}}\right)^{q}-1\right]\right] \frac{\Omega_{0}^{q}\Omega_{\cz}^{1-q}\tau_{nm}^{\ES}(\Omega_{0})}{t_{\sd}(\Omega_{0})},
\end{equation}
which, using Equations (\ref{eq:tESnm}), (\ref{eq:tesocz}), (\ref{eq:tsdocz}) and (\ref{eq:Fsimple}), reduces to
\begin{equation}
 V_{nm}(t) \rightarrow F_{nm}(t)\tau_{nm}^{\ES}(t) = V_{nm}^{\qs}(t),
\end{equation}
where $V_{nm}^{\qs}(t)$ is defined as the solution to Equation (\ref{eq:Tnm1}) without the $dV_{nm}/dt$ term, and so by definition is the projection of the quasi-steady solution (Equation \ref{eq:v1}) onto the horizontal and vertical eigenmodes.

In the case that $q=0$, %equation (\ref{eq:theintegral}) becomes
%\begin{equation}
% \int_{t_{0}}^{t}\frac{1}{\tau_{nm}^{\ES}(t')}dt' = -\frac{t_{\sd}(\Omega_{0})}{\tau_{nm}^{\ES}(\Omega_{0})}\int_{\Omega_{0}}^{\Omega_{\cz}(t)}\frac{d\Omega_{\cz} }{\Omega_{\cz}}= -\frac{t_{\sd}(\Omega_{0})}{\tau_{nm}^{\ES}(\Omega_{0})}\ln\left(\frac{\Omega_{\cz}(t)}{\Omega_{0}}\right),
%\end{equation}
%so
 in the limit of large $t$, Equation (\ref{eq:Vlim1}) becomes
\begin{align}
 V_{nm}(t) \rightarrow& \frac{f_n}{\mu(\Omega_{\cz}(t))}\int_{\Omega_{0}}^{\Omega_{\cz}(t)}\left(\frac{\Omega_{\cz}}{\Omega_{0}}\right)^{-t_{\sd}(\Omega_{0})/\tau_{nm}^{\ES}(\Omega_{0})}d\Omega_{\cz} \nonumber \\
 &= \frac{f_{n}\Omega_{0}}{\mu(t)}\left[\frac{1}{1-\frac{t_{\sd}(\Omega_{0})}{\tau_{nm}^{\ES}(\Omega_{0})}} \left(\left(\frac{\Omega_{\cz}(t)}{\Omega_{0}}\right)^{1-t_{\sd}(\Omega_{0})/\tau_{nm}^{\ES}(\Omega_{0})}-1\right) \right] \nonumber \\
 &=  f_{n}\Omega_{0}\left[\frac{1}{1-\frac{t_{\sd}(\Omega_{0})}{\tau_{nm}^{\ES}(\Omega_{0})}} \left(\left( \frac{\Omega_{\cz}(t)}{\Omega_{0}}\right)-\left(\frac{\Omega_{\cz}(t)}{\Omega_{0}}\right)^{t_{\sd}(\Omega_{0})/\tau_{nm}^{\ES}(\Omega_{0})} \right) \right]. 
\end{align}
Since $q=0$, the ratio of $t_{\sd}$ to $\tau_{nm}^{\ES}$ is a constant, for each $n$, $m$ (see Equation \ref{eq:ratio}), and whether the transient solution tends to the quasi-steady solution depends on this ratio.  If $t_{\sd}\gg \tau_{nm}^{\ES}$ (or equivalently $t_{\sd}\gg t_{\ES}$), then
\begin{equation}
V_{nm}(t) \rightarrow f_{n}\Omega_{0} \frac{\tau_{nm}^{\ES}(\Omega_{0})}{t_{\sd}(\Omega_{0})} = F_{nm}(t)\tau_{nm}^{\ES}(t) = V_{nm}^{\qs}(t).
\end{equation}
If, on the other hand, $t_{\sd}\ll \tau_{nm}^{\ES}$, then
\begin{equation}
 V_{nm}(t)\rightarrow - f_{n}\Omega_{0}\left(\frac{\Omega_{\cz}(t)}{\Omega_{0}}\right)^{t_{\sd}(\Omega_{0})/\tau_{nm}^{\ES}(\Omega_{0})} = F_{nm}(t)t_{\sd}(t)\left(\frac{\Omega_{\cz}(t)}{\Omega_{0}}\right)^{t_{\sd}(\Omega_{0})/\tau_{nm}^{\ES}(\Omega_{0})-1} \neq V_{nm}^{\qs}(t).
\end{equation}
Hence, the system only relaxes to the quasi-steady solution when $q=0$ provided $t_{\rm ES}(t_0) \ll t_{\rm sd}(t_0)$.

\section{DERIVATION OF THE EKMAN JUMP CONDITION IN A SPINNING-DOWN FRAME}

In this Section we derive the viscous jump condition across the Ekman layer reported in Equation (\ref{eq:jump}). Assuming that the Ekman layer is sufficiently thin to always be in balance, we apply the quasi-steady approximation to the momentum equation, which now reads
\begin{equation}
 2\mathbf{\Omega}_{\rm cz}\times\mathbf{u} + \dot{\mathbf{\Omega}}_{\rm cz}\times\mathbf{r} = -\frac{1}{\bar{\rho}_{\rm tc}}\nabla p  + \nu_{\tc} \frac{\partial^{2}\mathbf{u}}{\partial z^{2}},
 \label{eq:blmom} %+ \frac{g}{T}\hat{T}\hat{\mathbf{e}}_{z}
\end{equation}
where $\nu_{\tc}$ is the local viscosity, and where we have approximated the Laplacian in the viscous term by keeping only the vertical derivatives. Combining this momentum equation with conservation of mass gives
\begin{equation}
 -\frac{\partial v}{\partial z} = \frac{\nu_{\tc}^{2}}{4\Omega_{\rm cz}^{2}}\frac{\partial^{5}v}{\partial z^{5}}.
\end{equation}
To solve this equation, we first define $\delta_{\rm E} = \sqrt{\nu_{\tc}/2\Omega_{\rm cz}}$ and introduce the boundary-layer variable $\zeta= (z-z_{\rm tc})/\delta_{\rm E}$. Hence 
\begin{equation}
 -\frac{\partial v}{\partial \zeta} = \frac{\partial^{5}v}{\partial \zeta^{5}}.
\end{equation}
By construction, the variable $\zeta$ remains of order unity within the tachopause, and rapidly tends to infinity above it, or in other words, as $z$ enters the tachocline. We therefore have $v\rightarrow v_{0}(s)$ as $\zeta \rightarrow +\infty$, where $v_{0}(s)$ is the azimuthal velocity profile near the base of the tachocline. We also have $v(s, \zeta) = s\Omega_{\rm c}$ at $\zeta =0$ assuming a no-slip boundary condition with the core. Applying these two conditions yields
\begin{equation}
 v(s,\zeta) = v_{0}(s) + e^{-\zeta/\sqrt{2}}\left[ (s \Omega_{\rm c}  - v_{0}(s)) \cos\left(\frac{\zeta}{\sqrt{2}}\right) + c(s) \sin\left(\frac{\zeta}{\sqrt{2}}\right) \right],
\end{equation}
where $c(s)$ is an integrating function that remains to be determined.

A no-slip boundary condition also applies to the radial velocity, so $u(s,\zeta) = 0$ at $\zeta = 0$. Using the azimuthal component of the momentum equation to find $u(s,\zeta)$, we have 
\begin{equation}
u(s,\zeta) = -  \frac{\dot{\Omega}_{\rm cz}}{2 \Omega_{\rm cz}} s  + e^{-\zeta/\sqrt{2}} \left[  \left( s \Omega_{\rm c} -  v_0(s)\right) \sin \left(\frac{\zeta}{\sqrt{2}}\right) - c(s)  \cos \left(\frac{\zeta}{\sqrt{2}}\right) \right] ,
\end{equation}
so 
\begin{equation} 
c(s) = -  \frac{\dot{\Omega}_{\rm cz}}{2 \Omega_{\rm cz}} s .
\end{equation}

Finally, we require that the core be impermeable. To do so, we apply mass conservation to find $\partial w(s,\zeta)/\partial \zeta$:
\begin{equation}
\frac{\partial w}{\partial \zeta}  = \delta_{\rm E} \frac{\dot{\Omega}_{\rm cz}}{\Omega_{\rm cz}} - \delta_{\rm E}  e^{-\zeta/\sqrt{2}} \left[ \frac{1}{s} \frac{\partial}{\partial s} \left( s^2 \Omega_{\rm c} - s v_0(s)\right) \sin \left(\frac{\zeta}{\sqrt{2}}\right) +  \frac{\dot{\Omega}_{\rm cz}}{ \Omega_{\rm cz}}  \cos \left(\frac{\zeta}{\sqrt{2}}\right) \right] .
\label{eq:massconsbl}
\end{equation}

Integrating Equation \eqref{eq:massconsbl} from $\zeta = 0$ upward, and requiring $w(s,\zeta) = 0$ at $\zeta = 0$ yields 
\begin{align}
w(s,\zeta) = \delta_{\rm E} \frac{\dot{\Omega}_{\rm cz}}{\Omega_{\rm cz}} \zeta & + \frac{\delta_{\rm E}}{s} \frac{\partial}{\partial s} \left( s^2 \Omega_{\rm c} - s v_0(s)\right) \left[ e^{-\zeta/\sqrt{2}} \frac{ \sin \left(\frac{\zeta}{\sqrt{2}}\right)  + \cos\left(\frac{\zeta}{\sqrt{2}}\right) }{\sqrt{2}} \right]_0^\zeta \nonumber \\ 
& - \delta_{\rm E} \frac{\dot{\Omega}_{\rm cz}}{ \Omega_{\rm cz}} \left[ e^{-\zeta/\sqrt{2}} \frac{ \sin \left(\frac{\zeta}{\sqrt{2}}\right)  - \cos\left(\frac{\zeta}{\sqrt{2}}\right) }{\sqrt{2}} \right]_0^\zeta .
\end{align}
We see that, as $\zeta \rightarrow \infty$, 
\begin{equation}
w(s,\zeta) \rightarrow \frac{\dot{\Omega}_{\rm cz}}{\Omega_{\rm cz}} \delta_{\rm E} \left( \zeta  - \frac{ 1}{\sqrt{2}} \right) - \frac{\delta_{\rm E}}{\sqrt{2} s} \frac{\partial}{\partial s} \left( s^2 \Omega_{\rm c} - s v_0(s)\right) .
\end{equation}
Matching this onto the tachocline solution (see Equation \ref{eq:appw1}) yields the jump condition (\ref{eq:jump})
\begin{equation}
w_0(s)  = \frac{\dot{\Omega}_{\rm cz}}{\Omega_{\rm cz}} \delta_{\rm E} \left( 1 - \frac{ 1}{\sqrt{2}} \right) - \frac{\delta_{\rm E}}{\sqrt{2} s} \frac{\partial}{\partial s} \left( s^2 \Omega_{\rm c} - s v_0(s)\right) .
\end{equation}

\section{DERIVATION OF THE EVOLUTION EQUATION FOR $J_{\rm core}(t)$}

In this Appendix, we derive the evolution equation for the angular momentum of the rigidly rotating core, reported in Equation (\ref{eq:Omegabstrat}).
As in Appendix A, we solve the set of governing equations separately in the convection zone and in the tachocline, and match these solutions to the boundary conditions (at the top and side-walls of the domain), to the jump condition (at the base of the tachocline at $z=z_{\tc} + \delta$), and to each other (at the radiative--convective interface at $z = z_{\rm cz}$).

In the convection zone ($z_{\cz}<z<H$), assuming a Boussinesq system (which was proved to yield the same results as in the anelastic case in Appendix A), $w$ and $p$ are given by Equations (\ref{eq:w1}) and (\ref{eq:P1}), where $\{B_{n}\}$ are integration constants that need to be determined by matching these solutions to the tachocline. 

In the tachocline ($z_{\tc}+\delta<z<z_{\cz}$), we still have $\partial w/\partial z = \dot{\Omega}_{\rm cz}/\Omega_{\rm cz}$ (see Equation (\ref{eq:u1app}) and using mass conservation). However, we can no longer directly apply the impermeability condition at $z = z_{\tc}$, since $w$ must first be matched onto the tachopause solution. Hence, we write instead that 
\begin{equation}
 w = w_{0}(s) + \frac{\dot{\Omega}_{\rm cz}}{\Omega_{\rm cz}}(z-(z_{\tc}+\delta)), \label{eq:appw1}%, \quad v = v_{0}(s), \quad  p = p_{0}(s). \label{eq:appw1}
%checked
\end{equation}
where $w_0(s)$ is an integration function, that remains to be determined. 

Requiring continuity of $w$ at the radiative-convective interface ($z=z_{\cz}$) yields
\begin{equation}
 w_{0}(s) = \sum_{n}J_{0}\left(\lambda_{n}\frac{s}{R}\right)B_{n}\sinh\left(\lambda_{n}\frac{z_{\cz}-H}{R}\right) - \frac{\dot{\Omega}_{\rm cz}}{\Omega_{\rm cz}}\Delta. 
 \label{eq:appw02}
%checked
\end{equation}
Substituting this into \eqref{eq:amw0} then gives
\begin{equation}
 \frac{dJ_{\rm core}}{dt} = - \dot{\Omega}_{\cz}I_{\tc} + \frac{16 \Omega_{\rm cz} I_{\tc}}{\Delta}\sum_{n}\frac{J_{1}\left(\lambda_{n}\right)}{\lambda_{n}^{3}} B_{n}  \sinh\left(\lambda_{n}\frac{z_{\cz}-H}{R}\right), \label{eq:appOmegab1}
%Checked
\end{equation}
where $I_{\tc}$ is the moment of inertia of the tachocline defined in Equation \eqref{eq:tachoinertia} and where the only remaining unknowns are the $\{B_n\}$. The following calculations show how to derive them. 

Solving the thermal equilibrium equation \eqref{eq:thermaleq} for $T$ in the tachocline, with $w$ given by \eqref{eq:appw1} yields
\begin{align}
 T = \sum_{n}J_{0} & \left(\lambda_{n}\frac{s}{R}\right)\left[ \alpha_{n}\sinh\left( \lambda_{n}\frac{z-(z_{\tc}+\delta)}{R}\right) + \beta_{n}\cosh\left( \lambda_{n}\frac{z-(z_{\tc}+\delta)}{R} \right)\right. \nonumber \\
 & \left. \quad - \frac{\bar N^{2}_{\tc}R^{2}\bar{T}}{\kappa_{\tc}\lambda_{n}^{2}\bar g_{\rm tc}}\left(B_{n}\sinh\left( \lambda_{n} \frac{z_{\cz}-H}{R}\right) + \frac{2(z-z_{\cz})  }{\lambda_{n}J_{1}(\lambda_{n})} \frac{\dot{\Omega}_{\cz}}{\Omega_{\cz}} \right) \right],
%Checked 
\end{align}
where $\alpha_{n}$ and $\beta_{n}$ are found by applying the following boundary conditions: $T = 0$ at $z=z_{\cz}$ and at $z=z_{\tc}+\delta$. The second of these two boundary conditions can be justified only when the tachopause is much thinner than the tachocline, and much thinner than a thermal diffusion length. This is usually the case so
\begin{align}
 \alpha_{n}\sinh \left(\lambda_{n}\frac{\Delta}{R} \right) = \frac{\bar N^{2}_{\tc}R^{2}\bar{T}_{\tc}}{\kappa_{\tc}\lambda_{n}^{2}\bar{g}_{\tc}}B_{n}\sinh\left(\lambda_{n}\frac{z_{\cz}-H}{R}\right) - \beta_{n}\cosh\left(\lambda_{n}\frac{\Delta}{R}\right) \\
 \beta_{n} = \frac{\bar N^{2}_{\tc}R^{2}\bar{T}_{\tc}}{\kappa_{\tc}\lambda_{n}^{2}\bar{g}_{\tc}}\left(B_{n}\sinh\left( \lambda_{n}\frac{z_{\cz}-H}{R}\right) - \frac{\dot{\Omega}_{\cz}}{\Omega_{\cz}}\frac{2\Delta}{\lambda_{n}J_{1}(\lambda_{n})}\right),
%Checked
\end{align}
where $\Delta = z_{\cz}-(z_{\tc}+\delta)$.  

The vertical component of the momentum equation expressed in \eqref{eq:appP1} can then be used to calculate $p$ in the tachocline, so that
\begin{align}
 \frac{1}{\bar{\rho}_{\tc}}p & = \frac{\bar{g}_{\rm tc}}{\bar{T}_{\tc}}\sum_{n}J_{0}\left(\lambda_{n}\frac{s}{R}\right)\left[\frac{\alpha_{n}R}{\lambda_{n}} \cosh\left(\lambda_{n}\frac{z-(z_{\tc}+\delta)}{R} \right)+ \frac{\beta_{n}R}{\lambda_{n}}\sinh\left( \lambda_{n}\frac{z-(z_{\tc}+\delta)}{R} \right) \right. \nonumber \\ 
 &\left.- \frac{\bar N_{\tc}^{2}R^{2}\bar{T}_{\tc}}{\kappa_{\tc}\lambda_{n}^{2}\bar{g}_{\tc}}\left((z-(z_{\tc}+\delta))B_{n}\sinh\left( \lambda_{n}\frac{z_{\cz}-H}{R}\right) + \frac{\dot{\Omega}_{\cz}}{\Omega_{\cz}}\frac{(z-z_{\cz})^{2}}{\lambda_{n}J_{1}(\lambda_{n})}\right) + P_{n}\right],
\end{align}
where the $\{P_{n}\}$ are found from matching this solution with that of the convection zone (see Equation \ref{eq:P1}) at $z=z_{\cz}$. After some algebra, we find that 
\begin{align}
 P_{n} =& B_{n} \frac{\bar{T}_{\tc}}{\bar{g}_{\tc}}\sinh\left(\lambda_{n}\frac{z_{\cz}-H}{R}\right)\left[\frac{\bar{N}_{\tc}^{2}R^{2}\Delta}{\kappa_{\tc}\lambda_{n}^{2}} - \frac{R}{\tau\lambda_{n}\tanh\left(\lambda_{n}\frac{z_{\cz}-H}{R}\right)}\right] \nonumber \\
 &- \frac{\alpha_{n}R}{\lambda_{n}}\cosh\left(\lambda_{n}\frac{\Delta}{R} \right) - \frac{\beta_{n}R}{\lambda_{n}}\sinh\left(\lambda_{n}\frac{\Delta}{R}\right).
%Checked
\end{align}

Finally, using the radial component of the momentum equation yields $v$, 
\begin{equation}
 v = \frac{1}{2 \bar{\rho}_{\tc} \Omega_{\rm cz} } \frac{\partial p}{\partial s},
%Checked
\end{equation}
which can be used to calculate $v_{0}(s) \equiv v(s,z_{\rm tc}+\delta)$. We find 
\begin{equation}
 v_{0}(s) = \frac{\bar{g}_{\tc}}{2\Omega_{\cz} \bar{T}_{\tc}}\sum_{n}\frac{dJ_{0}(\lambda_{n}\frac{s}{R})}{ds}\left[\frac{\alpha_{n}R}{\lambda_{n}}- \frac{\bar N^{2}_{\tc}R^{2}\bar{T}_{\tc}}{\kappa_{\tc}\lambda_{n}^{2}\bar{g}_{\tc}}\frac{\dot{\Omega}_{\cz}}{\Omega_{\cz}}\frac{\Delta^{2}}{\lambda_{n}J_{1}(\lambda_{n})} + P_{n}\right]. \label{eq:appv02}
%Checked
\end{equation}
Substituting \eqref{eq:appv02} and \eqref{eq:appw02} into the jump condition \eqref{eq:jumpgen} finally provides an equation for the $\{B_n\}$, which, after significant algebra, can be cast in the form
\begin{equation}
 B_{n}\sinh\left(\lambda_{n}\frac{z_{\cz}-H}{R}\right) = \frac{2}{\lambda_{n}J_{1}(\lambda_{n})}\frac{a_{n}}{b_{n}}, \label{eq:appB3}
\end{equation}
where $a_{n}$ and $b_{n}$ recover the formulae given by Equations \eqref{eq:an} and \eqref{eq:bn1} in the limit $\delta \ll \Delta$. Using \eqref{eq:appB3} in \eqref{eq:appOmegab1} then leads to \eqref{eq:Omegabstrat}.

Finally, if we want to calculate $w_0(s)$, we simply substitute $B_n$ back into \eqref{eq:appw02} to get: 
\begin{equation}
 w_{0}(s) = \sum_{n}J_{0}\left(\lambda_{n}\frac{s}{R}\right)\left[ \frac{2}{\lambda_{n}J_{1}(\lambda_{n})}\frac{a_{n}}{b_{n}} \right]  - \frac{\dot{\Omega}_{\rm cz}}{\Omega_{\rm cz}} \Delta. \label{eq:appw03}
%checked
\end{equation}

\section{THE TACHOCLINE THICKNESS}

In GM98 and \citet{Woodal11}, the radial mass flux downwelling into the tachocline is caused by the gyroscopic pumping associated with the turbulent torques that permanently drive the observed latitudinal shear in the convection zone, rather than by the spin-down torque. For this reason, their respective estimates of the tachocline thickness $\Delta$ as a function of other stellar parameters do not directly apply here. Nevertheless, we can apply a similar method to the one they use to infer $\Delta$ in the spin-down case. We now proceed to describe this method and its limitations, first applied to the steady-state solar case, and then applied to our own spin-down problem. 

In the solar case studied by GM98 and \citet{Woodal11}, the thickness of the tachocline $\Delta$ is obtained by matching the vertical mass flux through the tachocline to the vertical mass flux through the tachopause. The latter is estimated by assuming advection-diffusion balance of the flux of horizontal magnetic field across the tachopause (whose thickness is $\delta$):
\begin{equation}
w_{\rm tc} \simeq \frac{\eta_{\rm tc}}{\delta} \mbox{  , }
\label{eq:wtp98}
\end{equation}
while the former is obtained, as in Section \ref{sec:steadystate}, by considering thermal-wind balance and thermal equilibrium across the tachocline, which yields
\begin{equation}
w_{\rm tc} \simeq 2 \chi \frac{ \Omega_{\rm cz}^2}{\bar N_{\rm tc}^2}  \frac{r_{\rm cz}^2\kappa_{\rm tc} }{\Delta^3}  \mbox{  , }
\label{eq:wtc98}
\end{equation}
where $\chi \Omega_{\rm cz}$ is an estimate of the amplitude of the {\it latitudinal} differential rotation in the convection zone. In the Sun, $\chi \sim 0.1$. Combining \eqref{eq:wtp98} and \eqref{eq:wtc98} yields a relationship between $\delta$ and $\Delta$: 
\begin{equation}
\Delta^3  \simeq 2 \chi \frac{ \Omega_{\rm cz}^2}{\bar N_{\rm tc}^2}  \frac{\kappa_{\rm tc} }{\eta_{\rm tc}}   r_{\rm cz}^2 \delta \mbox{   .}
\label{eq:tachoGM98}
\end{equation}
This equation is quite robust, since it relies on basic balances that are not easily upset by additional dynamics, and has been verified against numerical simulations \citep{AAal13}. 

However, in order to obtain $\Delta$ as a function of known stellar parameters and independently of $\delta$, one must make further assumptions concerning the nature and structure of the tachopause. This final step, unfortunately, is quite model-dependent. GM98 and \citet{Woodal11} propose different scalings for $\delta$ as a function of $B_0$ and $\Omega_{\rm cz}$ for instance. Both assume that the tachopause is laminar, but disagree on its thermal properties, leading to 
\begin{equation}
\frac{\Delta}{r_{\rm cz}} \propto \left( \frac{|B_0|}{\sqrt{\bar \rho_{\rm tc} }} \frac{r_{\rm cz}}{\sqrt{\kappa_{\rm tc} \eta_{\rm tc}}}  \right)^{-1/9}  \left(\frac{\kappa_{\rm tc}}{\eta_{\rm tc}} \right)^{1/3} \left(\frac{\Omega_{\rm cz}}{\bar N_{\rm tc}} \right)^{7/9}  
\label{eq:GMeq}
%Checked ok.
\end{equation}
for GM98 (using their Equation 6 as a definition of $\delta$), and 
\begin{equation}
\frac{\Delta}{r_{\rm cz}}  \propto  \left(\frac{ \Omega_{\rm cz}}{\bar N_{\rm tc}}\right)^{2/3}  \left( \frac{\kappa_{\rm tc} }{\eta_{\rm tc}} \right)^{1/3} \left(\frac{ \bar{\rho}_{\rm tc} \eta_{\rm tc} \Omega_{\rm cz} }{B_0^2} \right)^{1/6} \mbox{   ,}
\label{eq:Waleq}
\end{equation}
for \citet{Woodal11} (using Equation \ref{eq:delta} as a definition of $\delta$). Moreover, neither of these scalings apply if turbulence also plays a role in the angular-momentum transport balance across the tachopause (which cannot a priori be ruled out). In short, while Equation \eqref{eq:tachoGM98} robustly relates $\Delta$ to $\delta$ in the solar steady-state model, it is not sufficient on its own to derive a reliable estimate of how $\Delta$ varies with stellar parameters without further constraints on the tachopause structure. The latter can only be obtained in direct numerical simulations of the system, which are not yet available at this point. Not surprisingly, we find that the same problem affects the determination of $\Delta$ in a spin-down model. 

In Section \ref{sec:blcase}, we found that $w$ in the tachocline is neither constant with distance from the rotation axis nor with depth, so that a direct application of Equation \eqref{eq:wtp98} is not possible. Nevertheless, one can require advection-diffusion balance {\it on average} in the tachopause by setting (within the context of the cylindrical model used throughout this work)
\begin{equation}
\left| \frac{2\pi }{\pi R^2} \int_0^R w_0(s) s ds  \right| = \frac{\eta_{\rm tc}}{  \delta}  \mbox{  .}
\end{equation} 
Substituting $w_0(s)$ given in \eqref{eq:appw03} into this equation and evaluating the integral yields: 
\begin{equation}
\left| \sum_{n} \frac{ 4}{\lambda^2_n} \frac{a_{n}}{b_{n}}  -  \frac{\dot{\Omega}_{\rm cz}}{\Omega_{\rm cz}} \Delta  \right| =  \left| \sum_{n}  \frac{ 4}{\lambda^2_n}  \left( \frac{a_{n}}{b_{n}}  - \frac{8}{\lambda_n^2}  \frac{\dot{\Omega}_{\rm cz}}{\Omega_{\rm cz}} \Delta  \right) \right| =   \frac{\eta_{\rm tc}}{  \delta}  \mbox{ ,}
\end{equation}
where we have once again used the property $\sum_n (32/\lambda_n^4) = 1$. Dropping all but the first term in this sum (as in Section \ref{sec:globalsol}), and substituting $a_1$ and $b_1$ given in \eqref{eq:an_simple} yields 
\begin{equation}
 \frac{ 4}{\lambda^2_1}  \left| \frac{\frac{\dot{\Omega}_{\rm cz}}{\Omega_{\rm cz}} \Delta  - 2C\delta \Omega_{\rm c} + \frac{C\lambda^2_{1} }{24} \dot{\Omega}_{\cz}t_{\rm ES}(\Omega_{\cz})  \delta}{1+ \frac{C\lambda_{1}^{2}}{12}\Omega_{\cz}t_{\rm ES}(\Omega_{\cz})\frac{\delta}{\Delta}}  - \frac{8}{\lambda_1^2}  \frac{\dot{\Omega}_{\rm cz}}{\Omega_{\rm cz}} \Delta \right|  =   \frac{\eta_{\rm tc}}{  \delta}  \mbox{  .}
 \label{eq:temptacheq}
\end{equation}

While somewhat obscure, this expression is indeed the equivalent of \eqref{eq:tachoGM98} in the spin-down case. To see this, note that if $\dot{\Omega}_{\rm cz} = 0$, then 
\begin{equation}
 \frac{ 4}{\lambda^2_1}  \left[ \frac{  2C\delta \Omega_{\rm c} }{1+ \frac{C\lambda_{1}^{2}}{12}\Omega_{\cz}t_{\rm ES}(\Omega_{\cz})\frac{\delta}{\Delta}}   \right]  =   \frac{\eta_{\rm tc}}{  \delta}  \mbox{  ,}
\end{equation}
which recovers the same scalings as those of GM98 and \citet{Woodal11} (see Equation \ref{eq:tachoGM98}), as long as $\Omega_{\rm c}$ is re-interpreted as the latitudinal shear driving the large-scale flows $\chi \Omega_{\rm cz}$, and the second term in the denominator is much larger than 1 (which is usually the case unless $\delta / \Delta$ is unrealistically small). 

For young solar-type stars, however, spin-down dominates the dynamics of the system. Equation \eqref{eq:temptacheq} then takes a different form during the initial transient and the later quasi-steady state phases. During the transient, $\Omega_{\rm c} = \Omega_{\rm core} - \Omega_{\rm cz} \simeq - \Omega_{\rm cz}$, as discussed in Section \ref{sec:transient2}. In that case, and using the fact that $t_{\rm ES}(\Omega_{\rm cz}) \ll t_{\rm sd}(\Omega_{\rm cz})$ for our model to apply anyway, Equation \eqref{eq:temptacheq} can be approximated as 
\begin{equation} 
- c_1 \frac{\dot{\Omega}_{\rm cz}}{\Omega_{\rm cz}} \Delta  \simeq  \frac{\eta_{\rm tc}}{  \delta}  \mbox{  ,}
\label{eq:temp2}
\end{equation}
where $c_1$ is a constant of order unity. Using this equation in conjunction with a given tachopause model (as in Equation \ref{eq:delta} or as in Equation 6 of GM98) does yield an estimate for $\Delta$ as a function of stellar parameters. However, that estimate is very sensitive to any model uncertainty on the nature and structure of the tachopause, as discussed above in the context of the Sun.  

%Using \eqref{eq:delta} for instance, we get  
%\begin{equation}
%\frac{\Delta}{r_{\rm cz}}  \simeq  \frac{1}{c} \left| \frac{\Omega^2_{\rm cz}}{\dot{\Omega}_{\rm cz}}\right| \frac{ \eta_{\rm tc}}{r^2_{\rm cz} \Omega_{\rm cz}}   \sqrt{\frac{B_0^2}{2\pi\bar{\rho}_{\rm tc} \eta_{\rm tc} \Omega_{\rm cz} }} \mbox{  ,}
%\end{equation}
%while using the GM98 tachopause structure instead would yield 
%\begin{equation}
%\frac{\Delta}{r_{\rm cz}}  \simeq   \frac{1}{c} \left| \frac{\Omega^2_{\rm cz}}{\dot{\Omega}_{\rm cz}}\right|   \frac{\eta_{\rm tc}}{ r^2_{\rm cz}  \Omega_{\rm cz}}  \left(\frac{B_0^2 r_{\rm cz}^2 \bar N_{\rm tc }^2      }{8\pi \rho_{\rm tc} \eta_{\rm tc} \kappa_{\rm tc} \Omega_{\rm cz}^2 }  \right)^{1/6}   \mbox{  ,}
%\end{equation}
%All in all,  we see that $\Delta$ in the transient phase of spin-down is likely to have the following functional form: 
%\begin{equation}
%\frac{\Delta}{r_{\rm cz}} \propto \left| \frac{\Omega^2_{\rm cz}}{\dot{\Omega}_{\rm cz}}\right|  \frac{\eta_{\rm tc}}{ r^2_{\rm cz}  \Omega_{\rm cz}}  \left( \frac{\Omega_{\rm cz}}{\Omega_0}\right)^{\beta_1} \left( \frac{r_{\rm cz}}{R_\star}\right)^{\beta_2} \left( \frac{B_0^2}{2\pi\bar{\rho}_{\rm tc} \eta_{\rm tc} \Omega_{\rm cz} } \right)^{\beta_3}
%\end{equation}
%where the first two terms are explicitly contained in \eqref{eq:temp2}, and should therefore appear as such here as well, and where  the other two terms contain any remaining dependence on $B_0$, $\Omega_{\rm cz}$ and the thermodynamical properties of the star around $r_{\rm cz}$, via $\beta_1$, $\beta_2$ and $\beta_3$. 

In the quasi-steady phase, the problem is even worse. Since the core-envelope lag $\Omega_{\rm c} = \Omega_{\rm core} - \Omega_{\rm cz} $ is now given by Equation \eqref{eq:celag1}, Equation \eqref{eq:temptacheq} becomes
\begin{equation}
- c_2 \frac{\dot{\Omega}_{\rm cz}}{\Omega_{\rm cz}} \Delta \frac{I_{\rm core}}{I_{\rm tc}} \simeq  \frac{\eta_{\rm tc}}{  \delta}  \mbox{  ,}
\label{eq:temp10}
\end{equation}
where $c_2$ is also a constant of order unity. We then see that the dependence on $\Delta$ on the left-hand side vanishes altogether (since $I_{\rm tc} \propto \Delta$), which implies that this method {\it cannot} be used to constrain $\Delta$ directly. Instead, Equation \eqref{eq:temp10} provides a second constrain on $\delta$ -- the first one being given by various balances within the tachopause, leading for instance to Equation \eqref{eq:delta}, or  Equation (6) of GM98 -- and therefore defines the {\it position} of the tachopause to be the radius where the amplitude of the primordial field $B_0$ is such that the two definitions of $\delta$ coincide. Since the tachopause lies by construction at the bottom of the tachocline, one could in principle use this method to determine $\Delta$ if the radial variation of $B_0$ is known. However, any estimate of $\Delta$ based on this method will, once again, be uncomfortably model-dependent. 

In summary, we conclude that theory alone cannot robustly predict how the thickness of the tachocline varies with stellar parameters. Any estimate of $\Delta$ made by applying mass continuity across the interface between the tachocline and the tachopause, as in GM98, relies sensitively on the assumed structure of the tachopause, which is itself sensitively dependent on the nature and balance of forces, thermal energy transport and angular-momentum transport within. For this reason, until such a time where the tachopause is better understood (through direct numerical simulations for instance), we advocate the use of a more general tachocline law, as in Equation \eqref{eq:deltaparam} for instance. 

%%%%%%%%%%%%%%%%%%%%%%%%%%%%%%%%%%%%%%%%%%%%%%%%%%%%%%%%%%%%%%%%%%%%
\bibliography{StellarInterior2}

\end{document}